\documentclass{aa}  

\usepackage{graphicx}
\usepackage{txfonts}
\usepackage{lscape}             
\usepackage{placeins}           
                                
\usepackage{newtxtext}
\usepackage[varvw]{newtxmath} 

\usepackage[T1]{fontenc}
\usepackage{amsmath}
\usepackage{tabularx}

\usepackage[colorlinks=true,linkcolor=blue,urlcolor=blue,citecolor=blue,breaklinks=true]{hyperref}

\begin{document}

   \title{Magnetohydrodynamics of charged interstellar dust} \subtitle{Multifluid models and study of the linear modes}

   \author{G. Verrier$^{1}$
   \and P. Hennebelle$^{1}$ \and U. Lebreuilly$^{1}$ \and V. Vallucci-Goy$^{1,2}$
}
   \institute{
   $^{1}$Universit\'{e} Paris-Saclay, Universit\'{e} Paris Cité, CEA, CNRS, AIM, 91191, Gif-sur-Yvette, France
\email{gabriel.verrier@cea.fr}    \\
$^{2}$Institute of Space Sciences (ICE), CSIC, Campus UAB, Barcelona, Spain
}

   \date{Received January 27, 2026; accepted August 30, 2026}

  \abstract{
  Interstellar grains play key roles in star and planet formation, including the coupling of the gas to the magnetic field during the protostellar collapse. These roles depend on the local grain size distribution, which requires a multifluid treatment of charged dust.
  }{
  We aim to understand the fundamental physics of the dynamics of a dust distribution in interaction with the gas and the magnetic field. In particular, the purpose is to characterize the (de)coupling conditions of these different components.
  }{We provide a multifluid model of charged dust which accounts for the inertia of the grains. A chemical network is used to simulate the charge equilibrium in collapsing protostellar cores. We compute the Alfvén modes and the magnetosonic modes to understand the coupling regimes between the gas, the dust fluids, and the magnetic field. We also analyze and compare to the predictions of existing models, that are the neutral dust multifluid and the standard non-ideal magnetohydrodynamics.}{The charged multifluid model agrees with non-ideal magnetohydrodynamics on
  the larger scales of a collapsing dense core and the forming disk, while we successfully extend to new regimes where the inertia of dust grains matters. 
  We found that high charge-to-mass dust grains carry the propagation of magnetohydrodynamical waves in protostellar envelopes. We provide analytical expressions of the speed of these waves depending on the dust distribution. The magnetocompressive perturbations lead to local dust-to-gas ratio variations at au scales.
  }{
  A theoretical understanding of the dynamics of a charged dust distribution is provided in the linear regime. The closed set of 
  magnetohydrodynamics 
  equations can be implemented in numerical codes to
  explore nonlinear effects during the protostellar collapse such as 
  turbulence, angular momentum transport and magnetic dust clumping.
  }
   \keywords{ Magnetohydrodynamics (MHD);
    Plasmas;
    Waves;
    ISM: dust, extinction; 
    Stars: formation;
    methods: analytical; }
      \authorrunning{Verrier, G., Hennebelle, P., Lebreuilly, U. and Vallucci-Goy, V.}
  \maketitle

\nolinenumbers 
  
\section{Introduction}

Magnetic fields play a fundamental role in star formation, in particular during the protostellar collapse \citep{1999ASIC..540..305M,2007ARA&A..45..565M,2022FrASS...9.9223M,2023ASPC..534..317T}. As matter collapses, the magnetic field lines are bent and twisted, resulting in the propagation of Alfvén waves from the core to its surroundings. Angular momentum is transported away, resulting in a magnetic braking which affects the formation of a protostar and its circumstellar disk. This process relies on a theoretical description of the interaction between the magnetic field and the matter, known as magnetohydrodynamics (MHD). 
Assuming that the magnetic flux is frozen in the collapsing matter (i.e., ideal MHD) produces a very efficient braking that prevents the formation of a disk. On the other hand, ignoring the magnetic field (i.e.,  hydrodynamics) produces very large disks that are inconsistent with the observations of compact young disks \citep{Lebreuilly2021,2024A&A...682A..30L,2025MNRAS.543.3321M,2019A&A...621A..76M}. Intermediate regimes of coupling between the matter and the magnetic field are usually achieved by including non-ideal terms in the induction of the magnetic field, controlled by magnetic resistivities.  Such a non-ideal description is particularly relevant because molecular clouds, and in particular the densest regions, are very weakly ionized \citep{1998ApJ...499..234C,2014A&A...571A..33P}.

Even though dust grains represent about one percent of the total mass, because they collect charges at their surface, they become a significant intermediary to couple the gas (i.e., the bulk mass) to the magnetic field during the protostellar collapse \citep{2016A&A...592A..18M,2016MNRAS.460.2050Z}. However, the evolution of the dust and its properties during the protostellar phase is poorly known. One key aspect is the grain size distribution, controlling the available surface for grain charging. While dust grains are typically submicron in the diffuse interstellar medium \citep{1977MRN,2011A&A...525A.103C}, they reach millimeter and centimeter sizes in protoplanetary disks (see \citet{2023ASPC..534..717D}, and references therein). The population of small grains, depending on the dust growth processes at stake, deeply affects the magnetic resistivities during the protostellar collapse \citep{2023PASJ...75..835T,2024A&A...690A..23V}. 
 
Several authors have investigated the dynamics of a distribution of neutral grains using a dust multifluid approach both theoretically and numerically (see, e.g., \citet{2017ascl.soft09002P,2019ApJS..241...25B} for numerical codes). Multifluid models account for the inertia of grains and therefore handle a large range of dynamical coupling regimes between the grains and the gas. The differential dynamics is of primary importance in triggering resonant drag instabilities \citep{2018MNRAS.477.5011S}, which are possible paths to the formation of planetesimals in protoplanetary disks. On the other hand, at the scales of the interstellar medium, even tough the charging of the grains and their inertia are accounted for, their backreaction on the induction of the magnetic field is generally neglected in favor of an ideal coupling of the latter to the gas (see e.g. \citet{2021MNRAS.502.2630S} for the study of the propagation of cosmic rays and \citet{2025MNRAS.542.1011M} for the grain drift velocities in turbulence). However, this becomes invalid for dense clouds and collapsing cores in which dust grains represent important charge carriers. Accounting for the inertia of charged grains with their backreaction on the magnetic field gives access to small-scale physics such as the chemical separation induced by shocks in dense clouds \citep{2002ApJ...567..947C,2007A&A...476..263G}, the propagation of Alfvén waves \citep{1987ApJ...314..341P,2023A&A...674A.149H} and magnetosonic waves \citep{2004ApJ...610..781C}, and dust clumping in protostellar envelopes \citep{2025A&A...704A.142V}. In addition to these works, sophisticated multifluid models have also been designed and investigated in the context of solar partially ionized plasmas (see \citet{2024mpsp.book..203S}, and references therein). The closure of the MHD system by the derivation of an Ohm law often relies on an assumption on the dynamics of the charged species; most of the time, the inertia of the electrons is neglected. It can be convenient or necessary to neglect the inertia of ions and electrons together, because of the dynamical scales at stake, and because chemistry on grains can dominate their mass budget over hydrodynamical advection (see Fig. 5 in \citet{2020A&A...643A..17G}). In this case, chemical networks compute the local chemical equilibrium (charges and abundances) depending on the grain size distribution \citep{2016A&A...592A..18M,2016MNRAS.460.2050Z}. They are implemented in non-ideal MHD codes to obtain the standard magnetic resistivities but their use could be extended to simulations of a multifluid of charged dust. In that respect, we propose in this paper to bridge the gap between the standard resistivity-based non-ideal MHD approach \citep{1999MNRAS.303..239W}, which accounts for an arbitrary distribution of charges that couple to the gas and the magnetic field, and the dust multifluid approach.

The paper is organized as follows. 
We present the novel multifluid MHD model of charged dust in Sect. \ref{sec:MHD models} and the method to study the modes of the system in Sect. \ref{sec:methods}. 
Section \ref{sec:multifluid_physics} presents the physics of the dust multifluid. More precisely, Sect. \ref{sec:multidust_hydro_waves} describes the coupling conditions of dust fluids to the gas in the absence of a magnetic field. Section \ref{sec:multidust_plasma_waves} presents the magnetocompressive mode carried by multiple dust fluids. Section \ref{sec:single_multifluid_MHD_ms_modes_analysis} details the coupling regimes of the dust with the gas and the magnetic field. We end up with the full multifluid description of the propagation of the Alfvén waves and the magnetosonic waves in Sect. \ref{sec:multifluid_MHD_modes}. In Sect. \ref{sec:astrophysical_discussions}, we discuss the local dust enrichment in protostellar envelopes due to the propagation of magnetocompressive waves (Sect.  \ref{sec:single_multifluid_MHD_ms_modes_experiment}) and we discuss the role of grain inertia on the magnetic braking during the disk formation due to the propagation of Alfvén waves (Sect. \ref{sec:whistler_magnetic_braking}). Section \ref{sec:conclusion} presents our conclusions. Most of the physical results are based on derivations that can be found in the appendix.

\section{Dusty magnetohydrodynamics models} \label{sec:MHD models}

\subsection{Basic equations}

We present the equations of the dynamics of the neutral gas, the charged dust, the ions, the electrons, and the magnetic field. Charged species are characterized by their mass and number of charges, denoted by $(m_k, Z_k)$. We define two groups of charged species, whether their inertia is accounted for or not. This depends on the dynamical scales of interest. If the gyration time and the collision time of a charged species is short compared to the dynamical scales of interest, we neglect the inertia of this species (typical values in Sect. \ref{app:physical_setup}). These species that we refer as 'light' species are denoted by $l \in L$. If not, they are fully treated as fluids denoted by $d \in I$. In the practical case of this paper, we consider the inertia of dust grains $d \in I=\{1,..., \mathcal{N}\}$, where $\mathcal{N}$ is the number of fluids representing a dust distribution, and we neglect the inertia of ions and electrons $l \in L =\{i,e\}$. Therefore, $C= L \cup I$ is the complete set of charged species, indexed by $k \in C$. The basic equations are

\begin{equation}
    \partial_t \rho_g + \nabla \cdot (\rho_g \mathbf{V}_g)=0,
\label{eq:continuity_gas}
\end{equation}
\begin{equation}
    \partial_t \rho_d+ \nabla \cdot(\rho_d \mathbf{V}_d)= S_d,
\label{eq:continuity_dust}
\end{equation}

\begin{equation}
    \partial_t (\rho_g \mathbf{V}_g) + \nabla \cdot (  \rho_g \mathbf{V}_g \mathbf{V}_g + P_g \mathbf{1}
    ) = \sum_{k \in C}  \mathbf{f}_{k \to g},
    \label{eq:gas_momentum_conservation}
\end{equation}
\begin{equation}
    \partial_t (\rho_d \mathbf{V}_d) + \nabla \cdot \left(  \rho_d \mathbf{V}_d \mathbf{V}_d \right) = \mathbf{f}_{g \to d} + n_dZ_de \left( \mathbf{E} + \frac{\mathbf{V}_d}{c} \times \mathbf{B} \right),
    \label{eq:dust_momentum_conservation}
\end{equation}
\begin{equation}
    0 = \mathbf{f}_{g \to l} + n_l Z_l e \left( \mathbf{E} + \frac{\mathbf{V}_l}{c} \times \mathbf{B} \right),
    \label{eq:light_momentum_conservation}
\end{equation}

\begin{equation}
    \partial_t \mathbf{B} = -c \nabla \times \mathbf{E},
    \label{eq:mawell_faraday}
\end{equation}
\begin{equation}
    \mathbf{J} = \frac{c }{4 \pi}\nabla \times \mathbf{B}.
    \label{eq:mawell_ampere}
\end{equation}

Equations \eqref{eq:continuity_gas} and \eqref{eq:continuity_dust} are respectively the equations of mass conservation of the gas and of the dust. Mass exchanges between charged species are modeled by the source terms $S_k$. They can be, for instance, dust growth and chemistry on the surfaces of the grains. In the present work, we use the chemical network of \citet{2021A&A...649A..50M} to compute the abundances of ions and electrons $n_i,n_e$ and the mean number of charges carried by the dust grains $Z_d$ for each dust species, complying with the electroneutrality condition $\sum_{k \in C} n_kZ_k=0$. A key parameter in star and planet formation is the dust-to-gas ratio $\theta_d=\rho_d/\rho_g$, or equivalently the dust ratio $\epsilon_d=\rho_d/\rho$, where $\rho$ is the total density. By defining the total dust mass as $\varrho_d=\sum_{d \in I} \rho_d$ then $\rho=\rho_g+\varrho_d$. 

Equations \eqref{eq:gas_momentum_conservation} and  \eqref{eq:dust_momentum_conservation} are the equations of momentum conservations, and Eq. \eqref{eq:light_momentum_conservation} provides the velocity of the ions and the electrons based on the force balance. The gas is supported by the thermal pressure $P_g$. Each charged species is coupled to the gas via an individual drag force $\mathbf{f}_{g \to k}$. They back-react on the gas such that $\mathbf{f}_{g \to k} = - \mathbf{f}_{k \to g}  $ with 
\begin{equation}
    \mathbf{f}_{g \to k} = \rho_g \rho_k \gamma_k ( \mathbf{V}_g - \mathbf{V}_k),
\end{equation}
where $\gamma_k$ is the drag coefficient, which can be written as a function of the collision rate $\nu_k$ or the stopping time $t_{\mathrm{s},k}$, often used for dust grains,
\begin{equation}
    \rho_g \gamma_k = \nu_k = 1/t_{\mathrm{s},k}.
\end{equation}
Finally, charged species are coupled to the magnetic field by individual Lorentz forces. It is necessary to define the gyration frequency $\omega_k = (Z_k e B)/(m_k c)$ and the Hall factor 
$\Gamma_k=\omega_k/\nu_k$. 

The evolution of the magnetic field $\mathbf{B}$ is given by the Maxwell-Faraday equation in Eq. \eqref{eq:mawell_faraday}. It can be written as an induction equation by deriving an Ohm law between the electric field $\mathbf{E}$ and the electric current $\mathbf{J}=\sum_{k \in C} n_kZ_ke \mathbf{V}_k$. By neglecting the displacement current in the Maxwell-Ampere equation, which leads to filtering high frequency modes, the set of equations can be closed using Eq. \eqref{eq:mawell_ampere}. A closed set of equations constitutes a magnetohydrodynamics (MHD) model. We present three models hereafter, with the more complete one in Sect. \ref{sec:multifluid_MHD_equations}.

We could include other forces within the momentum equations that are significant during the protostellar collapse such as gravity, and inelastic collisions of the ions and the electrons with the grains when grains of same sizes but different charges are split into distinct fluids. We refer to \citet{1993ApJ...418..774C,1996ASIC..481..505M} for more details\footnote{This model requires to set three fluids of charge $Z_d \in \{-1,0,1\}$ in Eq. \eqref{eq:dust_momentum_conservation} for each grain size. In this paper, we consider that, for any grain, charge fluctuations at its surface occur on timescales shorter than the gyration time and the dynamical timescale, as in \citet{2020A&A...643A..17G}. Therefore, by filtering the charging timescale, we model grains of a given size by a unique dust fluid carrying a time-averaged 
mean charge.}. We do not include them in this paper to focus on the physics induced by the modeling of the interactions between the charges through the gas and the magnetic field.

\subsection{Standard non-ideal MHD} \label{sec:non_ideal_MHD_equations}

Standard non-ideal MHD equations describe the dynamics a single magnetized fluid which is non-ideally coupled to the magnetic field. The assumptions and the limits of this approach are discussed in \citet{2025OJAp....8E.141H}. The standard non-ideal MHD equations can be derived using the method of \citet{1999MNRAS.303..239W}, by neglecting the inertia of all charged species (including dust). Regarding our notations, $L=C$ and $I=\varnothing$ . The gas, carrying the bulk mass, is magnetized through collisions. It feels the Laplace force
\begin{equation}
    \frac{\mathbf{J} \times \mathbf{B}}{c} = \sum_{k \in C}  \mathbf{f}_{k \to g}.
    \label{eq:NIMHD_Laplace_force}
\end{equation}

We use this name to distinguish from individual Lorentz forces felt by the charged species, and to emphasize that the gas acts as an electric conductor through collective friction forces of embedded charges.

The generalized Ohm law of standard non-ideal MHD is
\begin{equation}
    \mathbf{E}_g = \eta_{\mathrm{O}}^C \mathbf{J} + \eta_{\mathrm{H}}^C \mathbf{J} \times \mathbf{b} + \eta_{\mathrm{AD}}^C \mathbf{b} \times ( \mathbf{J} \times \mathbf{b}),
    \label{eq:Ohm_NIMHD}
\end{equation}
where $\mathbf{E}_g = \mathbf{E} + \frac{\mathbf{V}_g}{c} \times \mathbf{B}  $ includes the non-ideal coupling terms, decomposed with respect to the direction of the magnetic field $\mathbf{b} = \mathbf{B} /B$ and the electric current $\mathbf{J}$. 
The non-ideal terms model velocity drifts of charged species relative to the gas. It includes Ohmic diffusion, Hall effect, and ambipolar drift with respective magnetic resistivities $\eta_{\mathrm{O}}^C$, $\eta_{\mathrm{H}}^C$ and $\eta_{\mathrm{AD}}^C$. They are related to the conductivities (and thus to the abundances and the charges) of all charged species (indicated by $C$) that are the dust species, the ions and the electrons, as defined in \citet{1999MNRAS.303..239W}. We detail in the next section (Sect. \ref{sec:multifluid_MHD_equations}).

\subsection{Multifluid magnetohydrodynamics} \label{sec:multifluid_MHD_equations}

\begin{figure}
    \centering
    \includegraphics[width=0.49\textwidth]{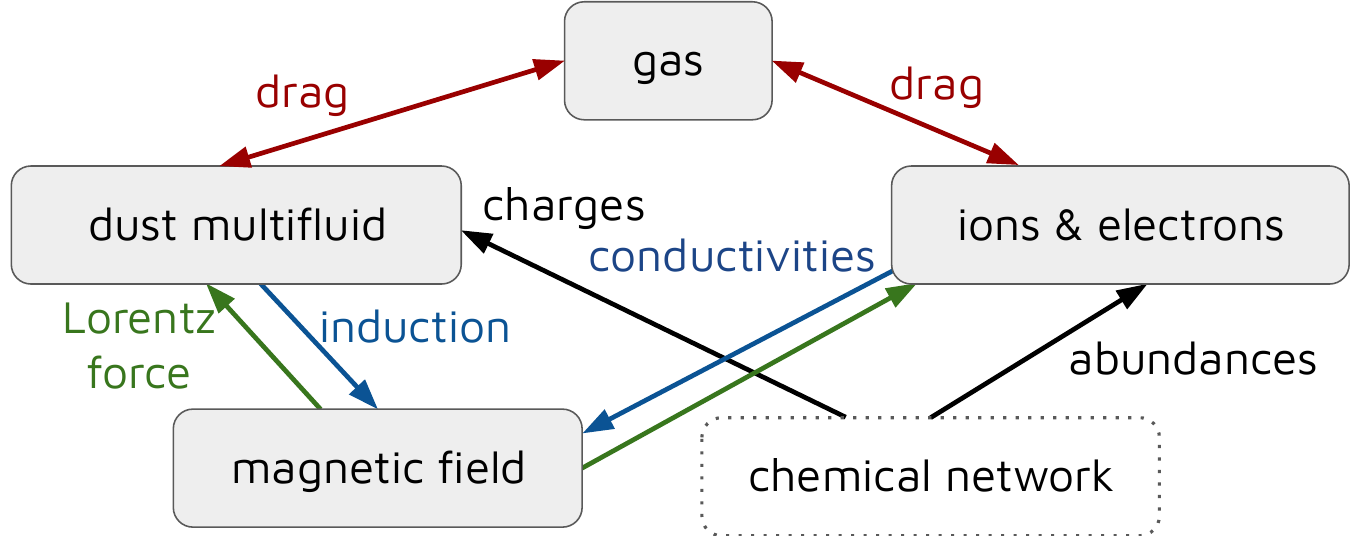}

    \caption{Diagram of the interactions between the gas, the dust multifluid and the magnetic field following the multifluid MHD model (Sect. \ref{sec:multifluid_MHD_equations}).}
    \label{fig:multifluid_MHD_interactions}
\end{figure}

We present novel MHD equations including a charged dust multifluid from Eqs. \eqref{eq:continuity_gas}-\eqref{eq:mawell_ampere}. They extend the neutral dust multifluid equations, that model a dust distribution and account for the dust inertia, and the standard non-ideal MHD equations, that couple the hydrodynamics with the magnetic field. The derivation is in Appendix \ref{app:multifluid_MHD_derivation}. The generalized Ohm law is
\begin{equation}
    \mathbf{E}_g = \eta_{\mathrm{O}}^L \Delta \mathbf{J}_L + \eta_{\mathrm{H}}^L \Delta \mathbf{J}_L \times \mathbf{b} + \eta_{\mathrm{AD}}^L \mathbf{b} \times (\Delta \mathbf{J}_L \times \mathbf{b}),
    \label{eq:Ohm_multifluid_MHD}
\end{equation}
with
\begin{equation}
    \Delta \mathbf{J}_L= \sum_{l \in L} n_lZ_le ( \mathbf{V}_l- \mathbf{V}_g).
\end{equation}

This Ohm law only involves the dynamics of the light species (ions and electrons, $l \in L$) relative to the gas. Indeed, the magnetic resistivities are here defined as
  \begin{equation}
    \eta_{\mathrm{O}}^L=\frac{1}{\sigma_{\mathrm{O}}^L}, \eta_{\mathrm{H}}^L=\frac{1}{\sigma_\perp^L} \frac{\sigma_{\mathrm{H}}^L}{\sigma_\perp^L}, \eta_{\mathrm{AD}}^L=\frac{1}{\sigma_\perp^L} \frac{\sigma_{\mathrm{P}}^L}{\sigma_\perp^L} - \frac{1}{\sigma_{\mathrm{O}}^L},
    \label{eq:resistivities_def}
  \end{equation}
where the conductivities, $\sigma_{\mathrm{O}}^L$, $\sigma_{\mathrm{H}}^L$ and $\sigma_{P}^L$, are linear combinations of the conductivities of the light charged species $l$:
\begin{equation}
    \sigma_{\mathrm{O}}^L=\frac{ec}{B}\sum_{l \in L} n_l Z_l \Gamma_l, \sigma_{\mathrm{H}}^L=-\frac{ec}{B}\sum_{l \in L} \frac{n_l Z_l}{1+\Gamma_l^2} \Gamma_l^2,
    \label{eq:conductivities_def_1}
\end{equation}
\begin{equation}
    \sigma_{\mathrm{P}}^L=\frac{ec}{B}\sum_{l \in L} \frac{n_l Z_l \Gamma_l}{1+\Gamma_l^2}, \sigma_\perp^L=\sqrt{(\sigma_{\mathrm{H}}^L)^2+(\sigma_{\mathrm{P}}^L)^2}.
    \label{eq:conductivities_def_2}
\end{equation}

The Ohm law of standard non-ideal MHD in Eq. \eqref{eq:Ohm_NIMHD}, with the use of the magnetic resistivities $\eta_{\mathrm{O}}^C$, $\eta_{\mathrm{H}}^C$ and $\eta_{\mathrm{AD}}^C$, is the special case of the generalized Ohm law in Eq. \eqref{eq:Ohm_multifluid_MHD} when $L=C$ (i.e., the case for which the inertia of all charges is neglected). In this paper, with the definition of $L \subset C$ (i.e., grain inertia is considered), the conductivities of the dust species are not included in the ion-electron magnetic resistivities $\eta_{\mathrm{O}}^L$, $\eta_{\mathrm{H}}^L$ and $\eta_{\mathrm{AD}}^L$ (but dust species still contribute to them via the charge equilibrium).

The system of equations is closed by Eq. \eqref{eq:mawell_ampere}, since $\mathbf{J}= \Delta \mathbf{J}_L + \Delta \mathbf{J}_I$, with $\Delta \mathbf{J}_I= \sum_{d \in I} n_dZ_de ( \mathbf{V}_d- \mathbf{V}_g)$. This model resembles the model of \citet{2025ApJ...985...55S} to study the dust battery mechanism, but thanks to the definitions of the resistivities  $\eta_{\mathrm{O}}^L$, $\eta_{\mathrm{H}}^L$ and $\eta_{\mathrm{AD}}^L$, we generalize to an arbitrary collection of light charged species. More details and discussions on the full set of equations can be found in Appendix \ref{app:multifluid_MHD}. In particular, the energy balance is described in Appendix \ref{app:multifluid_MHD_energy}. Fig. \ref{fig:multifluid_MHD_interactions} summarizes the interactions between the components of the multifluid MHD model.

\subsection{Multidust plasma} \label{sec:multidust_plasma_equations}

In this section, we present another model that helps us in understanding the fundamentals on the magnetic interactions between multiple dust fluids (Sect. \ref{sec:multidust_plasma_waves}). The multidust plasma model is a simplified version of the multifluid MHD model (Sect. \ref{sec:multifluid_MHD_equations}), for which we set the drag coefficients of the ions and the electrons, $\nu_i$ and $\nu_e$, to zero. Therefore, the ions and the electrons perfectly couple to the magnetic field (their Hall factors tend to infinity) and their back-reaction on the gas via drag forces is neglected. This assumption simplifies the resistivities. Indeed, in this limit, we obtain
\begin{equation}
   \eta_{\mathrm{O}}^L =0, \eta_{\mathrm{AD}}^L =0, \eta_{\mathrm{H}}^L = 1/\hat{\sigma}_{\mathrm{H}}^L,
\end{equation}
with the simplified expressions for the Hall conductivity:
\begin{equation}
    \hat{\sigma}_{\mathrm{H}}^L = - \sum_{l \in L} n_l Z_l \frac{ec}{B} = \sum_{d \in I} n_d Z_d \frac{ec}{B} = \frac{c^2}{4 \pi}\sum_{d\in I} \omega_d \frac{4 \pi \rho_d}{B^2}.
    \label{eq:multidust_plasma_Hall_conductivity}
\end{equation}

In other words, the electric field becomes
\begin{equation}
    \mathbf{E} = \frac{1}{c} \frac{(\mathbf{J}- \sum_{d \in I}  n_d Z_d e \mathbf{V}_d) \times \mathbf{B}}{\sum_{d \in I} n_d Z_d e} = - \frac{1}{c}  \sum_{d \in I} \xi_d  \mathbf{V}_d \times \mathbf{B} + \frac{1}{\hat{\sigma}_{\mathrm{H}}^L} \mathbf{J} \times \mathbf{b},
    \label{eq:multidust_plasma_ohm_law}
\end{equation}
where $\xi_d = n_d Z_d /(\sum_{i \in I } n_i Z_i)$ is the proportion of charges carried by the dust fluid $d$ relative to the charge carried by the whole dust distribution. From the form of Eq. \eqref{eq:multidust_plasma_ohm_law}, we note that the magnetic field is ideally coupled to the charge-weighted barycenter of the dust velocities corrected by a Hall effect term. This Ohm law is similar to the Five-fluid model of \citet{2024mpsp.book..203S}. For the case of a single dust fluid ($\mathcal{N}=1$), Eq. \eqref{eq:dust_momentum_conservation} becomes
\begin{equation}
    \partial_t (\rho_d \mathbf{V}_d) + \nabla \cdot \left(  \rho_d \mathbf{V}_d \mathbf{V}_d \right) = \mathbf{f}_{g \to d} + \frac{\mathbf{J} \times \mathbf{B}}{c}.
    \label{eq:multidust_plasma_one_fluid_magnetized}
\end{equation}

In this case, the dust fluid 
becomes magnetized, leading to the model name "multidust plasma".

\section{Methods} \label{sec:methods}

\subsection{Computation of the eigenmodes} \label{sec:eigenmodes_finding_method}

The dust multifluid model is high-dimensional. The dimension is even infinite when considering a continuous distribution (e.g., of sizes or charges). With an isothermal closure on the gas pressure, the 3D multifluid MHD system  consist in $4 \mathcal{N}+7$ equations, where $\mathcal{N}$ is the number of dust species. We therefore explore this system in the linear regime, consisting in studying its eigenmodes (or waves). We note that the dust density perturbations do not back-react on the rest of the system in the linear regime (but the dust velocity perturbations do). This is not the case for the gas, due to the pressure force. We study the propagation of the two types of one-dimensional magnetohydrodynamical waves in an uniform magnetic field $B_0 \mathbf{e_z} $, that are magnetosonic waves and Alfvén waves. 

For compressible (magneto)sonic waves, the perturbed fields along the direction perpendicular to the magnetic field $\mathbf{e}_x$ are
\begin{equation}
    \begin{cases}
        \rho_g(x,t)=\rho_g^0 + \delta \rho_g (x,t), \\
        \mathbf{V}_g(x,t)= v_{g,x}(x,t) \mathbf{e}_x, \\
        \mathbf{V}_d(x,t)= v_{d,x}(x,t) \mathbf{e}_x, \\
        \mathbf{B}(x,t) = B_0 \mathbf{e_z} + \delta b(x,t) \mathbf{e_z},
    \end{cases}
    \label{eq:perturbations_sonic}
\end{equation}
where the rightmost terms in the equations are the first-order terms. In the case of a charged multifluid, it is necessary to introduce velocity perturbations along $\mathbf{e}_y$ due to individual Lorentz forces (Appendices  \ref{app:multidust_plasma_ms} and \ref{app:multifluid_MHD_ms_modes}), and thus $\mathbf{V}_g(x,t)= v_{g,x}(x,t) \mathbf{e}_x + v_{g,y}(x,t) \mathbf{e}_y$, and similarly for dust species.

For incompressible Alfvén waves, the perturbed fields along the parallel direction to the magnetic field $\mathbf{e}_z$ are
\begin{equation}
    \begin{cases}
        \mathbf{V}_g(z,t)= v_{g,x}(z,t) \mathbf{e}_x + v_{g,y}(z,t) \mathbf{e}_y, \\
        \mathbf{V}_d(z,t)= v_{d,x}(z,t) \mathbf{e}_x + v_{d,y}(z,t) \mathbf{e}_y, \\
        \mathbf{B}(z,t) = B_0 \mathbf{e_z} +\delta b_{x}(z,t) \mathbf{e}_x + \delta b_{y}(z,t) \mathbf{e}_y. 
    \end{cases}
    \label{eq:perturbations_alfven}
\end{equation}
For studying circularly polarized Alfvén modes, we set new variables following the notation $u_{\sigma} = u_x + i \sigma u_y$, where $\sigma \in \{-1,1 \}$ is the sign of the rotation.

The high dimensionality of the system of perturbations makes the dispersion relation difficult to derive and to analyze. Our analysis partly relies on numerical solving, to remain as close as possible to the initial set of equations. Finding the eigenmodes consists in assuming that the perturbations $\delta \mathbf{w}$ in Eqs. \eqref{eq:perturbations_sonic} and \eqref{eq:perturbations_alfven} are proportional to $\exp(i \mathbf{k}.\mathbf{x} )$ (Fourier transform in space), with $\mathbf{k}= k \mathbf{e}_x \perp B_0 \mathbf{e}_z$ for (magneto)sonic modes and $\mathbf{k}= k \mathbf{e}_z$ for Alfvén modes (general magnetosonic modes that propagate at oblique angles are not considered here). We inject this solution form in the equations consisting of the partial differential equations (dynamics) and the algebraic constrains (intermediate variables and Ohm law). By combining these equations numerically, we obtain an autonomous linear ordinary differential system, which is formally 
\begin{equation}
    \frac{\mathrm{d}}{\mathrm{d}t} \delta \mathbf{w}(k,t)= \mathbb{A}(k)  \delta \mathbf{w}(k,t).
    \label{eq:formal_ODE_dynamics}
\end{equation}

The solutions are the linear combinations of the eigenstates of $\mathbb{A}(k)$. The eigenvalues satisfy $\det ( -i\omega \mathbf{1} -\mathbb{A}(k))=0$, which is the standard dispersion relation, according to the convention $\delta \mathbf{w} \propto \exp(i (\mathbf{k} \cdot \mathbf{x}- \omega t) )$. \citet{2023A&A...674A.149H} used a dichotomy algorithm, on a finding region, to invert a derived-by-hand relation $k=k(\omega)$. Here, we use the linear algebra methods of numpy and scipy, that can find the complete set of the solutions with reliability.

\subsection{Visualization of the eigenvectors} \label{app:eigenvectors_visualisation}

\begin{figure}
    \centering
    \includegraphics[width=0.45\textwidth]{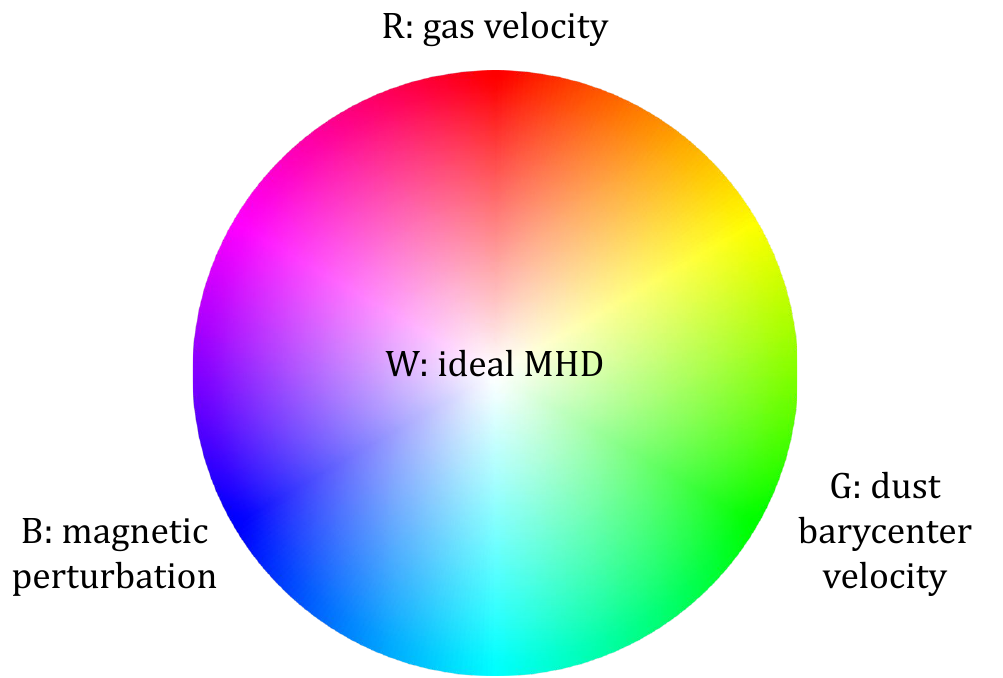}

    \caption{RGB color wheel encoding the components of the perturbations eigenvector relative to ideal MHD: red for the gas velocity, green for the dust barycenter velocity and blue for the magnetic perturbation. Ideal MHD appears in white, a preferential coupling of the dust with the magnetic field appears in cyan and a preferential coupling of the dust with the gas appears in yellow.}
    \label{fig:color_wheel}
\end{figure}

From the numerical computation of the eigenvectors, we can display their components to better understand the underlying physics. This is done for the solutions of the dispersion relation in Figs. \ref{fig:ms_wave_dispersion_17nm_color}, \ref{fig:alfven_wave_dispersion_color}, and \ref{fig:ms_wave_dispersion_5bins_100bins_color}. In this paper, we encode the intensity of the components with colors: red for the gas, green for the dust, blue for the magnetic field. Therefore, a coupling between these components produces a color mixture according to the RGB color wheel in Fig. \ref{fig:color_wheel}. These color intensities are normalized such that ideal MHD leads to equal amplitudes between the components, and thus to white. 

More precisely, to do so, we set the intensity of the RGB colors to $\mathcal{R}=\lVert \delta \mathbf{v}_g \rVert/c_{\mathrm{IMHD}}$, $\mathcal{G}=\lVert \delta \mathbf{v}_d \rVert /c_{\mathrm{IMHD}}$ and $\mathcal{B}= \lVert \delta \mathbf{b} \rVert/B_0$, where $\delta \mathbf{v}_g$ denotes the gas velocity perturbation, $\delta \mathbf{v}_d$ is the perturbation on the velocity barycenter of the dust fluids, $\delta \mathbf{b}$ is the magnetic field perturbation, and $c_{\mathrm{IMHD}}$ is the speed of the ideal MHD waves, depending on the propagation geometry (Sect. \ref{sec:eigenmodes_finding_method}). We set this wave speed to the bulk magnetosonic speed $c_{\mathrm{IMHD}}=c_{\mathrm{ms},gd}=(c_{\mathrm{s},gd}^2 + c_{\mathrm{a},gd}^2)^{1/2}$ for magnetosonic modes and to the bulk Alfvén speed $c_{\mathrm{IMHD}}= c_{\mathrm{a},gd}$ for Alfvén modes, where $c_{\mathrm{s},gd}= \sqrt{P_g/\rho}$ and $c_{\mathrm{a},gd}=B_0/\sqrt{4 \pi \rho}$. Then, the RGB colors are normalized by the highest component, $\max(\mathcal{R}, \mathcal{G}, \mathcal{B})$. 

This visualization can guide the physical assumptions for analyzing the eigenstates. However, this could be biased by the physical setup. For instance, for the case of a single dust fluid ($\mathcal{N}=1$), with a dust-to-gas ratio of $\theta_d= 1 \%$, the dust fluid has a dynamical response to a collision mode of $v_{d,x} =- v_{g,x}/\theta_d = -100 v_{g,x}$ (later presented in Sect. \ref{sec:multidust_hydro_waves}), or a response to the magnetocompressive wave $v_{d,x}= (c_{\mathrm{a},g}/\sqrt{\theta_d}) \delta b/B_0= 10 c_{\mathrm{a},g} \delta b/B_0$ (Sect. \ref{sec:multidust_plasma_waves}), which are significantly away from the ideal MHD reference. This is why we log-stretch the color amplitudes. This needs to be kept in mind to prevent misled conclusions, such as resonances.

\subsection{Numbering of the modes} 
\label{app:solution_numbering_system}

Colors can be difficult to read. To avoid any ambiguity in identifying a mode, we also number the solutions in figures and we refer to them in the analysis. It also allows us to verify that we identify the complete set of solutions by matching the number of solutions with the dimension of the linear system.

We need to clarify some choices regarding the numbering of the magnetosonic modes. We found that the magnetosonic modes are organized in pairs of left-propagating and right-propagating modes and individual purely damped modes. This could be investigated theoretically by studying the PT-symmetry of the dynamical matrix $\mathbb{A}(k)$
\citep{2024A&A...689A.237L}, which is out of the scope of this paper. Therefore, we only plot the real part of of the right-propagating modes, but we recall the existence of a symmetric left-propagating mode by indicating its number in parenthesis after the number of the right-propagating mode. Pairs of propagating modes can recombine when their propagating part tends to zero. The two resulting damped modes can then bifurcate. One of them can recombine with another mode at another scale. Because of these bifurcations, we have to make arbitrary choices on the mode numbering at a given scale $k$ that should not be interpreted as mode tracking through the scales. Moreover, the bifurcations depend on the dust distribution, and thus numbering can be inconsistent with changing the number of dust fluids $\mathcal{N}$.

\subsection{Summary of the analytical solutions and notations} \label{app:analytical solutions and notations}

\begin{table*}
\caption {Summary of the analytical solutions and notations}
\label{table:notations} 
\centering
\begin{tabular}{p{1.25cm} p{1.5cm} p{2.75cm}p{1.5cm} p{2.5cm} p{1.75cm} p{3.25cm}}
\hline \hline          
Geometry & Type &  Scale regime & Components
 & Figure & Expressions & Relevant quantities \\
\hline
  \multicolumn{7}{c}{\it Hydrodynamical modes}\\
\hline
   Sonic mode  & Sonic waves &   Large scales: \newline $k \ll k_{\mathrm{rec},d}$ & $\delta \rho_g$,  $ v_{g,x}$,  $ v_{d,x}$ & $\omega = \pm \tilde{c}_{g123}k -i \tau_{123} \tilde{c}_{g123}k^2$ in Fig. \ref{fig:dustywave} & Eq. \eqref{eq:dustywave_sonic_wave_large_scales} & $k_{\mathrm{rec},d}$: dust recoupling scale in Eq. \eqref{eq:hydro_sonic_wave_dust_recoupling_scale}, \newline $c_{\mathrm{s},gd}=\sqrt{P_g/\rho}$: bulk sound speed\\
     &   &  Small scales: \newline $k \gg k_{\mathrm{rec},d}$ & $\delta \rho_g$,  $v_{g,x}$ & $\omega = \pm c_{\mathrm{s},g} k - i \nu_{123}$ in Fig. \ref{fig:dustywave} & Eq. \eqref{eq:sonic_mode_small_scales} & $c_{\mathrm{s},g}=\sqrt{P_g/\rho_g}$: gas sound speed \\
    & Collision modes   & Large scales: \newline $k \ll k_{\mathrm{rec},d}$ & $ v_{g,x}$, $ v_{d,x}$ & $\omega = -i\Omega_j$ in \newline Fig. \ref{fig:dustywave} & Eq. \eqref{eq:dustywave_collision_mode} & \\
    &    & Small scales: \newline $k \gg k_{\mathrm{rec},d}$ & $v_{g,x}$, $v_{d,x}$ & $\omega = -i\nu_j$ in \newline Fig. \ref{fig:dustywave} & Eq. \eqref{eq:collision_mode_small_scales} & $t_{\mathrm{s},j}=1/\nu_j$: stopping time\\
\hline
  \multicolumn{7}{c}{\it Non-ideal MHD modes}\\
\hline
   Alfvén mode  & Ideal MHD waves   & Large scales: \newline $k \ll k_{\mathrm{D}}^C,k_{\mathrm{H}}^C$ & $v_{g,\sigma}$, $\delta b_{\sigma}$ &  NIMHD in Figs. \ref{fig:alfven_wave_dispersion_color} and \ref{fig:alfven_wave_dispersion_analytical} & $\omega \approx \pm c_{\mathrm{a},g}k$ & $c_{\mathrm{a},g}=B_0/\sqrt{4 \pi \rho_g}$ gas Alfvén speed, \newline
   $k_{\mathrm{D}}^C, k_{\mathrm{H}}^C$ in Eq. \eqref{eq:NIMHD_resistive_scales}\\
     & Whistler wave  & Small scales: \newline $k \gg k_{\mathrm{D}}^C,k_{\mathrm{H}}^C$ &  $\delta b_{\sigma}$ &  NIMHD in Figs. \ref{fig:alfven_wave_dispersion_color} and \ref{fig:alfven_wave_dispersion_analytical}
     & Eq. \eqref{eq:NIMHD_Alfvén_whistler}, \newline $\omega \propto k^2$ & \\
     & Left-hand polarized mode  & Small scales: \newline $k \gg k_{\mathrm{D}}^C,k_{\mathrm{H}}^C$ &  $ v_{g,\sigma}$ &  NIMHD in Figs. \ref{fig:alfven_wave_dispersion_color} and \ref{fig:alfven_wave_dispersion_analytical}
     & Eq. \eqref{eq:NIMHD_left_handed_mode}, \newline $\omega \propto k^0$ & \\
   Magneto- sonic mode  & Ideal MHD waves   & Large scales: \newline $k \ll k_{\mathrm{D}}^C$ & $\delta \rho_g$,  $ v_{g,x}
   $, $\delta b$ & NIMHD in \newline Figs. \ref{fig:ms_wave_dispersion_17nm_color}, \ref{fig:ms_wave_dispersion_5bins_100bins_color} and \ref{fig:ms_wave_dispersion_analytical} & Eq. \eqref{eq:NIMHD_ms_waves_large_scales}, \newline $\omega \approx \pm c_{\mathrm{ms},g} k$ & $c_{\mathrm{ms},g}=\sqrt{c_{\mathrm{a},g}^2+c_{\mathrm{s},g}^2}$ gas magnetosonic speed\\
   & Sonic waves   & Small scales: \newline $k \gg k_{\mathrm{D}}^C$ & $\delta \rho_g$,  $v_{g,x}
   $ & NIMHD in \newline Figs. \ref{fig:ms_wave_dispersion_17nm_color}, \ref{fig:ms_wave_dispersion_5bins_100bins_color} and \ref{fig:ms_wave_dispersion_analytical} & Eq. \eqref{eq:NIMHD_sound_waves_small_scales} & \\
    & Magnetic diffusion  & Small scales: \newline $k \gg k_{\mathrm{D}}^C$ & $\delta b
   $ & NIMHD in \newline Figs. \ref{fig:ms_wave_dispersion_17nm_color}, \ref{fig:ms_wave_dispersion_5bins_100bins_color} and \ref{fig:ms_wave_dispersion_analytical} & Eq. \eqref{eq:NIMHD_magnetic_diffusion_small_scales} & \\
\hline
  \multicolumn{7}{c}{\it Multidust MHD modes}\\
\hline
      Magneto- sonic mode & Dust magnetocompressive waves  & Small scales: \newline $ k \gg \nu_d/c_{\mathrm{a},d}$ & $v_{d,x}$, $\delta b$ & $\omega \approx \pm \tilde{c}_{\mathrm{ms},d} k$ in Fig. \ref{fig:ms_wave_dispersion_analytical} & Eqs. \eqref{eq:multidust_magnetosonic} and \eqref{eq:multidust_plasma_dust_magnetosonic_1bin} & $c_{\mathrm{a},d}=B_0/\sqrt{4 \pi \varrho_d}$: Alfvén speed of the dust bulk mass, \newline
      $\tilde{c}_{\mathrm{ms},d} $: dust magnetosonic speed in \newline Eq. \eqref{eq:multidust_magnetosonic} \\
\hline
\end{tabular}
\tablefoot{Analytical solutions and notations sorted by multifluid models: multifluid hydrodynamics (grain inertia included, no magnetic field), non-ideal MHD (inertia of charged grains, ions and electrons neglected, Sect. \ref{sec:non_ideal_MHD_equations}), multidust MHD (inertia of charged grains included, ions and electrons perfectly coupled to the magnetic field, Sect. \ref{sec:multidust_plasma_equations}), and multifluid MHD (inertia of charged grains included, ions and electrons coupled to the magnetic field and the gas, Sect. \ref{sec:multifluid_MHD_equations}). Non-exhaustive list of the set of solutions. The geometry refers to the propagating direction relative to the magnetic field and the perturbation components (see Sect. \ref{sec:eigenmodes_finding_method}). The type of the modes distinguishes the propagating waves (dispersive or not) and the collision modes. The scale regime provides information about the validity of the analytical solutions. We then give the main perturbation components of the mode. We emphasize that the dust density perturbations are not included contrary to the gas density perturbation because they are not part of the components of the eigenvectors, but compressive waves can induce them thanks the equation of mass conservation. It is illustrated for the dust magnetocompressible waves in Sect. \ref{sec:single_multifluid_MHD_ms_modes_experiment} with Eq. \eqref{eq:perturb_dust_density_mass_conversion}. We provide the reference to figures where the solutions are plotted. The equations of the explicit expressions are provided. In the last column, we introduce step by step some useful quantities and their notations. 
Table \ref{table:notations_continuation} continues this table.}
\end{table*}

\begin{table*}
\caption {Continuation of Table \ref{table:notations}.}
\label{table:notations_continuation} 
\centering
\begin{tabular}{p{1.25cm} p{1.5cm} p{2.75cm}p{1.5cm} p{2.5cm} p{1.75cm} p{3.25cm}}
\hline \hline          
Geometry & Type &  Scale regime & Components
 & Figure & Expressions & Relevant quantities \\
\hline
  \multicolumn{7}{c}{\it Multifluid MHD modes}\\
\hline
   Alfvén mode  & Ideal MHD waves   & Large scales: \newline $k \ll k_{\mathrm{rec},d},k_{\mathrm{D}}^C,k_{\mathrm{H}}^C$ & $ v_{g,\sigma}$, $v_{d,\sigma}$, $\delta b_{\sigma}$ &   & $\omega \approx \pm c_{\mathrm{a},gd} k$ & $c_{\mathrm{a},gd}=B_0/\sqrt{4 \pi \rho}$: bulk Alfvén speed\\
   & Dust Alfvén wave  & Intermediate scales &  $v_{d,\sigma}$, $\delta b_{\sigma}$  & $\omega_{db}^1$, $\omega_{bd}^1$ in \newline Fig. \ref{fig:alfven_wave_dispersion_analytical} & Eqs. \eqref{eq:multifluid_MHD_Alfven_wave_dust_intermediate_scales} and \eqref{eq:multifluid_Alfven_solution} & $\tilde{c}_{\mathrm{a},d}$: dust Alfvén speed in Eq. \eqref{eq:multifluid_MHD_dust_Alfven_speed}\\
   & Whistler wave  & Small scales &  $\delta b_{\sigma}$  & $\omega_{b}^\infty$ in Fig. \ref{fig:alfven_wave_dispersion_analytical} & Eq. \eqref{eq:multifluid_MHD_alfven_whistler_small} & \\
     & Collision modes  & Large scales: \newline $k \ll k_{\mathrm{rec},d},k_{\mathrm{D}}^C,k_{\mathrm{H}}^C$ & $ v_{g,\sigma}$, $v_{d,\sigma}$ &  $\omega_{dg}^0$ in Fig. \ref{fig:alfven_wave_dispersion_analytical} & Eq. \eqref{eq:multifluid_MHD_collision_mode_large_scales} & \\
    &   & Small scales & $ v_{g,\sigma}$, $v_{d,\sigma}$  & $\omega_{dg}^\infty, \omega_{gd}^\infty$ in \newline Fig. \ref{fig:alfven_wave_dispersion_analytical}  & Eq. \eqref{eq:multifluid_MHD_collision_modes_small_scales_1bin} & $\Omega_{d,\sigma}=\sigma \omega_d-i \nu_d$:  resonant frequency, 
    \newline
    $\omega_d$: gyration pulsation \\
   Magneto- sonic  mode  & Ideal MHD waves   & Large scales: \newline $k \ll k_{\mathrm{rec},d},k_{\mathrm{D}}^C$ & $\delta \rho_g$, $v_{g,x}$, $v_{d,x}$, $\delta b$& & $\omega \approx \pm c_{\mathrm{ms},gd}k$ & $c_{\mathrm{ms},gd}=\sqrt{c_{\mathrm{a},gd}^2+c_{\mathrm{s},gd}^2}$ bulk magnetosonic speed\\
    & Sonic waves   & Small scales: \newline $k \gg k_{\mathrm{rec},d},k_{\mathrm{D}}^C$ & $\delta \rho_g$, $v_{g,x}$, & $\omega_g^\infty$ in Fig. \ref{fig:ms_wave_dispersion_analytical}  & Eq. \eqref{eq:multifluid_MHD_sonic_wave_small_scales} & \\
    & Dust magnetocompressive waves & Intermediate scales: \newline $k_{\mathrm{dp}}<k < k_{\mathrm{hd}}$ & $ v_{d,x}, \delta b$ & $\omega_{db}^1$ in Fig. \ref{fig:ms_wave_dispersion_analytical}  & Eq. \eqref{eq:multifluid_MHD_dust_ms_wave_1bin} & $k_{\mathrm{dp}},k_{\mathrm{hd}}$ in Eq. \eqref{eq:multifluid_MHD_kdp_khd},  $\tilde{k}_{\mathrm{dp}},\tilde{k}_{\mathrm{hd}}$ in Eqs. \eqref{eq:diffusion_scale_multifluid_MHD_approximation} and \eqref{eq:magnetized_dust_multifluid_MHD_approximation},
    and $k_{\mathrm{dp}}^{\mathrm{p}},k_{\mathrm{hd}}^{\mathrm{p}}$ in Eq. \eqref{eq:multifluid_MHD_kdp_khd_propagation}\\
    & Hydro- dynamical dust  & Small scales: \newline $k > k_{\mathrm{hd}}$ & $v_{d,x}$ & $\omega_d^\infty$ in Fig.\ref{fig:ms_wave_dispersion_analytical} 
     & Eq. \eqref{eq:multifluid_MHD_ms_small_scales_damped_modes} & \\
    & Magnetic diffusion   & Small scales: \newline $k > k_{\mathrm{hd}}$ & $\delta b$ & $\omega_b^\infty$ in Fig. \ref{fig:ms_wave_dispersion_analytical}   & Eqs. \eqref{eq:multifluid_MHD_ms_small_scales_damped_modes} and \eqref{eq:multifluid_MHD_ms_small_scales_magnetic_mode} & \\
    & Collision modes  &  Large scales: \newline $k \ll k_{\mathrm{rec},d},k_{\mathrm{D}}^C$ & $v_{g,x} \pm i v_{g,y}$, $v_{d,x} \pm i v_{d,y}$ & $\omega_{dg}^0$ in Fig. \ref{fig:ms_wave_dispersion_analytical}  & Eq. \eqref{eq:multifluid_MHD_collision_mode_large_scales} & $\Omega_{d,\pm}= \pm \omega_d - i \nu_d$: resonant frequency\\
\hline
\end{tabular}
\end{table*}

In complement to the visualization of the components of the eigenvectors when plotting the exact solutions of the dispersion relation (Sect. \ref{app:eigenvectors_visualisation}), we plot analytical solutions in the appendix. In Tables \ref{table:notations} and \ref{table:notations_continuation}, we summarized these solutions and redirect towards their expressions. We also gather the most useful notations (pulsations, damping rates, scales, phase velocities), which are used in the main body of the paper and in the figures. In particular, $\omega_p^r$ denotes an analytical solution in the scale regime $r \in \{0,1,\infty\}$ (respectively for large scales ($k \to 0$), intermediate scales, and small scales ($k \to \infty$)), and involving the perturbations $p \in \mathcal{P} (\{g,d,b\})$ (respectively for the gas, the dust and the magnetic field, and any combination of them with priority order in the mode). Similarly, $\tilde{c}_{\varphi,p} $ denotes the phase velocity of a wave of propagation geometry $\varphi \in \{\mathrm{s},\mathrm{a},\mathrm{ms}\}$ (respectively sonic modes, Alfvén modes, and magnetosonic modes) and involving the perturbations $p$.

\subsection{Physical setup} \label{app:physical_setup}

\begin{figure}
    \centering
    \includegraphics[width=0.49\textwidth]{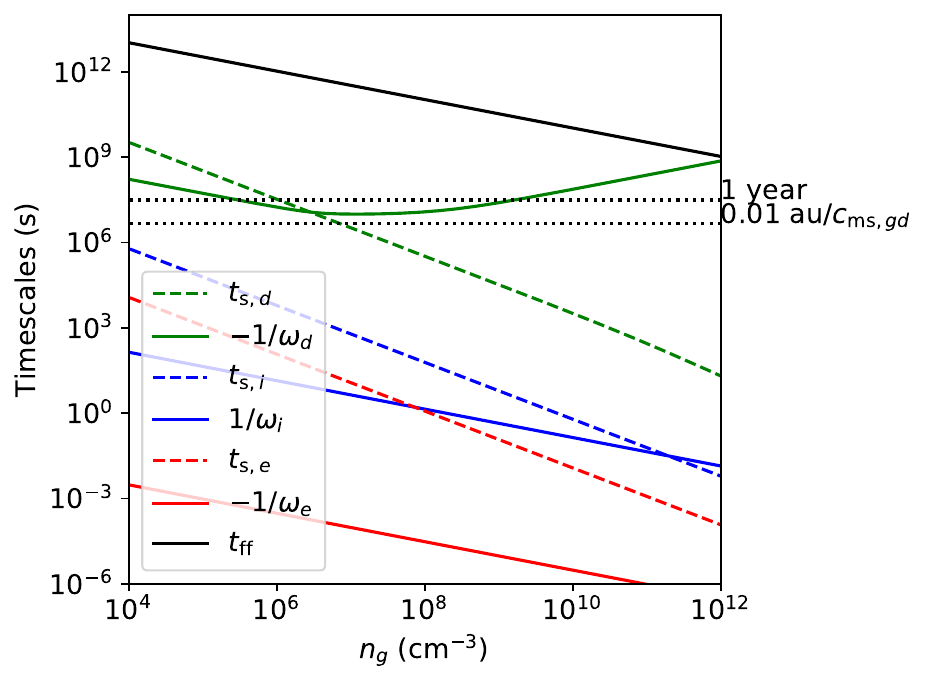}
    \caption{Dynamical timescales involved during a protostellar collapse, with stopping times $t_{\mathrm{s},k}$ and gyration times $1/|\omega_k|$ of the 17 nm dust grains (green), the ions (red) and the electrons (blue). The Hall factors of the charged species can be deduced from the ratio $ |\omega_k|t_{\mathrm{s},k}$ by using the log-scale. The freefall time $t_{\rm{ff}}= \sqrt{3\pi/(32 G \rho)}$ is in black. We indicate the one-year value and the typical crossing time of a standard magnetosonic wave traveling 0.01 au.}
    \label{fig:typical_timescales}
\end{figure}

\begin{figure}
    \centering
    \includegraphics[width=0.49\textwidth]{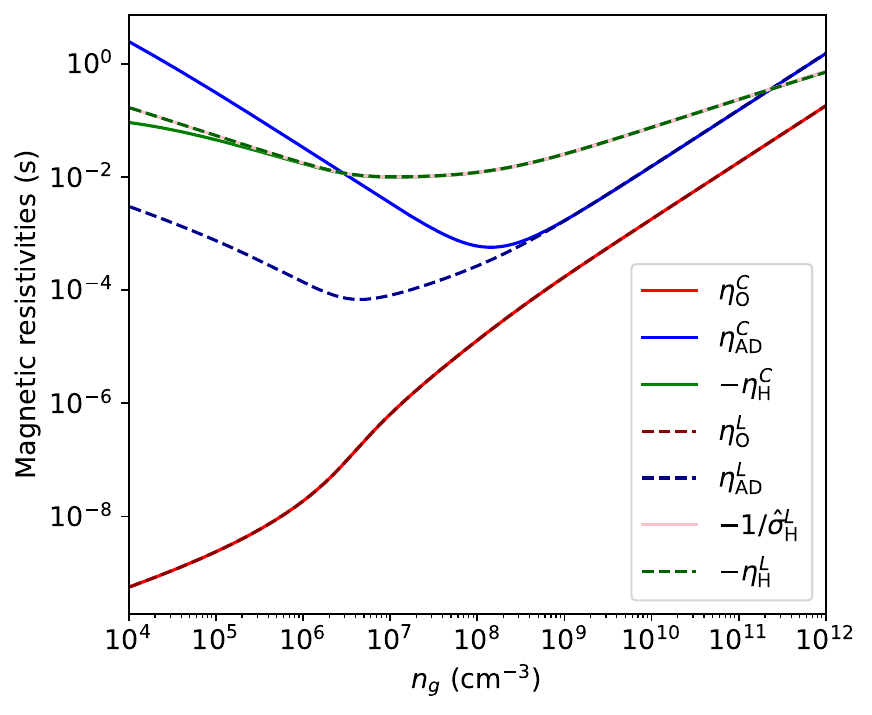}
    \caption{Magnetic resistivities for the three models presented in Sect. \ref{sec:MHD models}: non-ideal MHD in solid lines (dust, ions and electrons are included in $\eta_{\mathrm{O}}^C$, $\eta_{\mathrm{H}}^C$, and $\eta_{\mathrm{AD}}^C$), multifluid MHD in dotted lines (only the conductivities of the ions and the electrons are directly included in the computation of resistivities $\eta_{\mathrm{O}}^L$, $\eta_{\mathrm{H}}^L$, and $\eta_{\mathrm{AD}}^L$, but dust is included indirectly in the charge balance), and multidust plasma whose Hall resistivity ($1/\hat{\sigma}_{\mathrm{H}}^{L})$ is in pink. We note that $\eta_{\mathrm{O}}^L \approx \eta_{\mathrm{O}}^C$ and $\eta_{\mathrm{H}}^L \approx 1/\hat{\sigma}_{\mathrm{H}}^L$.}
    \label{fig:magnetic_resistivities}
\end{figure}

We consider a uniform cloud of gas, characterized by its density $\rho_g$ and its mean molecular weight $\mu_g = 2.31$ \citep{1987IAUS..115..417S}, compatible with an abundance of He relative to H$_{\rm{2}}$ of $\chi=0.183$. Therefore, we consider the particle number $n_g=\rho_g/(\mu_g m_{\mathrm{p}})=n_{\rm{H}} (1+\chi)/2 $, where $m_{\mathrm{p}}$ is the mass of the proton. We assume that it is an ideal gas of temperature $T$. The gas pressure is given by $P_g=c_{\mathrm{s},g}^2 \rho_g$ where the isothermal gas sound speed is therefore $c_{\mathrm{s},g} =\sqrt{k_{\mathrm{B}} T/(\mu_g m_{\mathrm{p}})}$. For the sake of simplicity, we assume in this paper that the compressive waves propagate isothermally (rather than adiabatically or following the gas energy equation in Eq. \eqref{eq:gas_energy_equation}). We follow the temperature law of \citet{2006ApJ...647L.151M,2016A&A...592A..18M}, that is
\begin{equation}
    T = T_0 \sqrt{1+ \left(\frac{n_g}{10^{11}~\rm{cm^{-3}}}\right)^{0.8}}. 
\end{equation}

We can analyze different regions of the collapse based on the gas density. The density criteria used by \citet{2020A&A...641A.112L}, and original references therein, are $n_g<10^7~{\rm{cm^{-3}}}$ for the protostellar envelope, $10^7~{\rm{cm^{-3}}}<n_g < 10^9 ~{\rm{cm^{-3}}}$ for the pseudo-disk, and $10^9 ~{\rm{cm^{-3}}}<n_g <10^{11}~{\rm{cm^{-3}}}$ for the disk, and $10^{11} ~{\rm{cm^{-3}}}<n_g <2.6\times 10^{12}~{\rm{cm^{-3}}}$ for the transition to the first hydrostatic. Above $2.6\times 10^{12}~{\rm{cm^{-3}}}$, the first hydrostatic core is formed.

From this setup, we can define the Jeans length, or equivalently the Jeans scale $k_{\mathrm{J}}=2 \pi/L_{\mathrm{J}}$, as
\begin{equation}
L_{\mathrm{J}} = c_{\mathrm{s},g} \sqrt{\frac{\pi}{G \rho}}.
\end{equation}

The plasma is weakly ionized. We consider one type of ions of mass $m_i=25 m_{\mathrm{p}}$ and of charge $Z_i=1$ \citep{2021A&A...649A..50M}. 
The dust multifluid samples the MRN distribution \citep{1977MRN} ranging from 5 nm grains to 250 nm grains with a total dust-to-gas ratio of $1 \%$. The exponent of the power law in terms of number of grains is $-3.5$ (see Appendix A of \citet{2023A&A...674A.149H} for details on the distribution sampling). The abundance of ions and electrons as well as the mean charge carried by each dust fluid are computed using the chemical network and the algorithm of \citet{2021A&A...649A..50M}, with an ionization rate $\zeta = 5 \times 10^{-17}~\rm{s^{-1}}$ (see its section 3.4 and its Figure 2 to see the resulting $Z_d$, $n_i$, and $n_e$ as a function of gas density). 

We assume that the uniform magnetic field follows \citet{2011ApJ...738..180L}, that is
\begin{equation}
    B_0= 1.43 \times 10^{-7} \sqrt{\frac{n_{\rm{H}}}{1 \ \rm{cm^{-3}} }} \ \rm{G}.
\end{equation}

To compute the dynamics of charged species,
we define their collision rates with the gas. From the grain size $s_{\mathrm{grain}}$ and its intrinsic density $\rho_{\rm{grain}}=2.3 \ \rm{g.cm^3}$, we can compute the stopping time of the grain in the Epstein drag regime \citep{1924PhRv...23..710E} with
\begin{equation}
    t_{\mathrm{s},d}=\sqrt{\frac{\pi }{8}} \frac{\rho_{\rm{grain}}}{\rho_g} \frac{s_{\rm{grain}}}{c_{\mathrm{s},g}}.
\end{equation}

For ions and electrons, we use the momentum transfer rate coefficients $\langle \sigma v \rangle_{gk}= (\mu_g m_{\mathrm{p}} + m_k) \gamma_k$. Following \citet{1983ApJ...264..485D}, 
we use 
\begin{equation}
    \langle \sigma v \rangle_{gi} = 1.9 \times 10^{-9} \ \rm{cm^3 s^{-1}},
\end{equation}
\begin{equation}
    \langle \sigma v \rangle_{ge} = 8.3 \times 10^{-9} \max \left(1, \sqrt{\frac{T}{100 \ \rm{K}}} \right) \ \rm{cm^3 s^{-1}}.
\end{equation}

We found that the mono-disperse distribution of $17$ nm grains, of gyration pulsation $\omega_{\rm{17nm}}$ and collision rate $\nu_{\rm{17nm}}$, provides resistivities similar to those computed using a MRN distribution. Therefore, we use this reference size when studying the case of a single dust fluid (i.e., when $\mathcal{N}=1$). The corresponding gyration times and stopping times of the charged species are shown in Fig. \ref{fig:typical_timescales}. On a one-year timescale, the inertia of dust grains cannot be neglected whereas the gyration times and stopping times of the ions and the electrons are short enough to neglect their inertia as a first approach. Based on these values, we arbitrarily define the reference scales $\omega_{\rm{ref}} = (\omega_{\rm{17nm}}^2 +\nu_{\rm{17nm}}^2)^{1/2}$ and $k_{\rm{ref}} = \omega_{\rm{ref}} /c_{\mathrm{a},gd} $ to plot the solutions of the dispersion relations.

In Fig \ref{fig:magnetic_resistivities}, we present the resulting resistivities for the three MHD models presented in Sect. \ref{sec:MHD models}. We note that $  \eta_{\mathrm{H}}^{C} \approx \eta_{\mathrm{H}}^{L} \approx 1/\hat{\sigma}_{\mathrm{H}}^{L} $ because of the high coupling of the electrons and the ions with the magnetic field (high Hall factor). By comparing the standard magnetic resistivities $(\eta_{\mathrm{O}}^C,\eta_{\mathrm{H}}^C,\eta_{\mathrm{AD}}^C)$ to the magnetic resistivities without the conductivities of dust grains $(\eta_{\mathrm{O}}^L,\eta_{\mathrm{H}}^L,\eta_{\mathrm{AD}}^L)$, we can infer the contribution of the conductivities of charged dust in standard non-ideal MHD resistivities, in particular in the ambipolar resistivity, $\eta_{\mathrm{AD}}^C$, which is the dominant resistivity at low density. For $n_g>10^8~\rm{cm^{-3}}$, the Hall factor of dust  grains drops significantly compared to the Hall factors of the ions and the electrons. Dust remains indirectly involved in the resistivities by regulating the number of electrons and ions in the medium.

\section{The dust multifluid physics} \label{sec:multifluid_physics}

\subsection{Collision modes and propagating modes: an illustration without magnetic fields} \label{sec:multidust_hydro_waves}

\begin{figure*}
    \centering
    \includegraphics[width=0.48\textwidth]{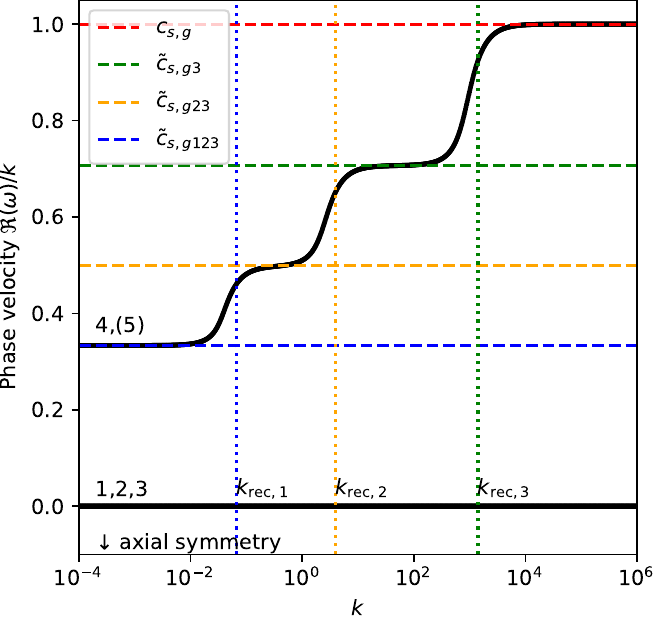}
    \includegraphics[width=0.49\textwidth]{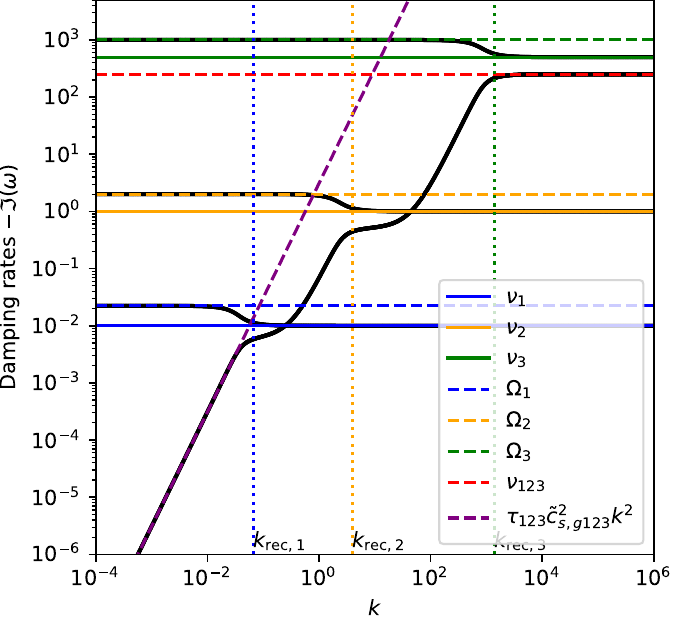}
    \caption{Dispersion relation of the hydrodynamical modes (in black, with numbers from 1 to 5 to identify the $\mathcal{N}+2=5$ modes) whose propagative part is shown on the left panel and the damping rate on the right panel. More precisely, on the left panel, the phase velocity $\Re(\omega)/k$ is shown instead of $\Re(\omega)$ for the sake of readability. Moreover, on the left panel, the left-propagating sound wave which is symmetric to the right-propagating sound wave (mode 4) is recalled by the number in parenthesis (mode 5). Details on the analytical solutions (colored lines) are in Table \ref{table:notations} and Appendix \ref{app:multidust_hydro_waves}. The dust species contributing to the sound wave speed in Eq. \eqref{eq:sound_wave_speed_mass_loading} are explicitly indicated in the subscript of $\tilde{c}_{s,g123}$, $\tilde{c}_{s,g23}$, $\tilde{c}_{s,g3}$ and $\tilde{c}_{\mathrm{s},g}$. The coupling scales $k_{\mathrm{rec},1}= \Omega_1 / \tilde{c}_{\mathrm{s},g123}$, $k_{\mathrm{rec},2}= \Omega_2 / \tilde{c}_{\mathrm{s},g23}$, and $k_{\mathrm{rec},3}= \Omega_1 / \tilde{c}_{\mathrm{s},g3}$ from Eq. \eqref{eq:hydro_sonic_wave_dust_recoupling_scale} are shown with vertical dotted lines. We use the notations $\nu_{123}=(\theta_1 \nu_1 + \theta_2 \nu_2 + \theta_3 \nu_3)/2$ and $\tau_{123}=(\epsilon_1 t_{\mathrm{s},1} + \epsilon_2 t_{\mathrm{s},2} + \epsilon_3 t_{\mathrm{s},3})/2$ in the legend.}
    \label{fig:dustywave}
\end{figure*}

Before analyzing the dynamics of the charged dust multifluid, we focus on the dynamics of a multifluid of neutral grains, which already contains some fundamental multifluid features.
In this section, we present the purely hydrodynamical modes in details. They are compressive modes along the direction $\mathbf{k}=k \mathbf{e_x}$, sometimes referred as sonic waves, or as the dustywave in multifluid code tests. It is treated in details for the case of a single dust fluid in the original work of \citet{2011MNRAS.418.1491L}. An analytical and numerical study of the multifluid case is provided in \citet{2019ApJS..241...25B}. In this paper, we compute the solutions assuming a hierarchy of stopping times.

To illustrate, we use $\mathcal{N}=3$ fluids of dust. We set the densities to $\rho_g=1$, $\rho_1=5$, $\rho_2=2$ and $\rho_3=1$, and the sound speed to $c_{\mathrm{s},g}=1$. We set a hierarchy of stopping times $\nu_1=1/t_{\mathrm{s},1}=0.01$, $\nu_2=1/t_{\mathrm{s},2}=1$, and $\nu_3=1/t_{\mathrm{s},3}=500$. 
The very distinct stopping times of the three dust fluids allow us to distinguish the coupling regimes. 
We present the solutions $\omega (k)$ in Fig. \ref{fig:dustywave}, with details in Appendix \ref{app:multidust_hydro_waves}. The set of perturbations is $(\delta \rho_g, v_{g,x}, v_{1,x}, ..., v_{\mathcal{N},x})$. Therefore, because of the dimension of the system, there are $\mathcal{N}+2=5$ modes to identify.

We identify a pair of damped sound waves (modes 4 and 5 in Fig. \ref{fig:dustywave}). Sound waves are driven by thermal pressure gradients, generated by gas density variations.
These two modes mainly depend on the phasing between the gas density perturbation and the gas velocity. Thanks to the propagation of the sound waves, variations on small spatial scales $1/k$ induce variations on small time scales $1/\omega$. This modifies the role of the dust inertia in the force balance, parametrized by the Stokes number $\mathrm{St}_d=|\Re(\omega)| t_{\mathrm{s},d}$, and thus the adaptation to the gas dynamics. Low-Stokes grains ($\mathrm{St}_d<1$) couple to the gas, whereas high-Stokes grains ($\mathrm{St}_d>1$) can decouple. Gas and low-Stokes grains constitute a well-coupled mixture that can carry sound waves of speed
\begin{equation}
    \tilde{c}_{\mathrm{s},gd}=c_{\mathrm{s},g}/\sqrt{1+\sum_{\mathrm{St}_d<1} \theta_d}.
    \label{eq:sound_wave_speed_mass_loading}
\end{equation}

This result is already known \citep{2012MNRAS.420.2345L,2020JPhCS1666a2050M}, and is referred to as mass loading in \citet{2025A&A...696L..23C}. In the left panel of Fig. \ref{fig:dustywave}, we show the possible wavespeed values by sorting the contribution of dust fluids by their stopping times (horizontal dashed lines in the left panel), depending on the coupling regimes. Equivalently to the Stokes number, we can see that the change of coupling regimes occur around specific scales $k_{\mathrm{rec},d}$ that we estimate later (vertical lines in Fig. \ref{fig:dustywave}). On the other hand, high-Stokes grains significantly drift relative to the gas, and thus, due to drag forces, they damp the sound waves (modes 4 and 5, right panel).

The remaining modes are $\mathcal{N}$ purely damped modes (modes 1, 2 and 3 in Fig. \ref{fig:dustywave}, right panel).
These additional modes result from the drag forces and the velocity drifts between the fluids. Because of the collisions, the fluids tend to relax towards the local velocity barycenter. Because this relaxation occurs locally in space, the eigenmode does not change with the spatial scale of the perturbation $1/k$, generating these constant values of $\omega$ except when the force balance changes, which occurs at the scales for which the grain inertia plays a role in the propagation of the sound waves, as explained above. For purely damped modes, it is more relevant to define the Stokes number as $\widehat{\mathrm{St}}_d=-\Im(\omega) t_{\mathrm{s},d}>0$.

For $k \to \infty$ (i.e., at small scales), the gas dynamics is dominated by the pressure force. In other words, the mechanical response of the gas thanks to the pressure support is faster than the dynamical adaptation to the colliding species. The gas becomes a poor intermediate for dust fluids to communicate. All fluids are decoupled, thus we obtain for each dust species the collision modes $\omega=-i\nu_d$. 

For $k \to 0$ (i.e., at large scales), considering that the pressure force vanishes, we obtain the same system of equations as the collision test, also known as the dustybox test \citep{2011MNRAS.418.1491L}. It is used as a numerical test for drag solvers and describes the relaxation of the velocities of the coupled fluids towards their barycenter. The solutions are the eigenvalues of the drag matrix. \citet{2019ApJS..241...25B} studied them for a monodisperse multifluid (same stopping time for all dust fluids). We present the solutions in the approximation of a hierarchy of stopping times. The dynamical responses of dust fluids to the frequency $\omega$ are
$v_{d,x} = v_{g,x} /(1- \Omega t_{\mathrm{s},d})$, where $\Omega=i \omega=-\Im(\omega)$, and thus $\widehat{\mathrm{St}}_d = \Omega t_{\mathrm{s},d}$. We identify by $\Omega_j$ the mode such that $\Omega \sim 1/t_{\mathrm{s},j}$, separating the dust species coupled to the gas at this timescale ($\widehat{\mathrm{St}}_d \ll 1$, i.e., $\Omega \ll 1/t_{\mathrm{s},d}$) and the dust species whose inertia dominate the force balance ($\widehat{\mathrm{St}}_d \gg 1$, i.e., $\Omega \gg 1/t_{\mathrm{s},d}$). At this timescale, the low-Stokes grains contribute to the bulk mass of a gas-dust mixture whereas the high-Stokes grains
drag this tightly coupled mixture (see Appendix \ref{app:multidust_hydro_waves_collision_modes}). Moreover, compared to the other fluids, the amplitude of the velocity of the dust species $j$ ($\widehat{\mathrm{St}}_j \sim 1)$ is high. We end up with the general value of $\Omega_j$ in Eq. \eqref{eq:dustywave_collision_mode} for $\mathcal{N}$ species (shown with horizontal dashed lines in Fig. \ref{fig:dustywave}, right panel), which simplifies to the known damping rate for the case of a single dust species $\Omega_d = (1+\theta_d)/t_{\mathrm{s},d}$ (i.e. for $\mathcal{N}=1$).

$\Omega_d$ corresponds to the shortest timescale for which the dust species $d$ is coupled to the gas. This provides an estimation of the recoupling length of the dust species to the mixture:
\begin{equation}
    k_{\mathrm{rec},d}= \Omega_d / \tilde{c}_{\mathrm{s},gd},
    \label{eq:hydro_sonic_wave_dust_recoupling_scale}
\end{equation}
where $\tilde{c}_{\mathrm{s},gd}$ is the speed of the sound waves carried by the gas, the dust fluid $d$ and the dust fluids with smaller stopping times. 
Neglecting the dust mass leads to the estimate $k_{\mathrm{rec},d} \approx 1/(c_{\mathrm{s},g} t_{\mathrm{s},d})$, which was already introduced by \citet{2012MNRAS.420.2345L} and \citet{2019MNRAS.488.5290L} in the context of dustywaves and shocks.

To conclude, we have identified, in pure hydrodynamics, the two fundamental types of mode induced by a dust multifluid approach: classical sonic waves that become mass-loaded and multiple collision modes. In both cases, dust species communicate via drag forces with the gas. The adaptation of a grain to the gas dynamics depends on the importance of the inertia term in the force balance, characterized by the Stokes number. Collision modes model these drifts. These conclusions are extended to the case of a distribution of charged dust in next sections.

\subsection{Collective behaviour of dust grains via the magnetic field: waves in a multidust plasma} \label{sec:multidust_plasma_waves}

Dust fluids can use two intermediate fields to communicate between each other: the gas and the magnetic field. In pure hydrodynamics, high-Stokes grains are decoupled from the gas and therefore they cannot influence each others.
We explore whether, thanks to the magnetic field, dust grains can recover a collective behaviour. To do so, we use the multidust plasma model, presented in Sect. \ref{sec:multidust_plasma_equations}, for which ions and electrons ideally couple to the magnetic field. The complete derivations are in Appendix \ref{app:multidust_plasma_ms}. For $k \to + \infty$, magnetic pressure, coming from the Hall drift as summarized in Eq. \eqref{eq:multidust_plasma_vdx}, can dominate the dynamics over drag forces. For $\omega \gg \nu_d$, dust grains can decouple from the gas and they couple to the magnetic field. Thanks to its inertia, dust carry magnetocompressive waves at small scales. We note that these waves are driven by magnetic pressure and not by thermal pressure and thus we use the term of "magnetocompressive" preferentially to "magnetosonic" even though these two terms refer to the same propagation geometry. We find that the speed of the pair of magnetocompressive waves is
\begin{equation}
    \tilde{c}_{\mathrm{ms},d} =\sqrt{\sum_{d \in I} \xi_d^2 \dfrac{B_0^2}{4 \pi \rho_d}} = c_{\mathrm{a},g} \sqrt{\sum_{d \in I} \dfrac{\xi_d^2}{\theta_d}},
    \label{eq:multidust_magnetosonic}
\end{equation}
where $\xi_d = n_d Z_d /(\sum_{i \in I } n_i Z_i)$ is the proportion of charges carried by the dust fluid $d$ relative to the charge carried by the whole dust distribution and $c_{\mathrm{a},g}  = B_0/\sqrt{4 \pi \rho_g}$ the Alfvén speed associated with the gas mass. Therefore, the speed of the magnetocompressive wave depends on the dust distribution and how we sample it when using a multifluid approach with a few number of dust bins $\mathcal{N}$. For example, for $\mathcal{N}=1$ (case of the single dust fluid), $\xi_d=1$, and thus $\tilde{c}_{\mathrm{ms},d}=c_{\mathrm{a},d}=B_0/\sqrt{4 \pi \varrho_d}$. When increasing the number of dust bins, we account for the dynamics of grains that couple very well to the magnetic field, which increases the speed of the wave. Indeed, the dynamical response of each dust fluid is proportional to the charge-to-mass ratio:
\begin{equation}
    v_{d,x} =\frac{\xi_d}{\theta_d} \left( \frac{c_{\mathrm{a},g}^2}{\pm \tilde{c}_{\mathrm{ms},d}} \frac{\delta b}{B_0} \right) \propto \frac{Z_d}{m_d}.
    \label{eq:multidust_plasma_eigenvector}
\end{equation}
Therefore, for a typical MRN distribution of grains, a power-law size distribution ranging from 5 nm to 250 nm, the biggest contributors to this magnetic mode are the smallest grains. This is because they represent a small fraction of the mass but a significant fraction of the charges.  With the physical setup presented in Sect. \ref{app:physical_setup}, in the protostellar envelope, $\tilde{c}_{\mathrm{ms},d} \approx 37 c_{\mathrm{a},g}$, with $c_{\mathrm{a},g} \approx 56 \ \rm{au/kyr} \approx 270 \ \rm{m/s}$. This is equivalent to a mass carrying the magnetocompressive wave of 
$\rho_{\mathrm{ms},d}=\rho_g c_{\mathrm{a},g}^2/\tilde{c}_{\mathrm{ms},d}^2 \approx 8 \times 10^{-4} \rho_g = 8 \times 10^{-2}  \sum_{d \in I} \rho_d$. Grains smaller than 7 nm contribute to 90\% of the wave speed.

Equation \eqref{eq:multidust_magnetosonic} provides an estimate of the speed of the magnetocompressive wave carried by the dust when neglecting the friction of ions and electrons on the gas. Because of these friction forces, the magnetic field decouples from the dust on the smallest scales. We test the speed estimate for the full model in the regime of the magnetic coupling of the dust in Sect. \ref{sec:multifluid_MHD_ms_modes}.

\subsection{Conditions for the magnetic coupling of dust grains: the regimes of magnetosonic waves} 
\label{sec:single_multifluid_MHD_ms_modes_analysis}

\begin{figure*}
    \centering
    \includegraphics[width=0.49\textwidth]{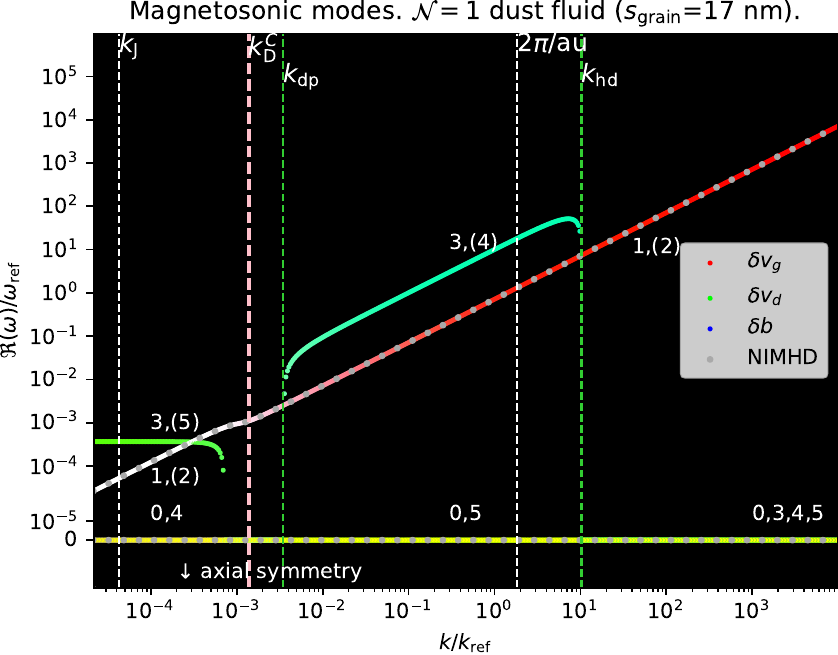}
    \includegraphics[width=0.49\textwidth]{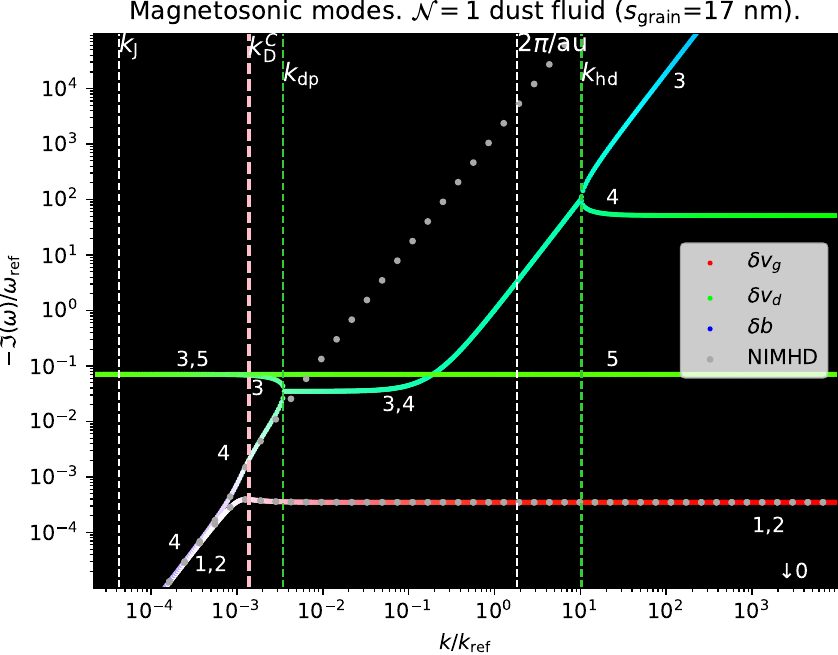}
    
    \caption{Dispersion relation of magnetosonic modes for one fluid of 17 nm grains, for the multifluid MHD model and for a local density of $n_g = 10^{4} \ \rm{cm^{-3}}$. The two panels show the real part and the imaginary part of $\omega/\omega_{\rm{ref}}$ as a function of $k/k_{\rm{ref}}$, where the reference values $\omega_{\rm{ref}}$ and $k_{\rm{ref}}$ are defined in Sect. \ref{app:physical_setup}. Relevant scales are indicated by vertical dashed lines, for instance, $k=2 \pi/\rm{au}$ and $k=k_{\mathrm{J}}$ indicate the 1 au scale and the Jeans scale respectively. The physical content of the eigenvectors is encoded by RGB colors as explained in Sect. \ref{app:eigenvectors_visualisation} and briefly recalled in the legend. The non-ideal MHD solution is over-plotted (gray dots denoted as NIMHD). Numbers from 0 to 5 identify the $2 \mathcal{N}+4=6$ solutions, including the left-propagating solutions with parenthesis, which are symmetric to the right-propagating modes.
    }
    \label{fig:ms_wave_dispersion_17nm_color}
\end{figure*}

\begin{figure}
    \centering
    \includegraphics[width=0.49\textwidth]{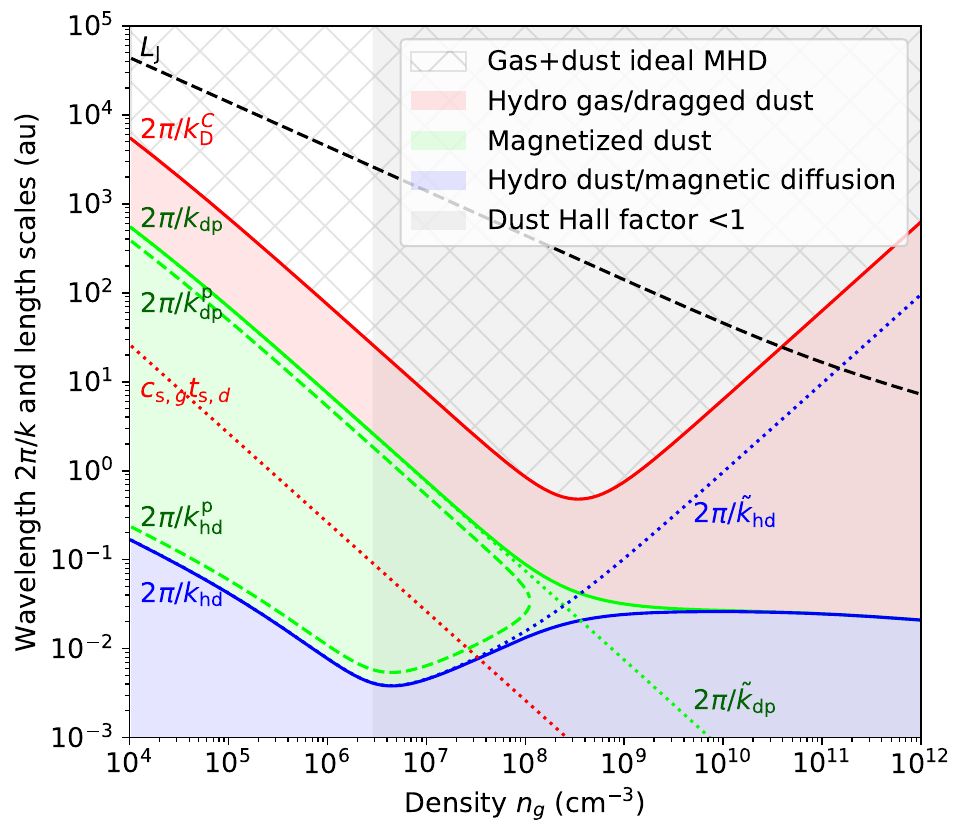}
    
    \caption{Regimes of magnetosonic modes for the case of a single dust fluid of 17 nm grains, as a function of the local gas density $n_g$. These regimes are shown in terms of physical scales by the corresponding wavelength $2 \pi/k$. The different coupling regimes are indicated in the legend: $k<k_{\mathrm{D}}^C$ in white with hatches,  $k_{\mathrm{D}}^C<k<k_{\mathrm{dp}}$ in red, $k_{\mathrm{dp}}<k<k_{\mathrm{hd}}$ in green with $k_{\mathrm{dp}}^{\mathrm{p}}$ and $k_{\mathrm{hd}}^{\mathrm{p}}$ represented with green dashed lines, $k>k_{\mathrm{hd}}$ in blue. The approximations of $k_{\mathrm{dp}}$ and $k_{\mathrm{hd}}$ presented in Eqs. \eqref{eq:diffusion_scale_multifluid_MHD_approximation} and \eqref{eq:magnetized_dust_multifluid_MHD_approximation} are shown with green and blue dots. The decoupling scale $c_{\mathrm{s},g} t_{\mathrm{s},d}$ of a 17 nm grain from the gas (Sect. \ref{sec:multidust_hydro_waves}, in the absence of a magnetic field) is shown with red dots. The Jeans length $L_{\mathrm{J}}=2 \pi/k_{\mathrm{J}}$ is the black dashed line. The density region for which the Hall factor of the grains is smaller than unity is in dark.}
    \label{fig:magnetosonic_coupling_regimes_17nm}
\end{figure}

In this section, we analyze the magnetosonic modes propagating in the gas and dust mixture according to the complete multifluid MHD model (presented in Sect. \ref{sec:multifluid_MHD_equations}). We start with the case of a single dust fluid of 17 nm grains. 
The dispersion relation is presented in Fig. \ref{fig:ms_wave_dispersion_17nm_color}. The details of the derivations are in Appendix \ref{app:multifluid_MHD_ms_modes}.

We recover, on large scales, the ideal MHD regime, for which the gas, the dust and the magnetic field are tightly coupled (pair of modes 1 and 2 in white in Fig.  \ref{fig:ms_wave_dispersion_17nm_color}). The two waves propagate at the magnetosonic speed of the bulk mass. 

As predicted by standard non-ideal MHD, for $k$ larger than $k_{\mathrm{D}}^C$ (pink vertical dotted line), the gas decouples from the magnetic field, due to the magnetic resistivities from the dust, the ions and the electrons. $k_{\mathrm{D}}^C$ is the diffusion scale in non-ideal MHD, presented in Eq. \eqref{eq:NIMHD_resistive_scales}, that is consistently recovered by the multifluid approach (without dust conductivities involved). At these scales, the back-reaction of charged species on the gas via the drag forces is not sufficient to ensure its coupling with the magnetic field. The propagation of waves in the gas is carried by thermal pressure (sonic waves).

Even though the drag force is weak on the gas, compared to the thermal pressure force, the back-reaction on the charged species is strong. At least, this is the case for low-Stokes grains, such as 17 nm grains.

As we reach smaller scales, the effective magnetic pressure on dust increases, until grains couple to the magnetic field efficiently (in Fig. \ref{fig:ms_wave_dispersion_17nm_color}, $k>k_{\mathrm{dp}}$, standing for dustyplasma). Similarly to what we presented in Sect. \ref{sec:multidust_plasma_waves}, the dust fluid carries the propagation of magnetocompressive waves (pair of modes 3 and 4 in cyan).

At smaller scales, the dust decouples from the magnetic field perturbation (in Fig. \ref{fig:ms_wave_dispersion_17nm_color}, $k>k_{\mathrm{hd}}$, standing for hydrodynamical dust) contrary to the multidust plasma model (damped mode 4 in green). This is due to collisions of ions and electrons with the gas. The magnetic field is diffused (damped mode 3 in blue). 

We analyze these coupling regimes in details. We found that the dust fluid couples to the magnetic perturbation at intermediate scales, when $k_{\mathrm{dp}}<k <k_{\mathrm{hd}}$. The dynamical system coupling the dust velocity with the magnetic field, in Eq. \eqref{eq:multifluid_MHD_monograin_ODE}, is similar to the non-ideal MHD equations, but with the ion-electron resistivities, $\eta_{\mathrm{O}}^L$ and $\eta_{\mathrm{AD}}^L$, and the Alfvén speed associated with the dust mass, $c_{\mathrm{a},d}$. The Hall term restores the magnetic pressure (details in Appendix \ref{app:multifluid_feedback_Lorentz_forces}). The effective collision rate on the dust fluid is
\begin{equation}
    \nu_{\mathrm{D}} = \nu_d + \frac{\omega_d^2}{\omega_{\mathrm{D}}} \quad \text{with} \quad  \omega_{\mathrm{D}} = \frac{c_{\mathrm{a},d}^2}{\frac{c^2}{4 \pi} (\eta_{\mathrm{O}}^L + \eta_{\mathrm{AD}}^L)}.
\end{equation}

The exact expressions of the scales $k_{\mathrm{dp}}$ and $k_{\mathrm{hd}}$, that are given in Eq. \eqref{eq:multifluid_MHD_kdp_khd}, are difficult to analyze. We can provide an approximation assuming that the ions and the electrons are very well coupled to the magnetic field. In this case, $\nu_d, \omega_d \ll \omega_{\mathrm{D}}$, and we obtain
\begin{equation}
   k_{\mathrm{hd}} \approx \tilde{k}_{\mathrm{hd}} = \frac{2 c_{\mathrm{a},d}}{ \frac{c^2}{4 \pi} ( \eta_{\mathrm{O}}^L+ \eta_{\mathrm{AD}}^L )},
   \label{eq:diffusion_scale_multifluid_MHD_approximation}
\end{equation}
 \begin{equation}
     k_{\mathrm{dp}} \approx \tilde{k}_{\mathrm{dp}} = \frac{\nu_{\mathrm{D}}}{2 c_{\mathrm{a},d}}.
     \label{eq:magnetized_dust_multifluid_MHD_approximation}
 \end{equation}   
 
The approximation of $k_{\mathrm{hd}}$ is similar to the diffusion scale $k_{\mathrm{D}}^C$ in standard non-ideal MHD but with the wavespeed $c_{\mathrm{a},d}$ and the ion-electron resistivities. The approximation of $k_{\mathrm{dp}}$ is reminiscent of the propagation scale we found for the multidust plasma model (Appendix \ref{app:multidust_plasma_ms}). They are good estimates for $n_g<10^8~\rm{cm^{-3}}$. We also provide the scales ${k}^{\mathrm{p}}_{\mathrm{hd}}, {k}^{\mathrm{p}}_{\mathrm{dp}}$ in Eq. \eqref{eq:multifluid_MHD_kdp_khd_propagation} such that the propagative part of the magnetocompressive wave is higher than the damping part. 

All the scales delimiting the coupling regimes are plotted in Fig. \ref{fig:magnetosonic_coupling_regimes_17nm}. We can see that at low density, the range of scales for which 17 nm grains decouple to the magnetic field is wide. It ranges from 0.5 au to 500 au at $n_g = 10^4 \ \rm{cm^{-3}}$. These typical scales decreases as $1/n_g$. From $n_g > 3 \times 10^6 \ \rm{cm^{-3}}$, the range of the coupling regime narrows. This coincides with the Hall factor of the dust grains 
becoming smaller than unity.
The 17 nm dust grains cannot carry the magnetocompressive wave for $n_g > 10^{8} \ \rm{cm^{-3}}$. 

These scales are macroscopic and current numerical simulations can reach such corresponding spatial resolution. We should be able to see the effect of the grain inertia on the coupling with the gas and the magnetic field. One same grain can be either coupled to the gas or/and to the magnetic field, depending on the scales at stake. Its preferential coupling is not entirely defined by its Hall factor.

We note that, at the collapse scale, which is estimated by the Jeans length in Fig. \ref{fig:magnetosonic_coupling_regimes_17nm}, the gas and the charges remain tightly coupled in the magnetosonic modes at the typical densities of dense cores.
At this scale, the collapsing matter is ideally coupled to the magnetic field, and therefore, the standard criteria for triggering the collapse of a dense core based on the competition between gravity and thermal and magnetic support \citep{1976ApJ...210..326M} remain unchanged by a multifluid description (nor by a non-ideal MHD description at low densities). More precisely, the two-fluid model of \citet{2011MNRAS.415.1751M}, once linearized, enables to demonstrate that there is an additional intermediate regime for which an instability occurs on timescales longer than the classical thermal Jeans instability thanks to ambipolar diffusion.\footnote{At large scales, there are three modes involved: two ideal magnetosonic modes (1 and 2 in Fig. \ref{fig:magnetosonic_coupling_regimes_17nm}) and one purely damped pressure driven mode (4). The pressure driven mode becomes purely unstable at wavelengths larger than the thermal Jeans length and therefore turns into a gravitationally-driven ambipolar diffusion mode. At wavelengths larger than the magnetosonic Jeans length, the two propagating magnetosonic modes become purely damped and bifurcate. It affects the gravitationally-driven ambipolar diffusion mode that becomes the Jeans freefall mode against the magnetic pressure and the thermal pressure.} This mode is the linear counterpart of a core collapse process known as ambipolar-diffusion–induced gravitationally-driven fragmentation
\citep{1987ASIC..210..453M,1991ApJ...373..169M}. 
Even if it is not presented in this paper, we checked that this result remains valid for an arbitrary number of charged (dust) fluids when including the self-gravity of the gas and the dust fluids.

\subsection{Full description of the modes in multifluid magnetohydrodynamics} \label{sec:multifluid_MHD_modes}


\subsubsection{Alfvén modes} \label{sec:multifluid_MHD_alfven_modes}

\begin{figure*}
    \centering
    \includegraphics[width=0.49\textwidth]{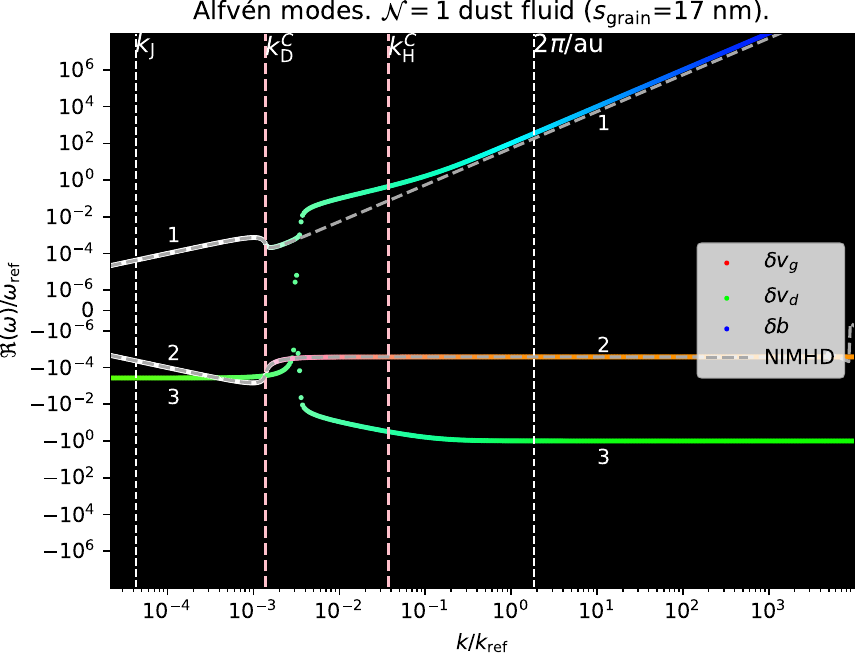}
    \includegraphics[width=0.49\textwidth]{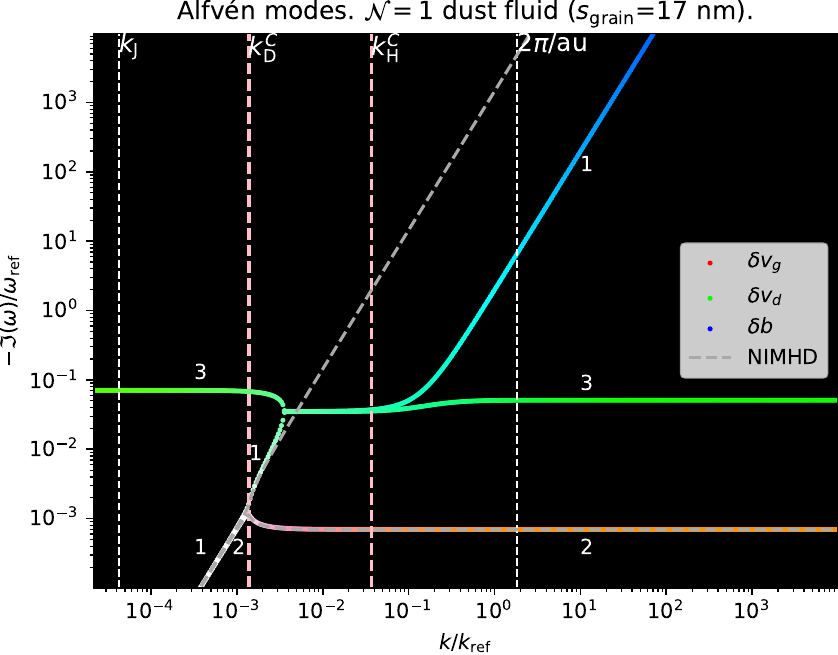}

    \includegraphics[width=0.49\textwidth]{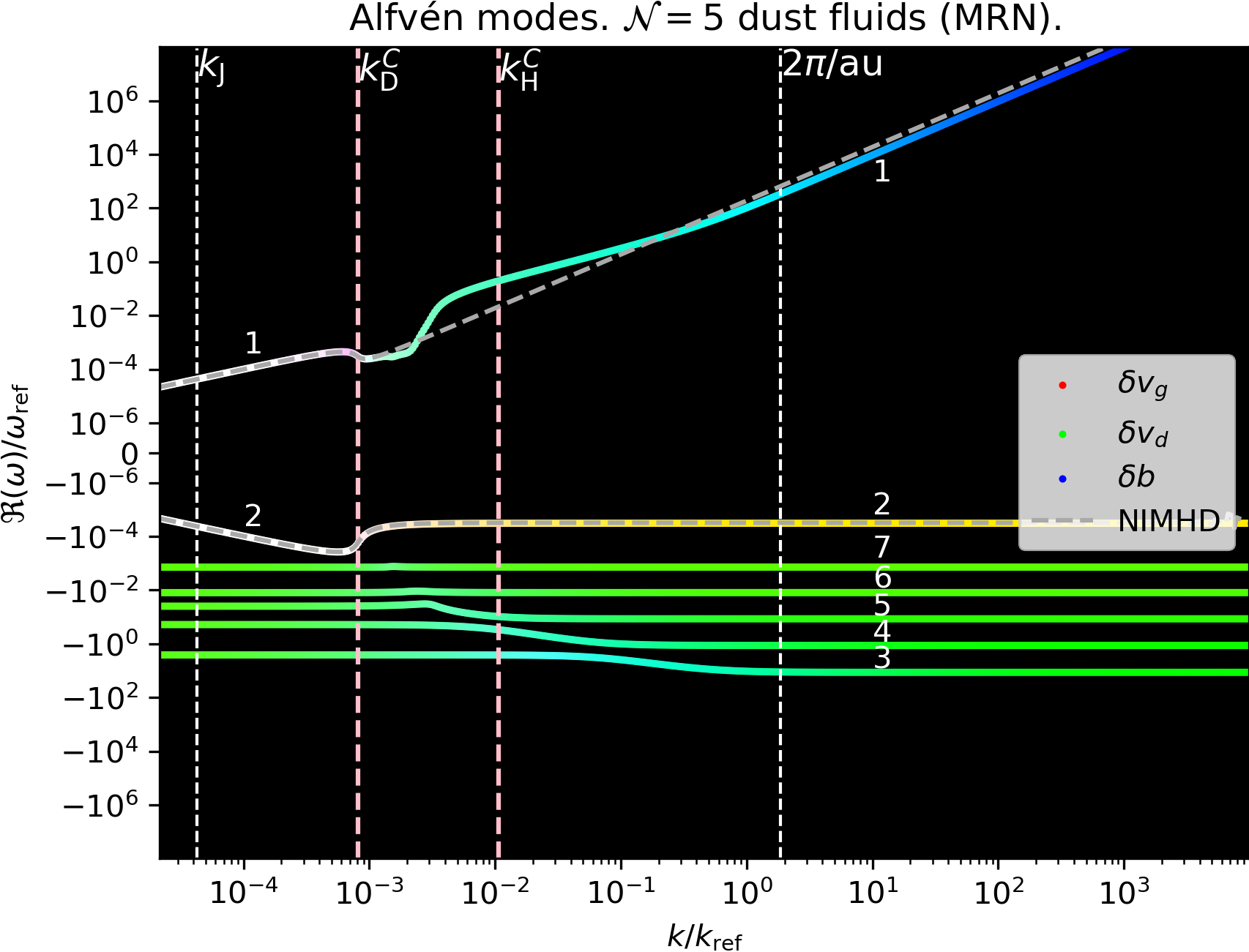}
    \includegraphics[width=0.49\textwidth]{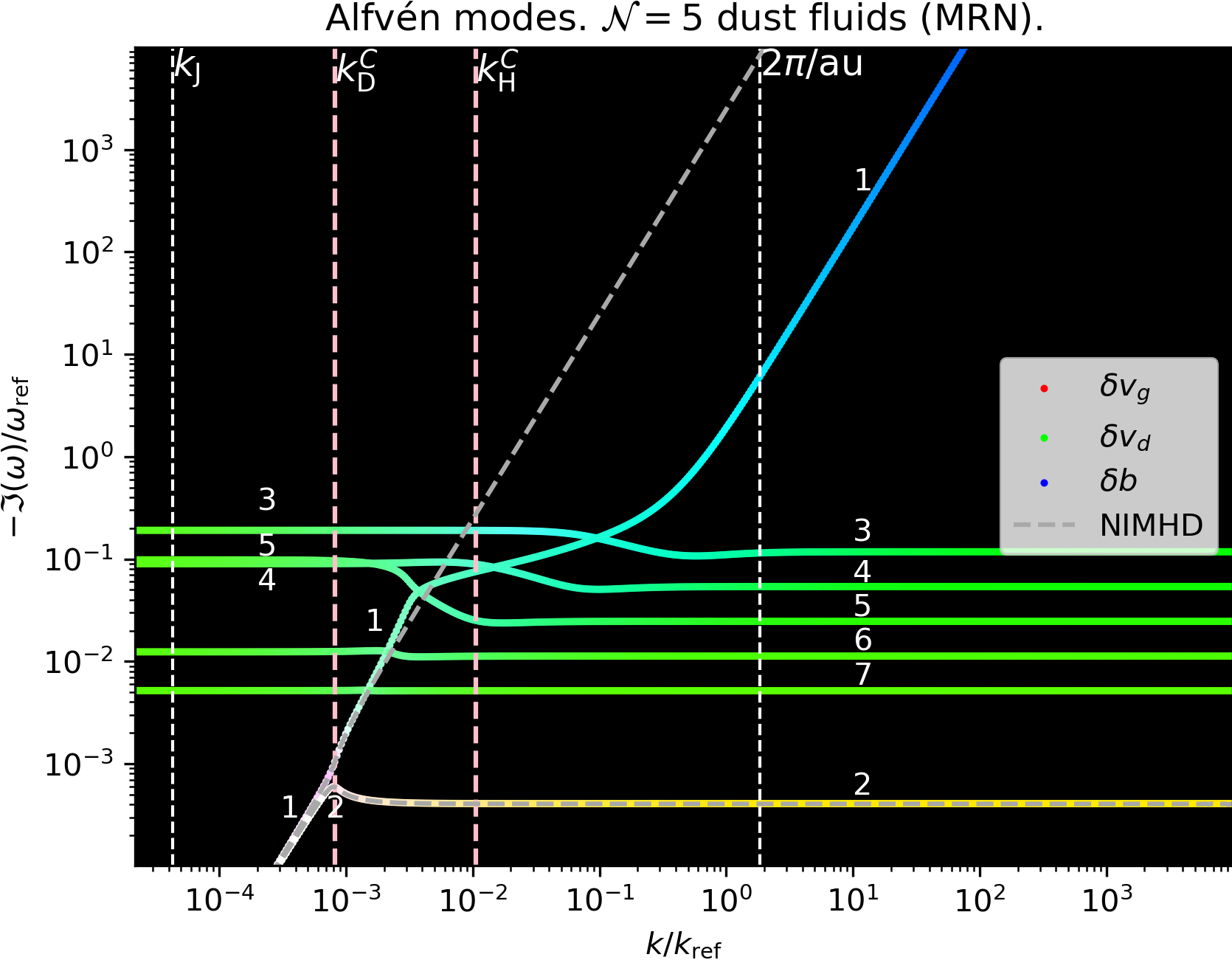}

    \includegraphics[width=0.49\textwidth]{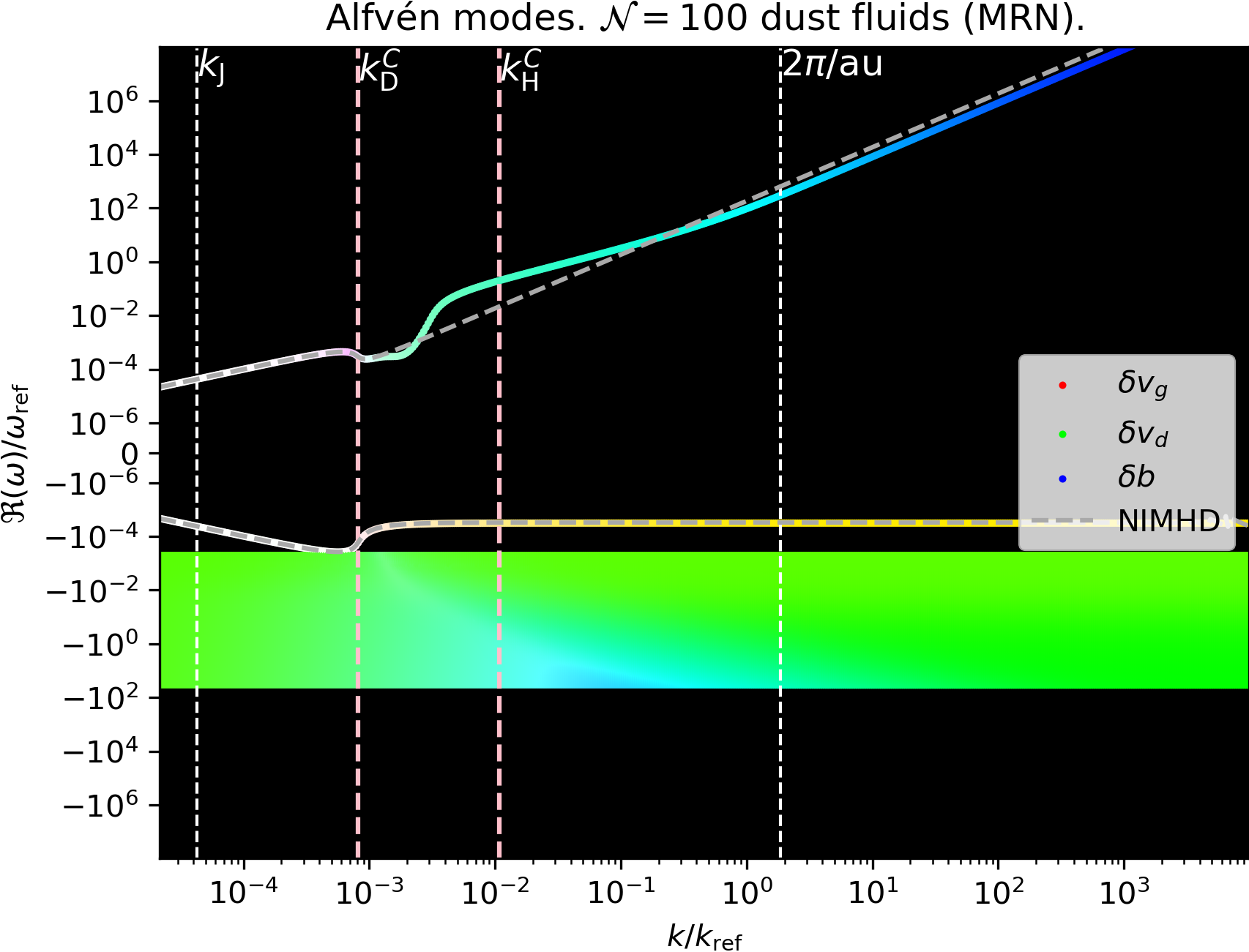}
    \includegraphics[width=0.49\textwidth]{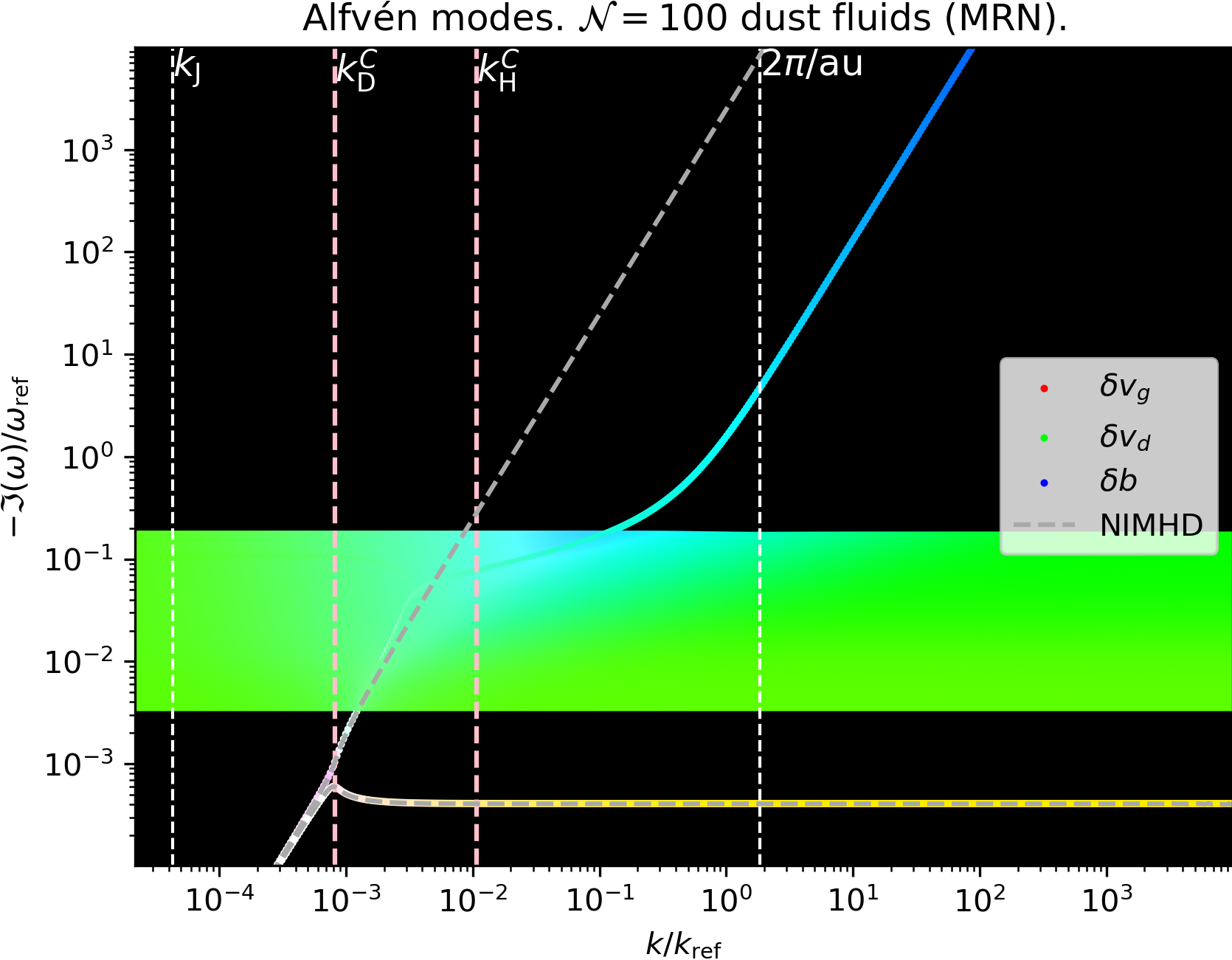}

    \caption{Dispersion relation of Alfvén waves ($\sigma=1$) for $\mathcal{N} =1,5,$ and $100$ dust fluids sampling a MRN distribution (each line). Numbers from $1$ to $\mathcal{N}+2$ identify the complete set of the multifluid MHD Alfvén modes.  Complementary information can be found in the description of Fig \ref{fig:ms_wave_dispersion_17nm_color}.}
    \label{fig:alfven_wave_dispersion_color}
\end{figure*}

In this section, we present the physics of the Alfvén waves, which are fundamental MHD waves, to complement the results of \citet{2023A&A...674A.149H}. We plot the dispersion relation for $\mathcal{N}=1,5,$ and $100$ dust fluids sampling the MRN distribution in Fig. \ref{fig:alfven_wave_dispersion_color}. The derivations are in Appendix \ref{app:multifluid_MHD_alfven_modes}. 

We recover modified versions of the two standard non-ideal MHD modes (gray dotted lines in Fig. \ref{fig:alfven_wave_dispersion_color}), that are the right-hand circularly polarized mode (identified as the mode 1 in Fig. \ref{fig:alfven_wave_dispersion_color}) turning to a  whistler mode at small scales ($\omega \propto k^2$ when $k \to + \infty$) and the left-hand circularly polarized mode (identified as the mode 2 in Fig. \ref{fig:alfven_wave_dispersion_color}, such that $\omega \propto k^0$ when $k \to + \infty$). In addition, we observe $\mathcal{N}$ new modes (modes 3 to $\mathcal{N}+2$ in Fig. \ref{fig:alfven_wave_dispersion_color}), which are reminiscent of the collision modes in pure hydrodynamics (Sect. \ref{sec:multidust_hydro_waves}), but with a real part for $\omega$.

We recover the standard non-ideal MHD solutions for large timescales, such that the dust inertia term is small. More precisely, we recover ideal MHD at large scales (modes 1 and 2 in white) and the standard left-handed mode at scales smaller than the resistive scale $1/k_{\mathrm{D}}^C$ (mode 2 in orange or yellow), which is when the gas decouples from the magnetic field (Appendix \ref{app:NIMHD_Alfvén} on non-ideal MHD). The whistler asymptote is modified by the multifluid approach: in multifluid MHD, thanks to grain inertia, the dust fluids can couple to the magnetic perturbation at intermediate scales (mode 1 in cyan) and decouple from the magnetic perturbation at small scales (mode 1 in blue) whereas in standard non-ideal MHD (in gray), the charged dust, the ions and the electrons remain coupled to the magnetic perturbation at small scales. We detail this regime later.

We now discuss the existence of the $\mathcal{N}$ other modes to clarify the results of \citet{2023A&A...674A.149H}. In their paper, they found that a charged dust distribution produces a series of left-hand-polarized Alfvén modes, but the underlying physics was missing.

Because of the inertia of the dust fluids, dust velocities do not necessarily satisfy the velocity drifts imposed in standard non-ideal MHD (Eq. \eqref{eq:light_drift} on light species). 
The possible velocity drifts between the fluids can be decomposed according to the eigenbasis, that are the solutions whose transport 
is constant over time, because they satisfy a specific force balance (including the inertia term).
There are $\mathcal{N}$ new modes, which can be interpreted as a number of degrees of freedom. For instance, we can choose the initial conditions of the $\mathcal{N}$ dust velocities. The time and space scales at play modify the dominating forces, so they control the possible force balances, degrees of coupling and back-reactions between the gas, the dust fluids and the magnetic field.

We go on to discuss the physics of these modes. The details of the solutions are in Appendices \ref{app:multifluid_MHD_alfven_modes_large_scales_collision} and \ref{app:multifluid_MHD_alfven_modes_small_scales_collision}. These modes are driven by the Lorentz forces, in addition to drag forces as presented in pure hydrodynamics. While drag forces tend to reduce velocity drifts with the gas, the Lorentz forces tend to separate the charges from the gas (typically drifting at the E-cross-B velocity), which affects the chemical separation and the electric currents. The possible velocity drifts depend on the Stokes number and the Hall factor of the grains. We require these two parameters that are usually used in distinct contexts: the Stokes number to model the dust grain inertia in multifluid hydrodynamics as in Sect. \ref{sec:multidust_hydro_waves} and the Hall factor to get the velocity drift of charged species in standard non-ideal MHD as in Eq. \eqref{eq:light_drift}.

The physics of the collision modes depends on the scale. At large scales ($k \to 0$), dust fluids can communicate via the electric field, but we can consider that the forces due to fluid velocity drifts (such as the drag forces) dominate the forces produced by the net electric current (forces restoring the magnetic tension in the monofluid formalism), contrary and complementary to the two propagating ideal Alfvén modes. 
At small scales ($k \to + \infty$), the magnetic field perturbation cannot induce any electric field perturbation, and therefore the dust fluids can only communicate via collisions with the gas. We note that, contrary to sonic waves where the gas dynamics is dominated by pressure forces at small scales (Sect. \ref{sec:multidust_hydro_waves}), there is no mechanism to fully decouple the gas from the dust fluids. Therefore, gas and dust fluids remain coupled in the left-handed modes (standard mode 2 and collision modes). 

In addition to this coupling through the electric field, another difference with the collision modes in pure hydrodynamics relies on the gyration term of the Lorentz force ($\propto \omega_d \mathbf{V}_d \times \mathbf{B}$). This term is spatial-scale independent in the linear regime, like drag forces, which explains why $\omega$ is mainly constant with $k$ for these modes. The gyration term does not produce any additional damping, contrary to drag forces, because it preserves the kinetic energy of the fluid. Ignoring the feedback of the gas and the electric field on the dust dynamics (decoupled grains) in Eq. \eqref{eq:multifluid_MHD_alfven_dust_dynamics} leads to the typical frequency $\Omega_{d, \sigma} = \sigma \omega_d - i \nu_d$ (small black dots in Fig. \ref{fig:alfven_wave_dispersion_analytical}), which justifies that the collision modes are left-handed ($\Re(\Omega_{d, \sigma})<0$ for $\sigma=1$). In this case, the collision modes appear to be analogous to dust-cyclotron modes (but with a damping part induced by drag forces).

We found that at some specific intermediate scales, the collision modes allow the propagation of Alfvén waves carried by the dust fluids (in cyan). More precisely, this propagation switches from one mode to another of increasing frequency with increasing $k$. The propagation in left-handed modes appears symmetric to the wave propagation in the right-handed mode (mode 1), as already identified in \citet{2023A&A...674A.149H}. It is obvious for the case $\mathcal{N}=1$. When we increase the number of dust fluids $\mathcal{N}$, the $\mathcal{N}$ individual collision modes are more narrow in frequencies $\omega$ and less varying with $k$ (such that the group velocity of these modes, if relevant for a distribution, seems to decrease with $\mathcal{N}$), until forming a continuous band when $\mathcal{N} \to + \infty$. 
This probably reflects the fact that each individual bin, which carries a very small fraction of the mass and of the charge, back-reacts very weakly on the gas and the magnetic field compared to the rest of the distribution. However, we note that the overall symmetry seems preserved when selecting the solutions for which the coupling of the dust fluids to the magnetic perturbation is the strongest (i.e., by following the pieces of the modes in cyan).

We analyze the coupling of the dust fluids with the magnetic perturbation. At scales smaller than the diffusion scale $1/k_{\mathrm{D}}^C$, for which the gas becomes weakly coupled to the charges, the dust fluids can carry the propagation of Alfvén waves driven by magnetic tension forces. At the larger scales of the coupling regime, the velocity of the Alfvén wave is a mass-loaded Alfvén speed:
\begin{equation}
    \tilde{c}_{\mathrm{a},d} = \frac{B_0}{\sqrt{4 \pi \sum_{d \in I, \mathrm{St_m}_{,d} \lesssim  1, | \Gamma_d|, \gtrsim 1} \rho_d}}.
    \label{eq:multifluid_MHD_dust_Alfven_speed}
\end{equation}

The contributors of the mass loading are the fluids with a high Hall factor $|\Gamma_d| = |\omega_d|/\nu_d \gtrsim 1$ and a low magnetic Stokes number $\mathrm{St_m}_{,d}=|\Re(\omega)/\omega_d| \lesssim 1$. More details on the solution can be found in Appendix \ref{app:multifluid_Alfven_all_inertia}. The velocity of the Alfvén wave is therefore higher than the Alfvén speed of the dust bulk mass (i.e., $\tilde{c}_{\mathrm{a},d}>{c}_{\mathrm{a},d}$), as for the dust magnetocompressional wave (even though the expression is different).  This mass loading is reminiscent of the mass loading of the sound wave by the low-Stokes neutral grains in Sect. \ref{sec:multidust_hydro_waves}. Charged fluids with a low magnetic Stokes number
and a high Hall factor can adapt their dynamics to the magnetic
field. Fore a MRN distribution, they correspond to the smallest grains.

At higher frequency, with the increase of the magnetic
Stokes number, dust fluids decouple from the magnetic perturbation, letting only the electrons and the ions coupled to the magnetic perturbation in the whistler asymptote. It generalizes the result in standard non-ideal MHD where the gas and the magnetic perturbation decouple at small scales (see Appendix \ref{app:NIMHD_Alfvén}, and the discussion on the consequences for designing numerical methods based on Riemann solvers). In multifluid MHD, because dust fluids decouple from the magnetic perturbation in the whistler mode, the asymptote (mode 1 in blue) relying on the ion-electron resistivities, $\eta_{\mathrm{O}}^L$, $\eta_{\mathrm{H}}^L$ and $\eta_{\mathrm{AD}}^L$, is different from the asymptote in non-ideal MHD (in gray), which relies on the standard resistivities $\eta_{\mathrm{O}}^C$, $\eta_{\mathrm{H}}^C$ and $\eta_{\mathrm{AD}}^C$ (see Eq. \eqref{eq:multifluid_MHD_alfven_whistler_small}). Therefore, the damping rate of whistler waves is lower than predicted by standard non-ideal MHD.

\subsubsection{Magnetosonic modes} \label{sec:multifluid_MHD_ms_modes}

\begin{figure*}
    \centering
    \includegraphics[width=0.49\textwidth]{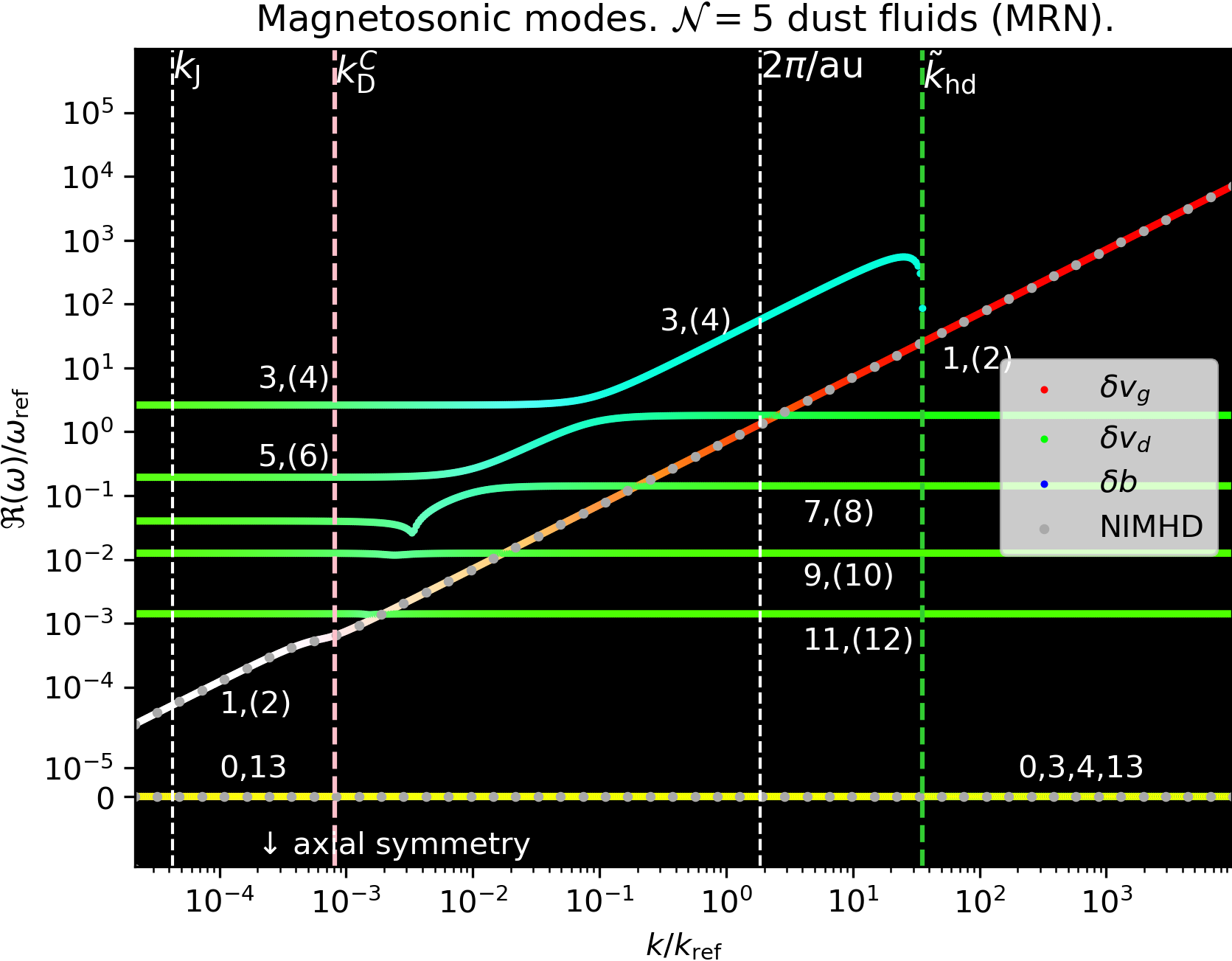}
    \includegraphics[width=0.49\textwidth]{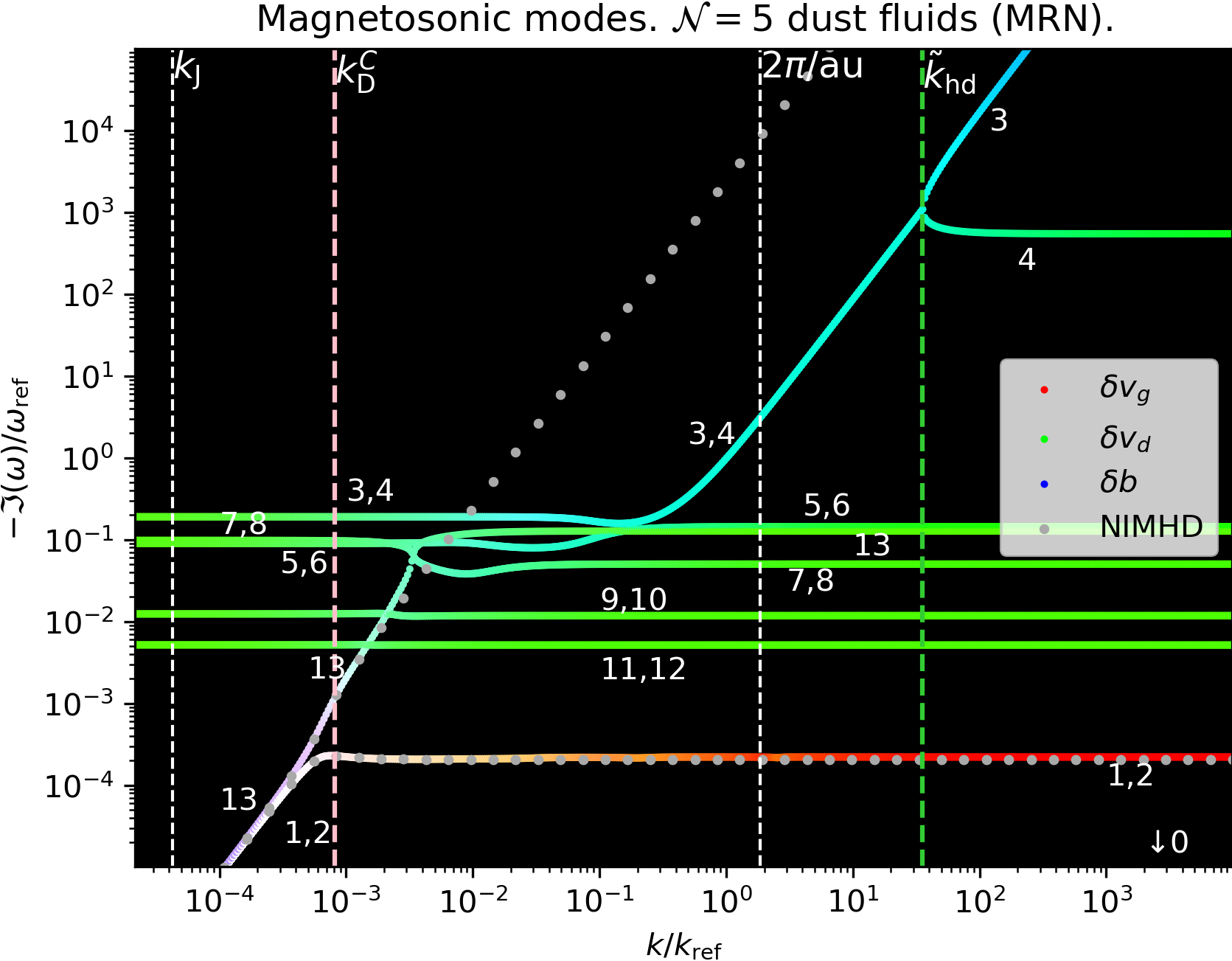}

    \includegraphics[width=0.49\textwidth]{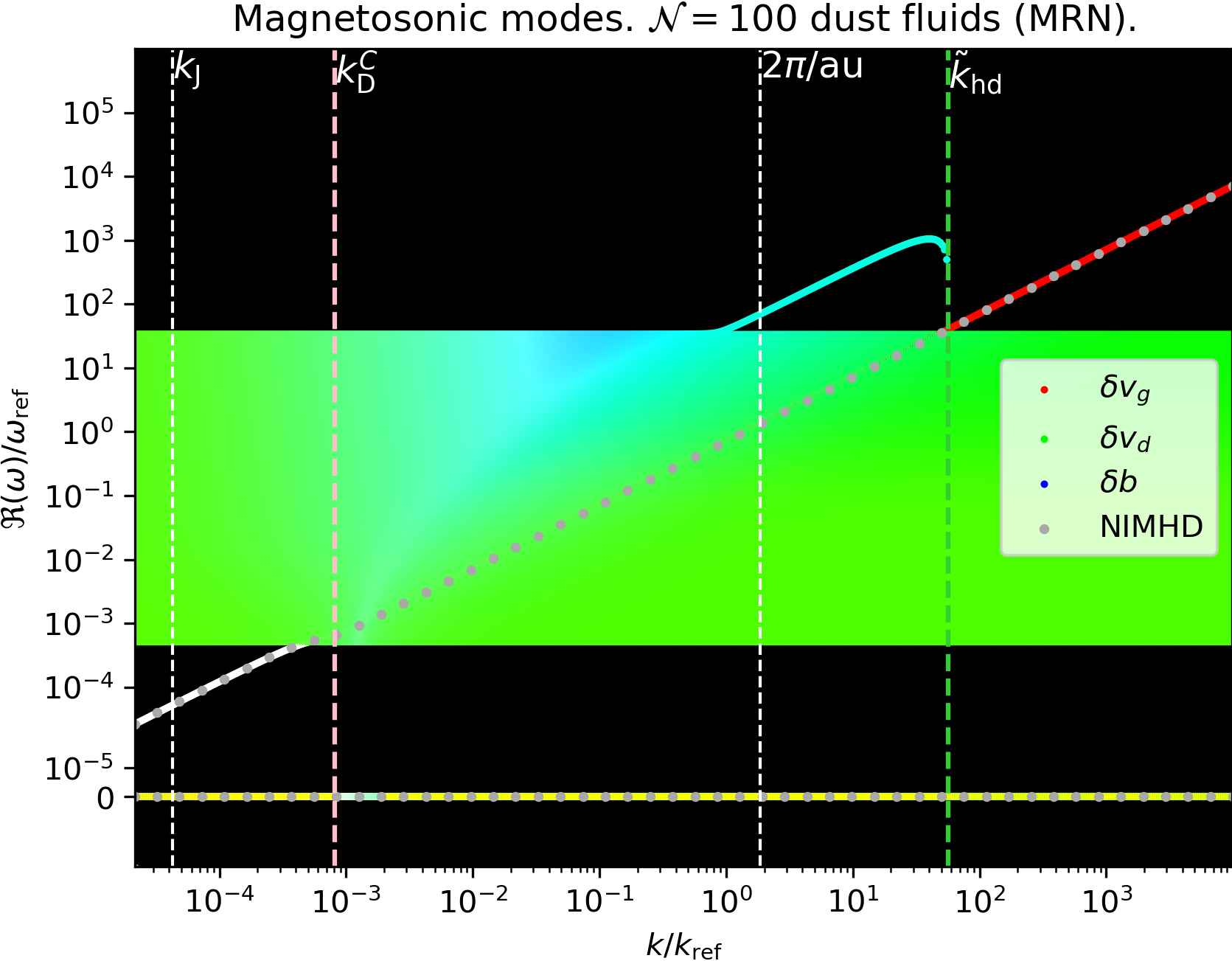}
    \includegraphics[width=0.49\textwidth]{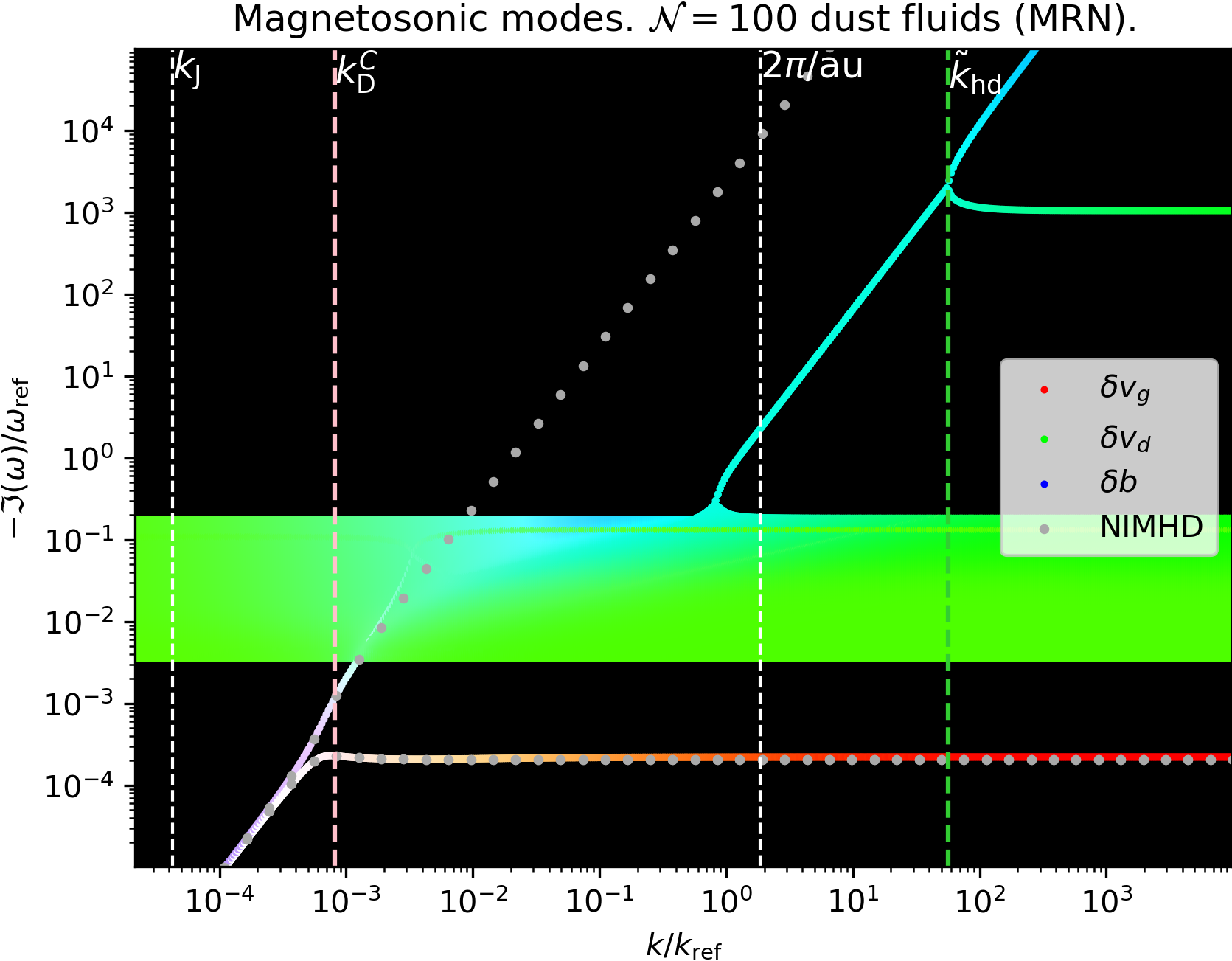}

    \caption{Dispersion relation of magnetosonic modes for $\mathcal{N} =5$ and $100$ dust fluids sampling a MRN distribution (each line). Numbers from $0$ to $2\mathcal{N}+3$ identify the complete set of multifluid MHD Alfvén modes. The case $\mathcal{N}=1$ is in Fig. \ref{fig:ms_wave_dispersion_17nm_color}. 
    Complementary information can be found in the description of Fig \ref{fig:ms_wave_dispersion_17nm_color}.}
    \label{fig:ms_wave_dispersion_5bins_100bins_color}
\end{figure*}

We now study the propagation of the magnetosonic waves and gather the preliminary results we obtained in previous sections.  We plot the dispersion relation for the case of a single dust fluid in Fig. \ref{fig:ms_wave_dispersion_17nm_color} (already presented in Sect. \ref{sec:single_multifluid_MHD_ms_modes_analysis}), and for $\mathcal{N}=5$ or $100$ dust bins sampling the MRN distribution in Fig. \ref{fig:ms_wave_dispersion_5bins_100bins_color}. The derivations are in Appendix \ref{app:multifluid_MHD_ms_modes}. 

As discussed in Sect. \ref{sec:single_multifluid_MHD_ms_modes_analysis}, the gas dynamics goes from the ideal MHD regime for $k<k_{\mathrm{D}}^C$ (modes 1 and 2 in white) to the hydrodynamical regime driven by the thermal pressure (modes 1 and 2 in red). As $\omega$ increases, dust fluids decouple from the sonic waves (shades from yellow to orange), as predicted in Sect. \ref{sec:multidust_hydro_waves}.

The direct consequence is that dust fluids can couple progressively to the magnetic field perturbation. The dust magnetocompressive modes we studied in Sections \ref{sec:multidust_plasma_waves} and \ref{sec:single_multifluid_MHD_ms_modes_analysis} are the continuation of the collision modes at large scales ($k \to 0$, modes in green). We note that studying the physics at $k \to 0$ with finite $\omega$  suppresses the geometry imposed by the propagation direction $\mathbf{k}$, that drives magnetic pressure and thermal pressure forces. Therefore, the physics of collision modes at large scales are the same as described in the section on Alfvén modes (Sect. \ref{sec:multifluid_MHD_alfven_modes}). At large scales, dust velocities already communicate via the electric field perturbations but there is no significant electric current. As $k$ increases, a net electric current appears and magnetic pressure forces can drive the dust magnetocompressive waves (modes in cyan). The velocity of the dust grains that are coupled to the right-propagating dust magnetocompressive wave tends to be in phase with the magnetic perturbation (and in phase opposition for the left-propagating wave),
until a new fluid of grains couple to the magnetic field perturbation at smaller scales. Therefore, the propagation of the magnetocompressive waves switches from one mode to another of increasing frequency when increasing $k$, similarly to what we found with the dust Alfvén waves (Sect. \ref{sec:multifluid_MHD_alfven_modes}). When all dust fluids couple to the magnetic field (modes 3 and 4 in cyan), we recover the magnetocompressive waves computed with the multidust plasma model (Sect. \ref{sec:multidust_plasma_waves}). In particular, the expression of the magnetocompressive wave speed $\tilde{c}_{\mathrm{ms},d}$ as a function of the dust distribution in Eq. \eqref{eq:multidust_magnetosonic} seems to be a very good estimate for the full model (blue dotted line in Fig. \ref{fig:ms_wave_dispersion_analytical}). This is probably because ions and electrons remain highly coupled to the magnetic field at densities for which grains can carry MHD waves.

At scales smaller than a characteristic scale, the collisions of the ions and the electrons with the gas matter. This can be interpreted as the separation of the charges due to collisions. The approximation of this cut-off scale in Eq. \eqref{eq:diffusion_scale_multifluid_MHD_approximation} denoted as $\tilde{k}_{\mathrm{hd}}$ in figures, consisting in a resistive scale based on the ion-electron resistivities and dust magnetosonic speed, seems to extend correctly to the multi-bin case.

\section{Astrophysical discussions} \label{sec:astrophysical_discussions}

\subsection{Dust enrichment within the protostellar envelope}
\label{sec:single_multifluid_MHD_ms_modes_experiment}

\begin{figure*}
    \centering

    \includegraphics[width=0.32\textwidth]{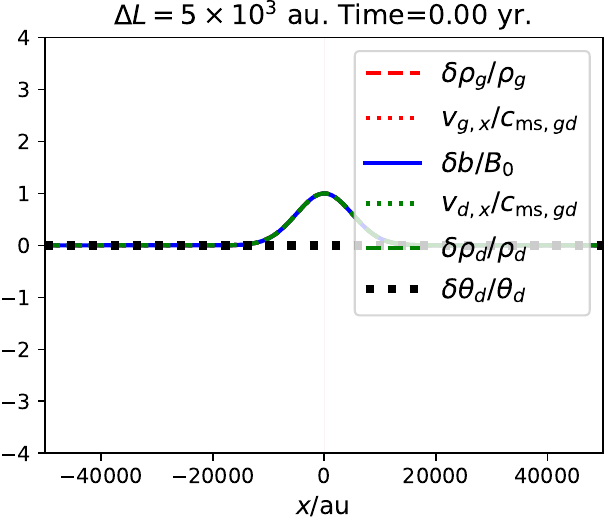}  
\includegraphics[width=0.32\textwidth]{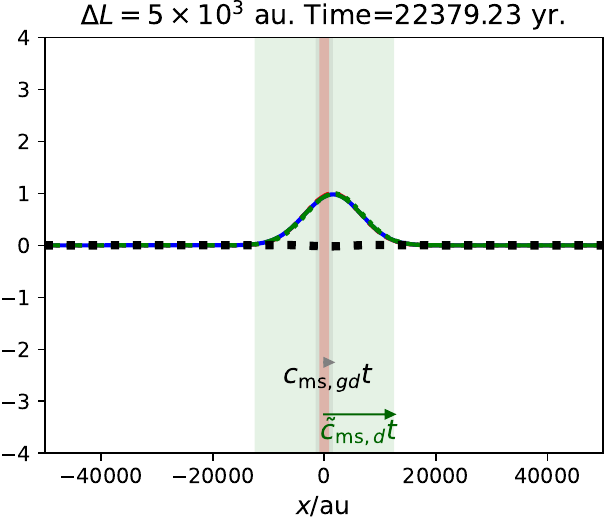}
\includegraphics[width=0.32\textwidth]{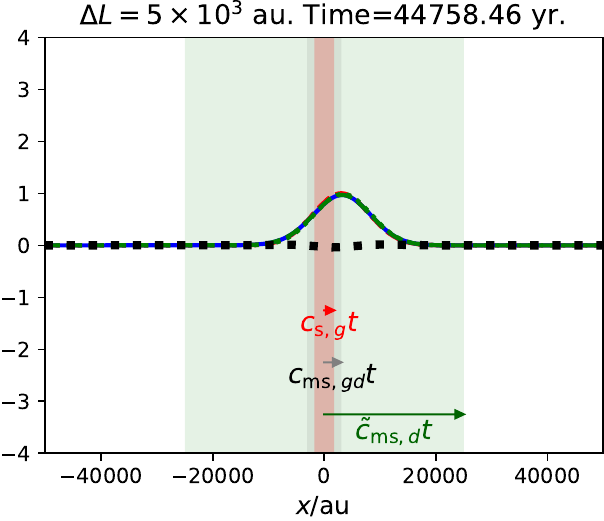}

    \includegraphics[width=0.32\textwidth]{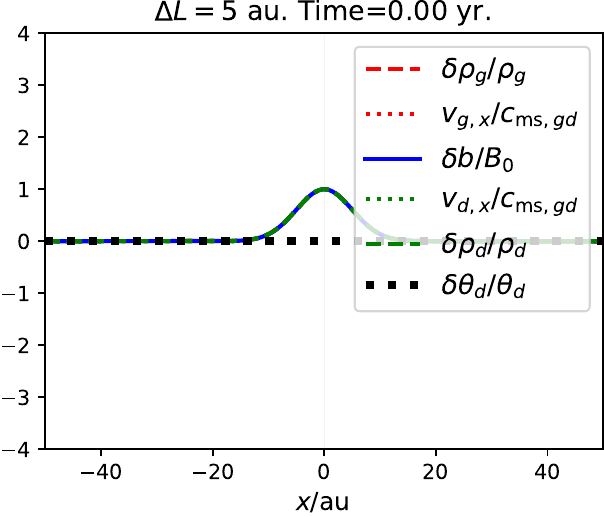}
\includegraphics[width=0.32\textwidth]{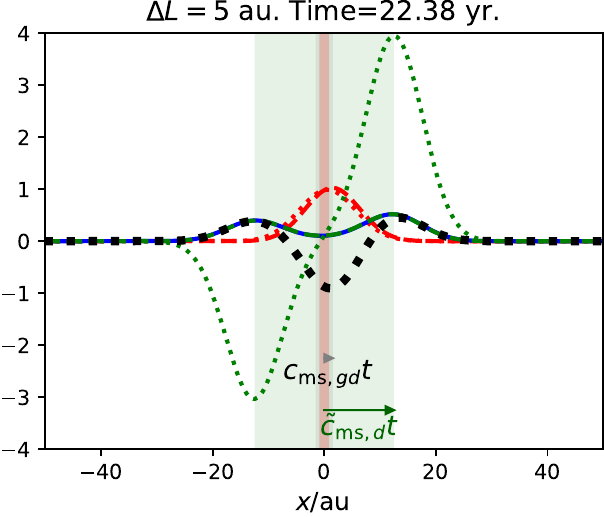}   \includegraphics[width=0.32\textwidth]{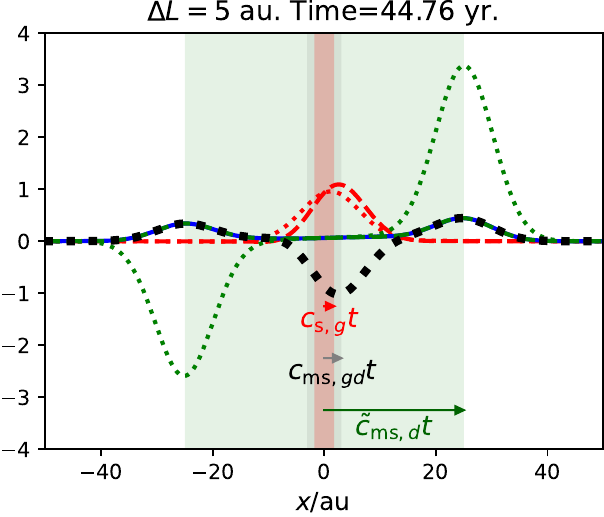}
    
    \includegraphics[width=0.32\textwidth]{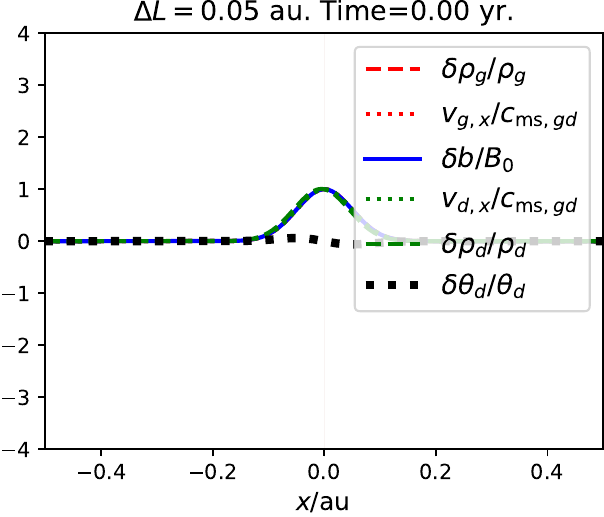}
\includegraphics[width=0.32\textwidth]{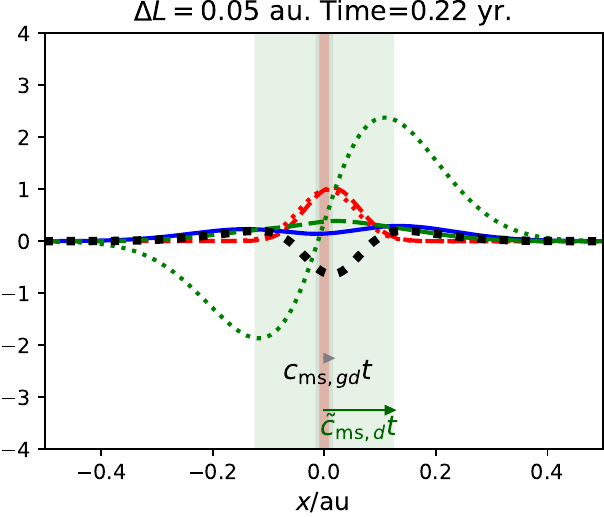}
\includegraphics[width=0.32\textwidth]{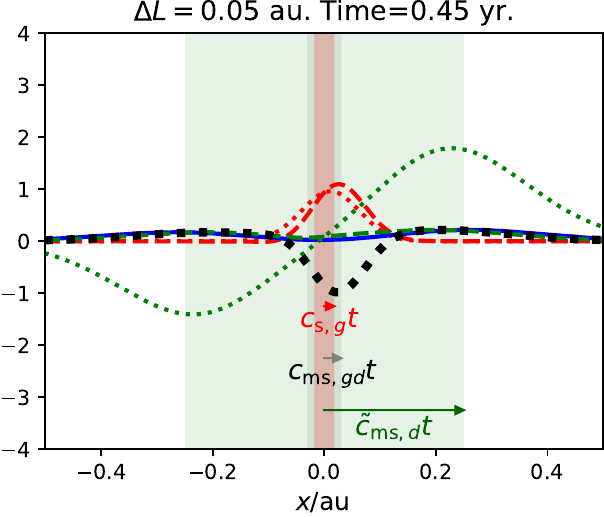}
    
    \caption{Evolution of a perturbation of spatial extent $\Delta L$ over a period of time $5 \Delta L/\tilde{c}_{\mathrm{ms},d}$ in the physical conditions at $n_g = 10^4 \ \rm{cm^{-3}}$. We show the amplitudes of the perturbations, relative to the ideal MHD solution. Upper panels for $\Delta L=5000$ au, middle panels for $\Delta L=5$ au, lower panels for $\Delta L=0.05$ au. The time of each snapshot is indicated in years. The waves propagating on distances $c_{\mathrm{s},g} t$, $c_{\mathrm{ms},gd} t$ and $\tilde{c}_{\mathrm{ms},d} t$ from the origin are represented in red, in gray and in green respectively.}
    \label{fig:perturb}
\end{figure*}

In this section, we demonstrate that the newly identified magnetocompressive waves do indeed produce variations in dust density. Because the dust decouples from the gas, the local dust-to-gas ratio varies. 

For a given mode $(k,\omega(k,n))$, where $n$ is an integer indexing a solution at $k$, the dust density variations and the dust-to-gas ratio variations are given by the mass conservation equations (neglecting the mass exchanges terms to focus on hydrodynamical decoupling):
\begin{equation}
    \delta \rho_d(k,n)  = \rho_d 
    \frac{v_{d,x}(k,n)}{\omega(k,n)/k},\label{eq:perturb_dust_density_mass_conversion}
\end{equation}
\begin{equation}
    \delta \theta_d(k,n)  = \theta_d \frac{v_{d,x} (k,n) - v_{g,x} (k,n)}{\omega(k,n)/k}.
    \label{eq:perturb_dust_to_gas_mass_conversion}
\end{equation}

Equation \eqref{eq:perturb_dust_to_gas_mass_conversion} means that relative dust-to-gas ratio  variations are equal to the dust velocity drift divided by the phase velocity of the wave. However, in physical conditions, the spatial extent of the perturbations is finite. The perturbations can be represented by a precise superposition of eigenmodes. In this case, the relation on the dust density and the dust-to-gas ratio become complex. 

We lead numerical experiments of the propagation of perturbations of spatial extent $\Delta L$. This is analytical and still in the linear regime. The dynamics of the perturbations can be obtained by superposing the dynamics of each spectral component of the Fourier transform of the initial condition, that are given by $\delta \mathbf{w}(k,t)=\exp(\mathbb{A}(k)t)\delta \mathbf{w}(k,0)$ according to Eq. \eqref{eq:formal_ODE_dynamics}. By decomposing each component of the Fourier transform in the basis of the eigenmodes of scale $k$, we can compute the dust density perturbation according to the equation of mass conversation in Eq. \eqref{eq:perturb_dust_density_mass_conversion}.

For our experiment, we define the initial conditions as
\begin{equation}
    \begin{cases}
        \delta \rho_g (x,0)=\rho_g^0 \exp(-x^2/(2 \Delta L^2)), \\
        v_{g,x}(x,0)= v_{d,x}(x,0) = c_{\mathrm{ms},gd} \exp(-x^2/(2 \Delta L^2)) , \\
        v_{g,y}(x,0)= v_{d,y}(x,0) = 0 , \\
 \delta b(x,t) = B_0 \exp(-x^2/(2 \Delta L^2)).
    \end{cases}
    \label{eq:perturbations_experiment_magnetosonic}
\end{equation}

All the perturbations are initially in phase, according to a right-propagating ideal magnetosonic mode. By choosing this initial condition, the initial dust-to-gas ratio variations are very small. Because we choose a Gaussian perturbation, its Fourier transform is also a Gaussian of typical extent $ 1/\Delta L$. Scales larger than $\Delta L$ dominate the wave packet. We consider the physical conditions at $n_g = 10^4 \ \rm{cm^{-3}}$, which corresponds to the protostellar envelope.

We present in Fig. \ref{fig:perturb} the evolution of the perturbations for $\Delta L \in \{ 0.05 \ \rm{au}, 5 \ \rm{au}, 5000 \ \rm{au} \}$ on a period of time $5 \Delta L/\tilde{c}_{\mathrm{ms},d}$, such that a dust magnetocompressive wave starting from the origin has time to cross the entire initial perturbation. This allows us to probe different scales and observe the degree of coupling between $\delta \rho_d$ and $\delta b$, and the dust-to-gas variations $\delta \theta_d$.

For a spatial extension $\Delta L = 5000$ au (upper panels), the scales at stake are large enough to respect an ideal coupling between the gas, the dust and the magnetic field. The three perturbations move in phase respecting $\delta \rho_g / \rho_g \approx \delta v_{g,x}/c_{\mathrm{ms},gd} \approx \delta b/B_0$. The dust-to-gas ratio variations (in black) are weak. 

For a spatial extension $\Delta L = 5$ au (second line of panels), the gas decouples from the dust. The evolution of the gas is close to a right-propagating sonic wave $\delta \rho_g / \rho_g \approx (c_{\mathrm{ms},gd}/c_{\mathrm{s},g}) \delta v_{g,x}/c_{\mathrm{ms},gd} $ with $c_{\mathrm{ms},gd}/c_{\mathrm{s},g} \approx 1.72$ (in red), but not perfectly because of the initial condition. The dust velocity splits into two parts propagating in opposite directions (left-propagating and right-propagating, in green), coupled to the magnetic field. It is close to the prediction of the dust magnetocompressive mode $\delta v_{d,x}/c_{\mathrm{ms},gd} \approx \pm \tilde{c}_{\mathrm{ms},d} /c_{\mathrm{ms},gd} \delta b/B_0$ with $\tilde{c}_{\mathrm{ms},d}/c_{\mathrm{ms},gd} \approx 8.1$ (green dots). The dust density is computed and we observe that the magnetic flux is frozen in the dust, similarly to an ideal coupling between the dust and the magnetic field. Indeed, by combining the mass conservation equation with the expression of the right-propagating dust magnetocompressive waves, that are Eqs. \eqref{eq:perturb_dust_density_mass_conversion} and \eqref{eq:multidust_plasma_eigenvector}, we predict that $\delta \rho_d/\rho_d = \delta b/B_0$. The dust density is transported at the dust magnetosonic speed, leading to dust depletion close to the origin ($\delta \theta_d<0$) and dust enrichment away ($\delta \theta_d>0$) (see the dust-to-gas ratio variations in black). The wave packet is progressively dispersed and damped.

For a spatial extension $\Delta L = 0.05$ au (lower panels), the evolution of the gas is similar, showing that the gas is decoupled from the rest. Dust and magnetic field are no longer well-coupled (in green and blue). Both perturbations are damped as predicted by the wave analysis. Only the few larger-scale modes of the wavepacket can propagate. Dust depletion is significant close to the origin but, contrary to the case of $\Delta L= 5$ au, dust enrichment is weak as the wavepacket propagates away from the origin.

To conclude, we recover the coupling regimes predicted in Sect. \ref{sec:single_multifluid_MHD_ms_modes_analysis}, depending on the spatial extent $\Delta L$ of the perturbation. This illustrates that dust enrichment due to magnetic effects is possible. The clumping condition is described by the green area in Fig. \ref{fig:magnetosonic_coupling_regimes_17nm}. It corresponds to low-density regions and small scales. This compressive mechanism could take place in protostellar envelopes and it can be investigated in the context of early planet formation scenarios. We can mention the one-dimensional numerical simulations of a magnetized dust fluid by \citet{2025A&A...704A.142V} that predict a high dust clumping when small-scale magnetocompressive modes are excited.

The MHD mechanism we found and the resulting dust enrichment could complement other dynamical mechanisms that produce dust clumping in disks, such as the streaming instability \citep{2005ApJ...620..459Y}, the magnetosonic resonant drag instability \citep{2018ApJ...856L..15S}, recently the background Hall drift instability \citep{2024ApJ...962..173W} and the
ambipolar streaming instability \citep{2026A&A...709A.252P}, and the magnetorotational instability driven turbulence \citep{2005ApJ...634.1353J}. Indeed, prior dust enrichment taking place during the protostellar phase can promote 
dynamical instabilities in disks. Moreover, it would be interesting to investigate whether the new multifluid MHD description modifies the evolution of the hydrodynamical and magnetodynamical instabilities in disks or generates new ones. Even though we found that the coupling regime between the dust and the magnetic field disappears at high density, we should rather use a chemical network which models the disk thermochemistry and account for the disk rotation in the background velocities and the inertial forces. On the other hand, even though our results seem promising for the more diffuse interstellar medium ($n_g < 10^{4} \ \rm{cm^{-3}}$, by extrapolating Fig. \ref{fig:magnetosonic_coupling_regimes_17nm}), because grains are not necessarily the main charge carriers, an adequate chemical network, which includes for example the positive charging of grains due to photoelectric emission \citep{2001ApJS..134..263W}, is required to conclude whether the backreaction of dust grains on the magnetic field can favor magnetic dust enrichment in molecular clouds.

\subsection{Role of grain inertia on magnetic braking during disk formation} \label{sec:whistler_magnetic_braking}

\begin{figure*}
    \centering
    \includegraphics[width=0.49\textwidth]{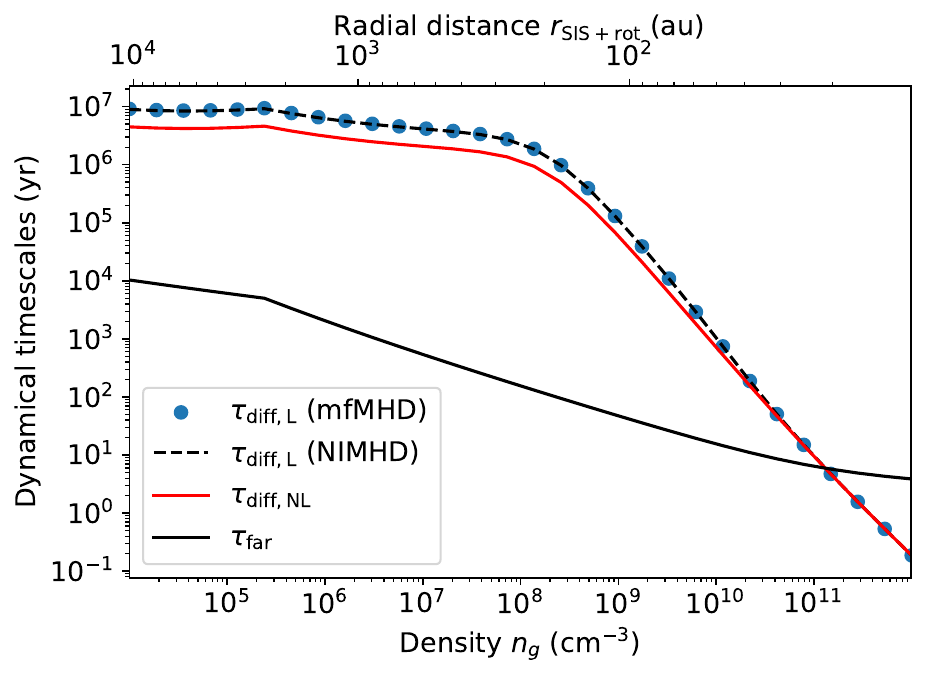}
    \includegraphics[width=0.49\textwidth]{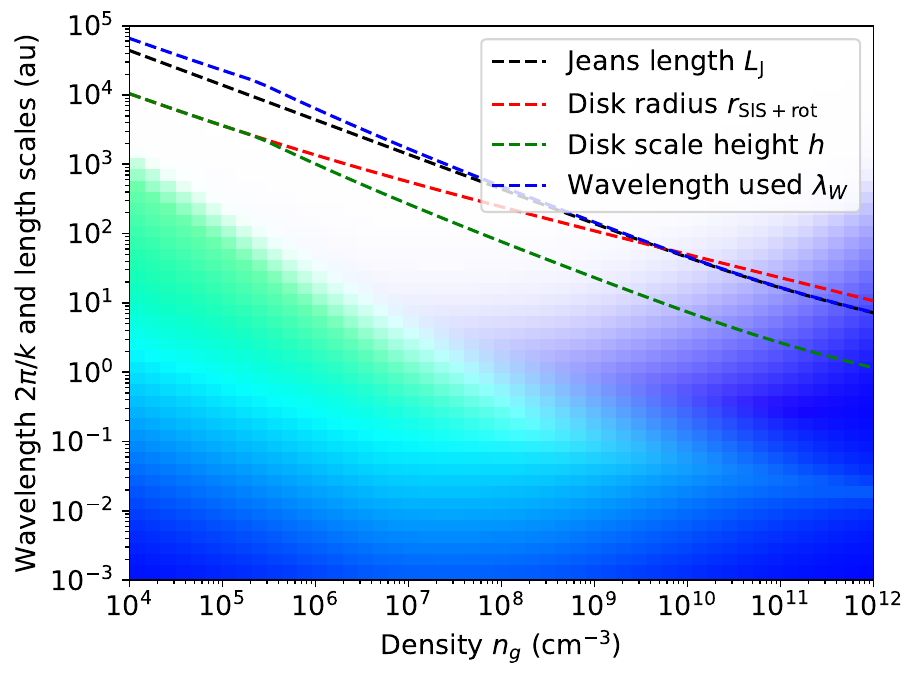}  
    \caption{Diffusion time (left panel) and coupling regimes (right panel) of the whistler mode for a single dust bin (17 nm grains) as a function of the density. Left: the dashed black line is the diffusion rate of the whistler mode in standard non-ideal MHD (NIMHD), and the blue dots are the diffusion rate in multifluid MHD (mfMHD), both evaluated at the scale $k=1/h$. The red line is the usual nonlinear diffusion time, which is also the diffusion time in the small-scale approximation in non-ideal MHD. The Faraday time is in black line. The radial distance $r_{\rm{SIS+rot}}$, along the secondary axis, as a function of the density, is obtained by inverting numerically Eq. \eqref{eq:SIS_rot_density_profile} assuming Keplerian rotation (Eq. \eqref{eq:Kelerian_velocity}). Right: The coupling regimes of the whistler mode in multifluid MHD as a function of the density and the wavelength. The coupling regimes are encoded by RGB colors as explained in Sect. \ref{app:eigenvectors_visualisation}. Dashed lines show the scales of interest. The regimes at these scales can be read as a consequence. In particular, the wavelength corresponding to $k=1/h$, which is the scale used for the left panel, is $\lambda_W=2 \pi h$ (blue dashed line).}
    \label{fig:whistler_diffusion_coupling_regimes_17nm}
\end{figure*}

\begin{figure}
    \centering
    \includegraphics[width=0.49\textwidth]{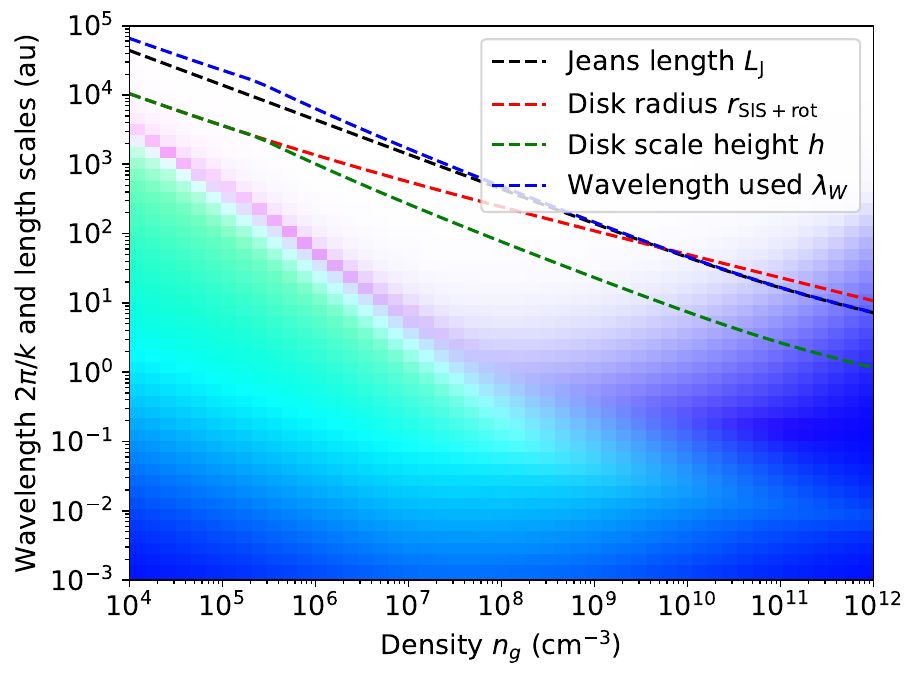}   
    \caption{Same as Fig. \ref{fig:whistler_diffusion_coupling_regimes_17nm} (right panel), but with 100 dust bins sampling the MRN distribution.}
    \label{fig:whistler_diffusion_coupling_regimes_MRN}
\end{figure}

In this section, we discuss the effect of the grain inertia on the coupling between the gas and the magnetic field during the protostellar collapse and the disk formation. Numerical simulations indicate that the degree of coupling between the gas and the magnetic field controls the size of the disk \citep{2022FrASS...9.9223M,2023ASPC..534..317T}. A weak coupling due to resistive effects leads to the formation of a large disk whereas a strong coupling prevents the formation of a disk. The Hall effect is also important. It leads to different disk sizes depending on the alignment between the core rotation and the magnetic field. The disk size can be explained by its underlying dynamical equilibrium. Many configurations, depending on the initial conditions of the collapse, have been identified and summarized in \citet{2021ApJ...922...36L}. The authors provided analytical disk sizes following the work of \citet{2016ApJ...830L...8H}. They compared nonlinear timescales based on the MHD equations to describe the dynamical equilibrium at the disk edge. In this section, we reproduce the analysis on the formation of the paradisk \citep{2015ApJ...810L..26T}, also known as  the rotationally supported disc \citep{2020MNRAS.492.3375Z}, that is described in the section "Strong-field Low-rotation Case" of \citet{2021ApJ...922...36L} and which corresponds to the magnetic field and the core rotation being anti-parallel. As explained later, we can include the linear timescale estimations of the non-ideal MHD and the multifluid MHD models in the analysis, compare them, and therefore evaluate the importance of the grain inertia.

We use the cylindrical coordinates $(r, \phi,z)$. $(B_r,B_\phi,B_z)$ are the components of the magnetic field and $(v_r,v_\phi, v_z)$ are the components of the velocity of the bulk mass (barycenter of the fluids). We denote the disk scale height as $h$ and the disk radius as $r$. We balance the rotation time $\tau_{\rm{rot}}$ with the magnetic braking time $\tau_{\rm{br}}$. Moreover, for the case of the paradisk, the induction time of $B_\phi$ from $B_z$, namely the Faraday time $\tau_{\rm{far}}$, is equal to the magnetic diffusion time $\tau_{\rm{diff,NL}}$, which depends on the magnetic resistivities. In \citet{2016ApJ...830L...8H}, these timescales are defined as

\begin{equation}
    \tau_{\rm{rot}} = \frac{2 \pi r}{v_\phi}, \tau_{\rm{br}} = \frac{\rho v_\phi 4 \pi h }{B_z B_\phi},
\end{equation}
\begin{equation}
   \tau_{\rm{far}} =  \frac{B_\phi h}{B_z v_\phi} , \tau_{\rm{diff,NL}} = \frac{h^2 B^2}{\frac{c^2}{4\pi} ( \eta_{\mathrm{O}}^C + \eta_{\mathrm{AD}}^C) B_z^2}.
\label{eq:nonlinear_faraday_and_diffusion_times}
\end{equation}

From $\tau_{\rm{rot}} \approx \tau_{\rm{br}}$, the Faraday time can be expressed as
\begin{equation}
   \tau_{\rm{far}} =  \frac{2 h^2 \rho }{r B_z^2} v_\phi.
\end{equation}

We note that, in the "Strong-field" approximation \citep{2021ApJ...922...36L}, the nonlinear diffusion time $\tau_{\rm{diff,NL}} $ corresponds to the diffusion rate of the whistler mode (Alfvén wave) in standard non-ideal MHD evaluated at $k=1/h$ in the small-scales approximation (as if $k \to +\infty)$ according to Eq. \eqref{eq:NIMHD_Alfvén_whistler}. The linear extrapolation is a delicate point but it seems reasonable to approximate the diffusion time of $B_\phi$ (smaller than $B_z$) by the diffusion time of the torsional magnetic modes, that are, unambiguously at small scales, the whistler waves (Sect. \ref{app:multifluid_MHD_alfven_whistler_small}). We note that the asymptotical estimation of the damping rate is not the same in multifluid MHD as long as dust conductivities are significant contributors to the standard resistivities (Eq. \eqref{eq:multifluid_MHD_alfven_whistler_small}). Thanks to the linear (and numerical) approach, we can avoid the small-scales approximation and generalize by defining the diffusion timescale as the damping rate of the whistler wave at $k=1/h$:
\begin{equation}
    \tau_{\rm{diff,L}} = -1/\Im(\omega(k=1/h)),
    \label{eq:linear_diffusion_time}
\end{equation}
whose value depends on the MHD model. It enables us to compare the standard non-ideal MHD with the multifluid MHD, and thus to measure the role of the grain inertia at this scale.

For our numerical application, we assume that the density profile is the one of the singular isothermal sphere (SIS) modified by rotation at high densities as in \citet{2016ApJ...830L...8H}, but other contribution factors are possible \citep{1984ApJ...286..529T,2021ApJ...922...36L} :
\begin{equation}
     \rho(r) =\frac{c_{\mathrm{s},g}^2}{2 \pi G r^2} \times \left( 1  + \frac{1}{2} \left( \frac{v_\phi}{c_{\mathrm{s},g}} \right)^2 \right).
     \label{eq:SIS_rot_density_profile}
\end{equation}

We also assume that the rotational velocity is close to Keplerian rotation 
\begin{equation}
    v_\phi \approx v_{\mathrm{K}} = \sqrt{\frac{G M_r}{r}},
    \label{eq:Kelerian_velocity}
\end{equation}
where $M_r$ is the mass enclosed within the radius $r$. For the sake of simplicity, we choose a constant mass of $M_r=0.1~M_{\odot}$.

Finally, the scale height $h$ of a disk of radius $r$ is obtained assuming vertical hydrostatic equilibrium
\begin{equation}
    h = \min \left( r,\frac{r c_{\mathrm{s},g}}{v_{\mathrm{K}}} \right).
\end{equation}

In the left panel of Fig. \ref{fig:whistler_diffusion_coupling_regimes_17nm}, we plot the Faraday time $\tau_{\rm{far}}$ and the diffusion time for different approximations ($\tau_{\rm{diff,NL}}$ in Eq. \eqref{eq:nonlinear_faraday_and_diffusion_times} for the commonly used nonlinear diffusion time, and $\tau_{\rm{diff,L}}$ in Eq. \eqref{eq:linear_diffusion_time} for the linear estimates of standard non-ideal MHD and multifluid MHD), as a function of the density. In Fig. \ref{fig:whistler_diffusion_coupling_regimes_17nm}, we start with a monodisperse distribution ($\mathcal{N}=1$, $s_{\rm{grain}}=17~\rm{nm}$). The intersection between the two timescales indicates the density for which the dynamical equilibrium for disk formation occurs. The corresponding disk size can be read on the secondary axis. 

The predictions of the diffusion time for the standard non-ideal MHD and for the multifluid MHD are very similar. The models predict a disk size of $21$ au at $ n_g \approx 1.3 \times 10^{11} \ \rm{cm^{-3}}$, which is compatible with what we expect from a moderate magnetic braking due to the anti-parallel alignment between the magnetic field and the core rotation. To understand the agreement between the two models, we investigate the coupling regime probed at the scale $k=1/h$. 

In the right panel of Fig. \ref{fig:whistler_diffusion_coupling_regimes_17nm}, we present the coupling regimes between the gas, the dust, and the magnetic field as a function of the density and the wavelength of the whistler mode. We note that the coupling regimes of the whistler wave are analogous to the coupling regimes we found for magnetosonic waves (Sect. \ref{sec:single_multifluid_MHD_ms_modes_analysis} with Fig. \ref{fig:magnetosonic_coupling_regimes_17nm}). 
The propagation of the whistler waves are also correctly described by ideal MHD (in white) at the Jeans scale and at the wavelength $\lambda_W=2 \pi h$ (or equivalently, $k=1/h$) at the beginning of the collapse. We note that $\lambda_W$ corresponds to the Jeans length when the Keplerian rotation dominates the expression of the density profile in Eq. \eqref{eq:SIS_rot_density_profile}. The resistive regime, for which the diffusion of the whistler wave is strong (in blue), is reached at higher density ($n_g >10^{10}~\rm{cm^{-3}}$). The balance between the diffusion time and the Faraday time is reached at these densities. First, the small-scale approximation stands at these densities and scales, which confirms the previous nonlinear estimates \citep{2016ApJ...830L...8H,2021ApJ...922...36L}. Secondly, the small-scale asymptote is correctly described by standard non-ideal MHD. This is because, at high densities ($n_g>10^8~\rm{cm^{-3}}$), the Hall factor of dust grains drops, and thus the standard non-ideal resistivities depend weakly on the dust conductivity (Sect. \ref{app:physical_setup} with Fig. \ref{fig:magnetic_resistivities}). To conclude, at the typical scales of the core and the disk, the standard non-ideal MHD correctly describes the coupling between the gas, the dust, and the magnetic field. The inertia, accounted for in the multifluid model, does not play a significant role.

To investigate the robustness of this conclusion, we test for a full MRN distribution, spanning a wider range of grain sizes (5 nm to 250 nm). We present in Fig. \ref{fig:whistler_diffusion_coupling_regimes_MRN} the result for 100 dust bins. The value of the diffusion time is affected by the changes of resistivities, leading to a slightly smaller disk ($18$ au). Surprisingly, despite a large range of grain sizes, the standard non-ideal MHD is still sufficient to estimate the diffusion rate at the typical scales of the disk. We varied the parameters of the distribution. We tested monodisperse distributions for several grain sizes. We tested power-law distributions with several minimal and maximum sizes and power exponents, which mimic dust growth. We also varied the grain density, the dust-to-gas ratio, and the ionization rate. We could not find a case where disk size predictions differ between non-ideal MHD and multifluid MHD. However, dust distribution impact the magnetic resistivities and thus the disk size, as already explored in \citet{2023PASJ...75..835T,2024A&A...690A..23V}. Even though the regime of coupling of the dust with the magnetic field (in green-blue in the right panels of Figs. \ref{fig:whistler_diffusion_coupling_regimes_17nm} and \ref{fig:whistler_diffusion_coupling_regimes_MRN}), which is not modeled by standard non-ideal MHD, is also affected, it vanishes at high density, when the disk starts forming. 
It is probably because the coupling conditions are difficult to reach. Indeed, dust grains probably require a minimal size to decouple from the gas (high Stokes number) and a maximum size to preferentially couple to the magnetic field (high Hall factor). Both the Stokes number and the Hall factor tend to drop with density. 

Our linear analysis concludes that the grain inertia does not play a significant role on the formation of a paradisk, or more precisely its size. However, its content could be affected by the evolution of the dust in low-density regions at small scales. One limitation of our analysis is that we do not model the transport of angular momentum from high-density regions to low-density regions, contrary to the analysis of \citet{2012A&A...543A.128J} who find that the magnetic braking of the disk depends on the density of the envelope. Even though we found it more delicate, it would be interesting to evaluate the role of the grain inertia for another collapse configuration. For instance, the parallel alignment between the magnetic field and the core rotation leads to another type of disk that is the orthodisk \citep{2021ApJ...922...36L}. Finally, the nonlinear physics of the dust multifluid during a protostellar collapse could be investigated in future numerical simulations.

\section{Conclusion} \label{sec:conclusion}

In this paper, we explore the multifluid physics of interstellar charged dust. We present a novel closed MHD system and we derive the waves propagating in the conditions of a protostellar collapse. In addition to finding the solutions of the dispersion relation, we describe the underlying physics of the eigenmodes. 
We compare to the multifluid hydrodynamics and the standard non-ideal MHD. We list below the main features of this new multifluid:

\begin{itemize}
    \item The dust species are treated as fluids that can couple to the magnetic field, instead of being included in magnetic resistivities. Thanks to this approach, the inertia of the charged dust grains is modeled. Moreover, the dust dynamics backreacts on the evolution of the magnetic field. 
    \item The main resulting features are the mass loading of the waves (coupling of the components, e.g. dust with sonic waves driven by the gas thermal pressure and dust with the Alfvén waves driven by magnetic tension) and the modeling of the velocity drifts between the components.
    \item As predicted by standard non-ideal MHD in the conditions of protostellar collapses, there is a resistive scale for which the gas decouples from the magnetic field, because the back-reactions of the charged species on the gas are too weak. At smaller scales, if the Lorentz force on the dust grains dominates the drag force, the dust grains carry magnetohydrodynamical waves. The inertia of the grains plays a primary role here. Indeed, the dynamical response of the grains to the magnetic perturbation is a function of the charge-to-mass ratio. 
    \item An analytical expression of the speed of the magnetocompressive wave carried by a dust size distribution is provided in Sect. \ref{sec:multidust_plasma_waves}. For a typical MRN distribution, the smallest grains mostly contribute to the propagation of the magnetocompressive waves. In this case, the speed of the magnetocompressive waves is between 3 and 4 times higher than the Alfvén speed of the dust bulk mass. The latter is already about 8 times and 10 times higher than the magnetosonic speed and the Alfvén speed predicted by ideal MHD respectively (assuming a dust-to-gas ratio of 1\%).
    \item We have also revised the analysis of the Alfvén waves by \citet{2023A&A...674A.149H} in Sect. \ref{sec:multifluid_MHD_alfven_modes}. We provide an analytical expression of the speed of the Alfvén waves carried by a dust size distribution, which is detailed in Appendix \ref{app:multifluid_Alfven_all_inertia}. Contrary to magnetosonic waves, the gas and dust mixture remains coupled at the very small scales (left-handed polarized modes), whereas the magnetic field perturbation decouples from the fluids (whistler mode). 
    \item Because gas and dust species are no longer treated as a single fluid, multifluid models exhibit a collection of collision modes. These modes arise from the interaction between the components of the system. Because of the collisions with the gas, the velocity of the fluids tend to relax towards the velocity barycenter (Sect. \ref{sec:multidust_hydro_waves}), minimizing their respective drifts, while individual Lorentz forces tend to decouple the charges from the gas (Sect. \ref{sec:multifluid_MHD_modes}). The underlying force balance and the role of inertia depend on the scales at stake. The details of the differential dynamics between the components and the wave speeds are essential to set the conditions to resonant instabilities \citep{2018ApJ...856L..15S}.
    \item When approximating a continuous dust distribution by an infinite number of bins, we obtain a continuum of collision modes, because of the infinite number of velocity configurations.
\end{itemize}

We also discussed some astrophysical consequences:
\begin{itemize}
    \item The dust decouples from the gas at au-scales in protostellar envelopes. The dust density relative variations are of the order of the amplitude of the magnetic perturbation causing the propagation of magnetocompressive waves relative to the parallel magnetic field (Sect. \ref{sec:single_multifluid_MHD_ms_modes_experiment}). Magnetic pressure produces local dust enrichment (i.e., dust-to-gas ratio variations) that can set favorable conditions for planet formation in denser regions. Interestingly, both the largest grains and the smallest grains can decouple from the gas but based on two different mechanisms that are respectively the grain inertia and the magnetic pressure. It leads to a complex differential dynamics within the grain distribution.
    \item The prediction of the magnetic diffusion time of standard non-ideal MHD is close to the prediction of multifluid MHD at the scales of the collapsing dense core and the forming disk (Sect. \ref{sec:whistler_magnetic_braking}). Consequently, the standard non-ideal MHD model seems sufficient to describe the magnetic braking and explain the size of paradisks. More generally, our linear analysis suggests that the role of the grain inertia at the scales of the collapse and the forming disk is weak.    
\end{itemize}

The multifluid physics of charged dust is very rich. We provide some theoretical understanding. Our multifluid model can be coupled to a chemical network and be implemented in numerical codes to simulate the evolution of a dust distribution and the transport of angular momentum consistently during a protostellar collapse, from the envelope to the disk. 
More generally, the dust multifluid framework, when extending the non-ideal magnetohydrodynamics, connects key roles of dust grains in star, disk and planet formation.

\begin{acknowledgements}
We thank the referee for their comments, which contribute in improving the clarity and the quality of the paper.
This research has received funding from the European Research Council synergy grant ECOGAL (Grant: 855130).
 
\end{acknowledgements}
\bibliographystyle{aa}
\bibliography{ref}

\appendix

\section{Complements on the multifluid magnetohydrodynamics} \label{app:multifluid_MHD}
\subsection{Derivation of the generalized Ohm law} \label{app:multifluid_MHD_derivation}

In this section, we derive the Ohm law of the multifluid MHD model presented in Sect. \ref{sec:multifluid_MHD_equations}. The derivation is similar to \citet{1999MNRAS.303..239W}, excepting that we need to be careful to the fact that light species (subset $L$ of the charged species $C$) do not respect electroneutrality alone.

We introduce 
\begin{equation}
     \mathbf{E}_g=\mathbf{E}+\frac{\mathbf{V}_g}{c} \times \mathbf{B}
\end{equation}
and the drift of charged species relative to the gas $\Delta \mathbf{V}_k = \mathbf{V}_k-\mathbf{V}_g$. Equation \eqref{eq:light_momentum_conservation} on light species (and extensions in Appendix \ref{app:multifluid_MHD_more_forces}) can be written as
\begin{equation}
    Z_l e \left( \frac{\Delta \mathbf{V}_l}{c} \times \mathbf{B} + \mathbf{E}_g \right) - \nu_l m_l \Delta \mathbf{V}_l= \mathbf{0}.
    \label{eq:light_momentum_conservation_drift}
\end{equation}

Thus, using the Hall factor $\Gamma_l$, it leads to the linear system on $\Delta \mathbf{V}_l$:
\begin{equation}
    \Delta \mathbf{V}_l - \Gamma_l \Delta \mathbf{V}_l \times \mathbf{b} = \frac{c \Gamma_l}{B} \mathbf{E}_g.
\end{equation}
Before inverting this system, we denote the perpendicular and parallel components to $\mathbf{b}= \mathbf{B}/B$ by the indices $\perp$ and $\parallel$ respectively:
\begin{equation}
    \mathbf{E}_g = 
    \underbrace{
    (\mathbf{E}_g \cdot \mathbf{b}) \mathbf{b} }_{(\mathbf{E}_g)_\parallel}
    + 
    \underbrace{
    \mathbf{b} \times (\mathbf{E}_g \times \mathbf{b})}_{(\mathbf{E}_g)_\perp}.
\end{equation}

By projecting along $\mathbf{b}$, and by computing the cross product with $\mathbf{b}$, and by gathering $\mathbf{V}_l = (\mathbf{V}_l)_\parallel + (\mathbf{V}_l)_\perp$, we obtain the velocity drift of light charged species as a function of the electric field:
\begin{equation}
    \Delta \mathbf{V}_l = \frac{c \Gamma_l}{B} \left( \frac{\Gamma_l}{1+\Gamma_l^2}\mathbf{E}_g \times \mathbf{b} + \frac{1}{1+\Gamma_l^2} (\mathbf{E}_g)_\perp + (\mathbf{E}_g)_\parallel \right).
    \label{eq:light_drift}
\end{equation}

Using electroneutrality, we define
\begin{equation}
    \mathbf{J} = \Delta \mathbf{J}_L + \Delta \mathbf{J}_I,
    \label{eq:current_splitting}
\end{equation}
where $\Delta \mathbf{J}_L =\sum_{l \in L} n_l  Z_l e \Delta \mathbf{V}_l$ and $\Delta \mathbf{J}_I =\sum_{d \in I} n_d  Z_d e \Delta \mathbf{V}_d$.

By injecting the velocity drifts in Eq. \eqref{eq:light_drift} in $\Delta \mathbf{J}_L $,  we obtain the electrical current as a function of the electric field, with the conductivities from light species defined in Eqs \eqref{eq:conductivities_def_1} and \eqref{eq:conductivities_def_2}:
\begin{equation}
    \Delta \mathbf{J}_L=\sigma_{\mathrm{O}}^L (\mathbf{E}_g)_\parallel + \sigma_{\mathrm{H}}^L \mathbf{b} \times \mathbf{E}_g  + \sigma_{\mathrm{P}}^L (\mathbf{E}_g)_\perp
\end{equation}

We invert this system by isolating $(\mathbf{E}_g)_\parallel $ and $(\mathbf{E}_g)_\perp $. We obtain
\begin{equation}
    \mathbf{E}_g = \frac{1}{\sigma_{\mathrm{O}}^L} (\Delta \mathbf{J}_L)_\parallel + \frac{\sigma_{\mathrm{H}}^L}{(\sigma_\perp^L)^2} \Delta \mathbf{J}_L \times \mathbf{b} + \frac{\sigma_{\mathrm{P}}^L}{(\sigma_\perp^L)^2} \Delta (\mathbf{J}_L)_\perp,
\end{equation}
which corresponds to the generalized Ohm law in Eq. \eqref{eq:Ohm_multifluid_MHD} with magnetic resistivities. The system of equations is successfully closed by combining Eq. \eqref{eq:current_splitting}, and Maxwell-Ampere equation recalled by Eq. \eqref{eq:mawell_ampere}
\begin{equation}
    \Delta \mathbf{J}_L = \frac{c }{4 \pi}\nabla \times \mathbf{B} - \sum_{d \in I} n_d  Z_d e ( \mathbf{V}_d - \mathbf{V}_g).
\end{equation}

\subsection{Additional forces on ions and electrons} \label{app:multifluid_MHD_more_forces}

We could add forces to the momentum balance of ions and the electrons in Eq. \eqref{eq:light_momentum_conservation}, such as internal pressure denoted by $\mathbf{f}_l$, or forces implied by mass exchanges denoted by $S_l \mathbf{V}_l$. Thus, the equation of momentum is 
\begin{equation}
    0 = S_l \mathbf{V}_l + \mathbf{f}_l - \rho_l \nu_l \Delta \mathbf{V}_l + n_l Z_l e( \mathbf{E} + \frac{\mathbf{V}_l}{c} \times \mathbf{B}).
\end{equation}

This can be rewritten in the same form as Eq. \eqref{eq:light_momentum_conservation_drift}:
\begin{equation}
    Z_l e \left( \frac{\Delta \mathbf{V}_l}{c} \times \mathbf{B} + \mathbf{E}' \right) - \nu_l' m_l \Delta \mathbf{V}_l= \mathbf{0},
\end{equation}
with variables $\mathbf{E}'=\mathbf{E}_g + \mathcal{E}_l$, with $\mathcal{E}_l=\dfrac{S_l \mathbf{V}_g+\mathbf{f}_l}{n_lZ_le}$ and $\nu_l'=\nu_l- \dfrac{S_l}{\rho_l}$. 

Thus, this modified linear relation can be inverted the same way as in Sect. \ref{app:multifluid_MHD_derivation}, leading to an Ohm law of the same form 
\begin{equation}
    \Delta \mathbf{J}_L' =\sigma_{\mathrm{O}}' (\mathbf{E}_g)_\parallel + \sigma_{\mathrm{H}}' \mathbf{b} \times \mathbf{E}_g  + \sigma_{\mathrm{P}}' (\mathbf{E}_g)_\perp,
\end{equation}
where $\Delta \mathbf{J}_L'= \Delta \mathbf{J}_L - \sum_{l \in L} n_l Z_l e \mathbf{W}_l$ with
\begin{equation}
        \mathbf{W}_l = \frac{c \Gamma_l'}{B} \left( \frac{\Gamma_l'}{1+\Gamma_l'^2}\mathcal{E}_l \times \mathbf{b} + \frac{1}{1+\Gamma_l'^2} (\mathcal{E}_l)_\perp + (\mathcal{E}_l)_\parallel \right).
\end{equation}
The modified Hall factors $\Gamma_l'=\omega_l/\nu_l'$, and subsequent conductivities $\sigma_{\mathrm{O}}'$, $\sigma_{\mathrm{H}}'$ and $\sigma_{\mathrm{P}}'$, and resistivities are computed with the modified collision rates $\nu_l'$. We note that the system is not properly closed if $\nu_l'$ depend on the velocities of the ions and the electrons.

The possibility of including other forces in the momentum equations in addition to the choice of splitting a distribution of charged species in $I$ and $L$ (including the inertia or not) with our multifluid MHD model offers a high degree of flexibility in terms of dynamical modeling.

\subsection{Feedback on the gas} 

After computing the electric field thanks to Ohm law, the electromagnetic back-reacts on the charged species via individual Lorentz forces. Then, charged species back-react on the gas via drag forces. The gas is no longer (perfectly) magnetized, contrary to Eq. \eqref{eq:NIMHD_Laplace_force} in standard non-ideal MHD. Indeed, the resulting contribution of ions and electrons to the drag force on the gas is

\begin{equation}
    \sum_{l \in L} \mathbf{f}_{l \to g}= \left(\sum_{l \in L} n_l Z_l e \right) \mathbf{E}_g + \frac{\Delta \mathbf{J}_L \times \mathbf{B}}{c} .
\end{equation}

\subsection{Feedback on the charged species} \label{app:multifluid_feedback_Lorentz_forces}

The magnetic field backreacts on the charges through individual Lorentz forces:
\begin{equation}
    \mathbf{f}_{\mathrm{Lor} \to k}= n_k Z_ke \left(\mathbf{E} + \frac{\mathbf{V}_k}{c} \times \mathbf{B} \right) = n_k Z_ke \left(\mathbf{E}_g + \frac{\mathbf{V}_k -\mathbf{V}_g}{c}  \times \mathbf{B} \right).
\end{equation}

From the Ohm law in Eq. \eqref{eq:Ohm_multifluid_MHD}, it becomes
\begin{equation}
\begin{aligned}
\mathbf{f}_{\mathrm{Lor} \to k} = n_k Z_ke ( &((\mathbf{V}_k -\mathbf{V}_g )/c) \times \mathbf{B} +  \eta_{\mathrm{H}}^L \Delta \mathbf{J}_L \times \mathbf{b} \\ &+ \eta_{\mathrm{O}}^L \Delta \mathbf{J}_L  + \eta_{\mathrm{AD}}^L \mathbf{b} \times (\Delta \mathbf{J}_L \times \mathbf{b})).
\end{aligned}
\end{equation}

The two first terms in the expression, which are dominant (Sect. \ref{app:physical_setup}), help in restoring a partial Laplace force on the charged species. Indeed, if we assume that the ions and the electrons are perfectly coupled to magnetic field (multidust model presented in Sect. \ref{sec:multidust_plasma_equations}), the Lorentz forces on the dust fluids become
\begin{equation}
    \mathbf{f}_{\mathrm{Lor} \to d} = n_d Z_de \left( \frac{\mathbf{V}_d  \times \mathbf{B}}{c} - \sum_{ i\in I} \xi_i \frac{\mathbf{V}_i  \times \mathbf{B}}{c} \right) 
    + \xi_d  \frac{\mathbf{J} \times \mathbf{B}}{c},
\end{equation}
where  $\xi_d = n_d Z_d /(\sum_{i \in I } n_i Z_i)$ is the proportion of charges carried by the dust fluid $d$ relative to the dust distribution. For the case of a single dust fluid ($\mathcal{N} = 1$), we recover the complete Laplace force as presented in Eq. \eqref{eq:multidust_plasma_one_fluid_magnetized}.

\subsection{Energy balance} \label{app:multifluid_MHD_energy}

\begin{figure*}
    \centering
    \includegraphics[width=0.99\textwidth]{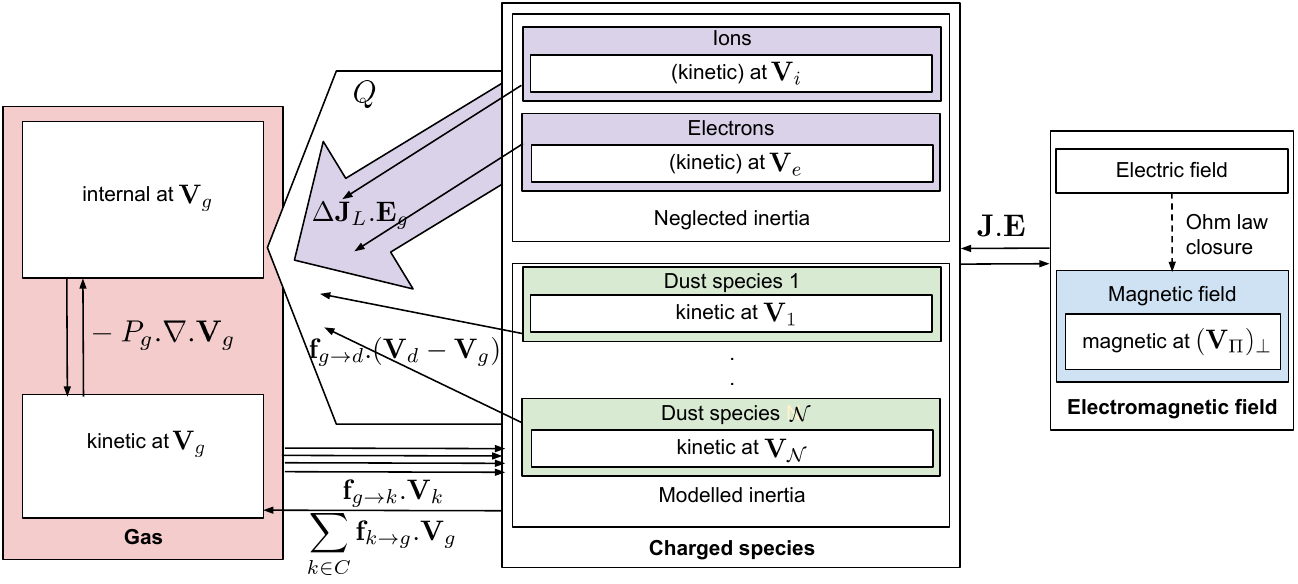}
    \caption{Energy balance between different reservoirs. The advection velocity of each energy reservoirs is indicated. Solid arrows represent energy transfers.}
    \label{fig:energy_balance}
\end{figure*}

The multifluid MHD describes the interactions between different components (gas, dust species, ions and electrons, magnetic field) and thus exchanges of energy between different reservoirs. Energy equations are necessary to close the MHD system. Indeed, the thermal pressure in Eq. \eqref{eq:gas_momentum_conservation} needs to be specified as a function of the other quantities. We present a simple energy balance, illustrated in Fig. \ref{fig:energy_balance}.

We define the Poynting vector $\mathbf{\Pi}$ and the magnetic energy $u_m$ as follows (the electric energy is not a significant energy reservoir when neglecting the current displacement):
\begin{equation}
    \mathbf{\Pi} = \frac{c}{4 \pi} \mathbf{E} \times \mathbf{B}, \quad u_m = \frac{B^2}{8 \pi}.
\end{equation}
From Maxwell's equations, we obtain a conservation of the magnetic energy, carried by the Poynting vector and communicating with the rest of the system by Joule effect
\begin{equation}
    \frac{\partial u_m}{\partial t} + \nabla \cdot \mathbf{\Pi} = - \mathbf{J} \cdot \mathbf{E}.
\end{equation}

We can see more explicitly that the transport of magnetic energy is coupled to matter by rewriting the Poynting vector thanks to the Ohm law:
\begin{equation}
    \mathbf{\Pi} =  \frac{1}{4 \pi} \mathbf{B} \times ( \mathbf{V}_\Pi \times \mathbf{B}) )=\frac{1}{4 \pi}  (\mathbf{V}_\Pi)_\perp B^2 =  \frac{B^2}{4 \pi} \mathbf{V}_\Pi  - \frac{1}{4 \pi}(\mathbf{V}_\Pi \cdot \mathbf{B}) \mathbf{B},
\end{equation}
with $\mathbf{V}_\Pi = \mathbf{V}_g - \frac{c}{B} ( \eta_{\mathrm{H}}^L \Delta \mathbf{J}_L + (\eta_{\mathrm{O}}^L + \eta_{\mathrm{AD}}^L) \mathbf{b} \times \Delta \mathbf{J}_L)$.

The evolution of the kinetic energies of the gas, the dust fluids, and equivalently for light species (ions and electrons) are respectively
\begin{equation}
\begin{aligned}
    \frac{\partial}{\partial t} \left( \frac{1}{2} \rho_g V_g^2  \right) = &- \nabla \cdot \left( \frac{1}{2} \rho_g V_g^2  \mathbf{V}_g + P_g \mathbf{V}_g  \right) \\ &+ P_g \nabla \cdot \mathbf{V}_g + \sum_{k \in C} \mathbf{f}_{k \to g} \cdot \mathbf{V}_g,    
\end{aligned}
\end{equation}
\begin{equation}
\begin{aligned}
    \frac{\partial}{\partial t} \left( \frac{1}{2} \rho_d V_d^2  \right) = &- \nabla \cdot \left( \frac{1}{2} \rho_d V_d^2  \mathbf{V}_d  \right) \\ &+ \mathbf{f}_{g \to d} \cdot \mathbf{V}_d + Z_d n_d e (\mathbf{E} + \frac{\mathbf{V}_d}{c} \times \mathbf{B}) \cdot \mathbf{V}_d,    
\end{aligned}
\end{equation}
\begin{equation}
    0 =\mathbf{f}_{g \to l} \cdot \mathbf{V}_l + Z_l n_l e (\mathbf{E} + \frac{\mathbf{V}_l}{c} \times \mathbf{B}) \cdot \mathbf{V}_l.
\end{equation}

By summing all the kinetic contributions within the plasma, we end up with
\begin{equation}
\begin{aligned}
    &\frac{\partial}{\partial t} \left( \sum_{p \in \{g,d \}} \frac{1}{2} \rho_p V_p^2  \right) +\nabla \cdot \left( \sum_{p \in \{g,d \}} \frac{1}{2} \rho_p V_p^2  \mathbf{V}_p + P_g \mathbf{V}_g  \right) \\ &= P_g \nabla \cdot \mathbf{V}_g + \sum_{k \in C} (\mathbf{f}_{k \to g} \cdot \mathbf{V}_g + \mathbf{f}_{g \to k} \cdot \mathbf{V}_k) 
    + \mathbf{J} \cdot \mathbf{E}.
\end{aligned}   
\end{equation}

Finally, for the internal energy of the gas, we model it as
\begin{equation}
    \frac{\partial \rho_g e_g}{\partial t} = - \nabla \cdot (\rho_g e_g \mathbf{V}_g) - P_g \nabla \cdot \mathbf{V}_g + \sum_{k \in C} Q_{kg}
    \label{eq:gas_energy_equation}
\end{equation}

By assuming that the thermal transfer towards the gas is given by the power of interactions between the gas and the charged species (by drag forces), we obtain
\begin{equation}
    Q_{kg} = - (\mathbf{f}_{k \to g} \cdot \mathbf{V}_g + \mathbf{f}_{g \to k} \cdot \mathbf{V}_k) = \rho_k \rho_g \gamma_k (\mathbf{V}_k-\mathbf{V}_g)^2 >0
\end{equation}
where $Q_{lg}$ (for ions and electrons) are computed from velocity drifts in Eq. \eqref{eq:light_drift}. Moreover, drag forces on light species balance individual Lorentz forces and thus
\begin{equation}
    \sum_{l \in L} Q_{lg}= \sum_{l \in L} Z_l n_l e \left( \mathbf{E}_g + \frac{\mathbf{V}_l-\mathbf{V}_g}{c} \times \mathbf{B} \right) \cdot (\mathbf{V}_l-\mathbf{V}_g) = \Delta \mathbf{J}_L \cdot \mathbf{E}_g,
\end{equation}
with
\begin{equation}
    \Delta \mathbf{J}_L \cdot \mathbf{E}_g = \frac{1}{\sigma_{\mathrm{O}}^L} \lVert  (\Delta \mathbf{J}_L)_{\parallel}\rVert^2 + \frac{\sigma_{\mathrm{P}}^L}{(\sigma_{\perp}^L)^2}  \lVert  (\Delta \mathbf{J}_L)_{\perp}\rVert^2 > 0.
\end{equation}
We recover what is known as resistive heating in standard non-ideal MHD. 

To conclude, by denoting the kinetic energies by $\varepsilon_p= \frac{1}{2} \rho_p V_p^2$, we obtain the conservation of the total energy within the system:
\begin{equation}
\begin{aligned}
    &\frac{\partial}{\partial t} \left( \sum_{p \in \{g,d \}} \varepsilon_p + \rho_g e_g + u_m  \right) \\
    &+\nabla \cdot \left( \sum_{p \in \{g,d \}} \varepsilon_p \mathbf{V}_p + (P_g + \rho_g e_g  )\mathbf{V}_g +  \frac{B^2}{4 \pi} \mathbf{V}_\Pi  - \frac{1}{4 \pi}(\mathbf{V}_\Pi \cdot \mathbf{B}) \mathbf{B} \right) \\ &= 0.    
\end{aligned}
\end{equation}

\section{Hydrodynamical modes} \label{app:multidust_hydro_waves}

\subsection{Dispersion relation}

The linearized equations on the $\mathcal{N}+2$ perturbations (with Fourier transform in space) lead to the differential system in time 

\begin{equation}
    \mathrm{d}_t \delta \rho_g + ik \rho_g v_{g,x} = 0,
\end{equation}
\begin{equation}
     \mathrm{d}_t v_{g,x} = -c_{\mathrm{s},g}^2 \frac{ik}{\rho_g} \delta \rho_g + \sum_{d \in I} \theta_d \nu_d (v_{d,x} - v_{g,x}),
\end{equation}
\begin{equation}
     \mathrm{d}_t v_{d,x} = \nu_d (v_{g,x} - v_{d,x}),
\end{equation}
with $\nu_d=1/t_{\mathrm{s},d}$. Therefore, the dispersion relation is
\begin{equation}
    \omega^2 = k^2 c_{\mathrm{s},g}^2 \dfrac{1}{1 + \sum_{d \in I} \dfrac{\theta_d}{1-i \omega t_{\mathrm{s},d}}}.
    \label{eq:dustywave_dispersion}
\end{equation}

The dispersion relation is a real-coefficient polynomial in $i \omega$, meaning that the solutions $\omega$ are imaginary numbers (purely damped modes here) and pairs of propagating waves (with opposite propagation directions (left-propagating and right-propagating), with the same speed and with the same damping rate). From this dispersion relation, we already note that $\omega t_{\mathrm{s},d}$ (compared to unity, which leads to the definition of the Stokes number) controls the mass-loading of the sound waves by the dust fluid $d$.

\subsection{Asymptotic solutions}

Sound waves are the modes verifying $\omega \to \infty$ when $k \to \infty$ and $\omega \to 0$ when $k \to 0$, whereas collision modes remain finite (but do not vanish) at all scales.

\subsubsection{Small scales}

For $k \to \infty$ (small scales), by assuming that the fluids are linearly decoupled, then we obtain, for the gas 
\begin{equation}
    \omega = \pm c_{\mathrm{s},g} k - i \frac{\sum_{d \in I} \theta_d \nu_d}{2},
    \label{eq:sonic_mode_small_scales}
\end{equation}
and for each dust species
\begin{equation}
    \omega = -i \nu_d.
    \label{eq:collision_mode_small_scales}
\end{equation}

We note that linear decoupling is not equivalent to assuming that the collision rates tend to zero (drag decoupling). Linear decoupling only means that we have independent eigenstates for each fluid, that are $(\delta \rho_g, v_{g,x},0,...,0)^T$, $(0, 0,v_{1,x},...,0)^T$, until $(0, 0,0,...,v_{\mathcal{N},x})^T$. 

\subsubsection{Sonic waves at large scales}

For $k \to 0$, for sonic waves, we assume that $\omega \to 0$. We denote the effective sound wave speed as
\begin{equation}
    c_{\mathrm{s},gd} = \frac{c_{\mathrm{s},g}}{\sqrt{1 + \sum_{d \in I} \theta_d}},
\end{equation}
and we define
\begin{equation}
    \tau = \frac{1}{2} \sum_{d \in I} \epsilon_d t_{\mathrm{s},d},
\end{equation}
with dust ratios $\epsilon_d=\rho_d/(\rho_g + \sum_{d \in I} \rho_d)$. 
We obtain from the dispersion relation the solutions for the sound waves at large scales
\begin{equation}
    \omega = \pm c_{\mathrm{s},gd} k  - i\tau c_{\mathrm{s},gd}^2 k^2 . 
    \label{eq:dustywave_sonic_wave_large_scales}
\end{equation}

\subsubsection{Collision modes at large scales} \label{app:multidust_hydro_waves_collision_modes}

The modes of the collision test are recovered by setting $c_{\mathrm{s},g}=0$ in Eq. \eqref{eq:dustywave_dispersion}, leading to
\begin{equation}
    1+ \sum_{d \in I} \frac{\theta_d}{1-i \omega t_{\mathrm{s},d}} = 0.
\end{equation}

If we assume a hierarchy of stopping times, by looking for a mode such that $\Omega = i\omega \sim 1/t_{\mathrm{s},j}$ we can split in low-Stokes dust species ($\Omega \ll 1/t_{\mathrm{s},l}$) and high-Stokes species ($\Omega \gg 1/t_{\mathrm{s},i}$):
\begin{equation}
    1+ \sum_{d \in I} \frac{\theta_d}{1- \Omega t_{\mathrm{s},d}} \approx 1+ \frac{\theta_j}{1- \Omega t_{\mathrm{s},j}} + \sum_{l \in I | t_{\mathrm{s},l} < t_{\mathrm{s},j}} \theta_l - \frac{1}{\Omega} \sum_{i \in I |  t_{\mathrm{s},i} > t_{\mathrm{s},j}} \frac{\theta_i}{t_{\mathrm{s},i}} .
\end{equation}

These two categories contribute to different characteristic quantities. There are the effective bulk mass with
\begin{equation}
    \theta_{L,j} = \sum_{l \in I | t_{\mathrm{s},l} < t_{\mathrm{s},j}} \theta_l
\end{equation}
for the lightest (low-Stokes) dust species, and the effective collision rate
\begin{equation}
    \nu_{I,j} = \sum_{i \in I |  t_{\mathrm{s},i} > t_{\mathrm{s},j}} \frac{\theta_i}{t_{\mathrm{s},i}} 
\end{equation}
for the heaviest (high-Stokes) dust species due to velocity drifts.

We notice that \citet{2019ApJS..241...25B} uses a very similar technique in its Appendix D to estimate a bound for the most damped mode. 

Under the hierarchy assumption, the dispersion relation of collision modes simplifies to
\begin{equation}
    1 + \theta_{L,j} + \frac{\theta_j}{1- \Omega t_{\mathrm{s},j}} - \frac{1}{\Omega} \nu_{I,j} = 0.
    \label{eq:dustybox_dispersion}
\end{equation}

We note that, for the case of a single dust fluid ($\mathcal{N}=1$), $\theta_{L,j}=0$ and $\nu_{I,j} =0$. In this case, we recover the exact dispersion relation, with $\Omega = (1+\theta_j)/t_{\mathrm{s},j}$ (the other root is $ \Omega=0$ for the barycenter motion, substituting the role of sonic waves), and thus the eigenmode is $v_{j,x}=v_{g,x}/(1-\Omega t_{\mathrm{s},j})=-v_{g,x}/\theta_j$.

Equation \eqref{eq:dustybox_dispersion} is a second-order polynomial: equation in $\Omega$
\begin{equation}
    \Omega^2 t_{\mathrm{s},j} (1+ \theta_{L,j}) -\Omega (1 + \theta_{L,j} +\theta_{j} + t_{\mathrm{s},j} \nu_{I,j}) + \nu_{I,j} = 0
\end{equation}
whose solution of interest (close to $1/t_{\mathrm{s},j}$) is 
\begin{equation}
    \Omega_j = \frac{ 1 + \theta_{L,j} +\theta_{j} + t_{\mathrm{s},j} \nu_{I,j} + \vartheta_j }{2t_{\mathrm{s},j} (1+ \theta_{L,j}) },
    \label{eq:dustywave_collision_mode}
\end{equation}
with $\vartheta_j=  \sqrt{ ( 1 + \theta_{L,j} +\theta_{j} + t_{\mathrm{s},j} \nu_{I,j})^2 - 4 t_{\mathrm{s},j} \nu_{I,j} (1+\theta_{L,j}) }$.


\section{Non-ideal MHD modes} 

The model, including the equations and the assumptions, is presented in Sect. \ref{sec:non_ideal_MHD_equations}. We remind that standard non-ideal MHD resistivities, denoted as $\eta_{\mathrm{O}}^C, \eta_{\mathrm{H}}^C$ and $\eta_{\mathrm{AD}}^C$, include the conductivities of the ions, the electrons, as well as the dust species (because the grain inertia is neglected), that are all the charged species $C$.

\subsection{Alfvén waves} \label{app:NIMHD_Alfvén}

The geometry of the perturbations is described in Eq. 
\eqref{eq:perturbations_alfven}. We denote the polarization of the Alfvén mode as $\sigma \in \{-1,1 \}$. 
The dynamics of the two perturbations $v_{g,\sigma} = v_{g,x} + i \sigma v_{g,y}$ and $\delta b_\sigma = \delta b_x + i \sigma \delta b_y$ is
\begin{equation}
    \mathrm{d}_t v_{g,\sigma}= \frac{B_0}{4 \pi \rho_g} ik \delta b_\sigma,
\end{equation}
\begin{equation}
    \mathrm{d}_t \delta b_\sigma = ik B_0 v_{g,\sigma} +(-ik)^2 \frac{c^2}{4 \pi} \eta_\sigma^C \delta b_\sigma,
\end{equation}
with $\eta_\sigma^C = \eta_{\mathrm{O}}^C+\eta_{\mathrm{AD}}^C-i\sigma \eta_{\mathrm{H}}^C$. This leads to the dispersion relation
\begin{equation}
    \omega^2 + \frac{c^2}{4 \pi} (\sigma \eta_{\mathrm{H}}^C + i (\eta_{\mathrm{O}}^C+\eta_{\mathrm{AD}}^C))k^2 \omega - c_{\mathrm{a},g}^2 k^2=0,
\end{equation}
with $c_{\mathrm{a},g}= B_0/\sqrt{4 \pi \rho_g}$ the gas Alfvén speed. The eigenvector is
\begin{equation}
    v_{g,\sigma}= - \frac{B_0}{4 \pi \rho_g} \frac{k}{\omega} \delta b_\sigma = - c_{\mathrm{a},g}^2 \frac{k}{\omega} \frac{\delta b_\sigma}{B_0}
    \label{eq:NIMHD_alfven_eigenvector}
\end{equation}

 The two solutions, denoted by $s\in \{-1,1 \}$ are
\begin{equation}
\begin{aligned}
    \omega   &= -\sigma s \frac{c_{\mathrm{a},g} k^2}{k_\eta} \left( ss_H \sqrt{\frac{1}{r_\eta}}+ \sqrt{\sqrt{\psi_\eta^2+1}+\psi_\eta}    \right) \\ & - i   \frac{c_{\mathrm{a},g} k^2}{k_\eta} \left(   \sqrt{ r_\eta} + s s_H \sqrt{\sqrt{\psi_\eta^2+1}-\psi_\eta}    \right).
\end{aligned}
\end{equation}

$s_H$ is the sign of the Hall resistivity. We define respectively the diffusion scale, the Hall scale and the resistive scale by
\begin{equation}
    k_{\mathrm{D}}^C = \frac{2 c_{\mathrm{a},g}}{ \frac{c^2}{4 \pi}(\eta_{\mathrm{O}}^C+\eta_{\mathrm{AD}}^C)}, \quad k_{\mathrm{H}}^C = \frac{2 c_{\mathrm{a},g}}{ \frac{c^2}{4 \pi}|\eta_{\mathrm{H}}^C|}, \quad k_{\eta} = \sqrt{k_{\mathrm{D}}^C k_{\mathrm{H}}^C}.
    \label{eq:NIMHD_resistive_scales}
\end{equation}
We also define the ratio between magnetic diffusion and magnetic dispersion by $
    r_\eta = (\eta_{\mathrm{O}}^C+ \eta_{\mathrm{AD}}^C)/|\eta_{\mathrm{H}}^C|
$ and we use the intermediate variable $\psi_\eta= (1/r_\eta - r_\eta+ (k_\eta/k)^2)/2$.

At large scales ($k \to 0$), it agrees with ideal MHD until $k_{\mathrm{D}}^C$ or $k_{\mathrm{H}}^C$ as presented in Sect. \ref{sec:single_multifluid_MHD_ms_modes_analysis}. At small scales ($k \to \infty$), we recover a whistler mode in $\omega \propto k^2$ and a left-hand polarized mode $\omega \propto k^0$. When injecting in the eigenvector relation given by the Eq. \eqref{eq:NIMHD_alfven_eigenvector}, we note that the magnetic field and the gas are linearly decoupled at small scales. Indeed, the whistler mode is a dispersive magnetic mode:
\begin{equation}
    \omega = - \frac{c^2}{4 \pi} (\sigma \eta_{\mathrm{H}}^C + i(\eta_{\mathrm{O}}^C+\eta_{\mathrm{AD}}^C))k^2 \Rightarrow  \delta b \propto  k v_{g,\sigma},
    \label{eq:NIMHD_Alfvén_whistler}
\end{equation}
and the left-hand polarized mode is a mode only on the gas velocity:
\begin{equation}
    \omega =  \dfrac{c_{\mathrm{a},g}^2}{\frac{c^2}{4 \pi} (\sigma \eta_{\mathrm{H}}^C + i(\eta_{\mathrm{O}}^C+\eta_{\mathrm{AD}}^C))} \Rightarrow v_{g,\sigma} \propto k \delta b .
    \label{eq:NIMHD_left_handed_mode}
\end{equation}

Understanding the wave physics is important to design the wan fan of Riemann solvers, a central tool of finite-volume codes. \citet{2019A&A...631A..66M} found that only using the whistler speed for the magnetic field variable, and not for the other variables, improves the conservation of the angular momentum during the protostellar collapse, especially at the formation of the first hydrostatic core, for which the mesh-refinement increases to resolve au scales. Our eigenvector analysis of the whistler mode in Eq. \eqref{eq:NIMHD_Alfvén_whistler} can possibly explain the origin of the problem. When $k \gg k_{\mathrm{H}}^C $, which corresponds to scales smaller than 100 au, the whistler mode is a purely magnetic mode. Therefore, when the mesh-refinement is sufficient to resolve the Hall scale ($\Delta x  \ll 1/k_{\mathrm{H}}^C$), using the whistler speed only for computing the hydrodynamical flux of the magnetic field is preferable. On the contrary, if the resolution is too low, the coupling between the magnetic field and the other variables needs to be considered. There is no reason to separate the solvers. However, the whistler speed in the solver of the magnetic field correctly converges towards the Alfvén speed in ideal MHD at low resolution, which is used for the other variables. Therefore, even though Riemann solvers are separate, this strategy recovers the ideal MHD regime.

\subsection{Magnetosonic waves} \label{app:NIMHD_ms}

\subsubsection{Dispersion relation}

The dynamics of the three perturbations are
\begin{equation}
    \frac{\mathrm{d}}{\mathrm{d}t}
    \begin{pmatrix}
        \delta \rho_g \\
        v_{g,x} \\
        \delta b 
    \end{pmatrix}
    = - i k
    \begin{pmatrix}
        0 & \rho_g & 0 \\
        c_{\mathrm{s},g}^2/\rho_g & 0 & B_0/(4 \pi \rho_g) \\
        0 & B_0 & -ik \frac{c^2}{4 \pi} (\eta_{\mathrm{O}}^C + \eta_{\mathrm{AD}}^C)
    \end{pmatrix}
        \begin{pmatrix}
        \delta \rho_g \\
        v_{g,x} \\
        \delta b 
    \end{pmatrix},
\end{equation}

leading to the dispersion relation
\begin{equation}
    \omega^3 + i\frac{c^2}{4 \pi} (\eta_{\mathrm{O}}^C + \eta_{\mathrm{AD}}^C) k^2 \omega^2 - c_{\mathrm{ms},g}^2 k^2 \omega - i \frac{c^2}{4 \pi} (\eta_{\mathrm{O}}^C + \eta_{\mathrm{AD}}^C) k^4 c_{\mathrm{s},g}^2 = 0,
\end{equation}
with the gas magnetosonic speed $c_{\mathrm{ms},g}^2=c_{\mathrm{a},g}^2 +c_{\mathrm{s},g}^2$. This is a third-order polynomial equation in $i \omega$, meaning that there are one pair of propagating waves (with opposite propagation directions (left-propagating and right-propagating), with the same speed and the same damping rate) and one non-propagating damped mode. The roots can be determined explicitly but they are difficult to analyze. The full system is solved numerically and the solutions are presented in the figures of Sects. \ref{sec:single_multifluid_MHD_ms_modes_analysis} and \ref{sec:multifluid_MHD_ms_modes}. We analyze the asymptotes hereafter.

\subsubsection{Small scales} \label{app:NIMHD_ms_small_scales}

At small scales ($k \to \infty)$, the gas and the magnetic field perturbation decouples. We recover a damped sound wave for the gas and the resistive diffusion of the magnetic field. 

Indeed, assuming that there is only a magnetic perturbation provides the solution
\begin{equation}
    \omega = -ik^2 \frac{c^2}{4 \pi} (\eta_{\mathrm{O}}^C + \eta_{\mathrm{AD}}^C).
    \label{eq:NIMHD_magnetic_diffusion_small_scales}
\end{equation}

For the sound wave, we assume that $\omega = \Re(\omega) + i \Im(\omega)$ with $\Re(\omega) \sim c_{\mathrm{s},g} k$ and $\Im(\omega) $ constant, as $k$ tends to infinity.  By expanding the dispersion relation to the first order in $\Im(\omega)/\Re(\omega)$, we find that
\begin{equation}
    \Im(\omega) \to - \frac{c_{\mathrm{a},g}^2}{2\frac{c^2}{4 \pi} (\eta_{\mathrm{O}}^C + \eta_{\mathrm{AD}}^C)  } = - \frac{1}{4}c_{\mathrm{a},g} k_{\mathrm{D}}^C.
\end{equation}

Therefore, the asymptotic solutions of sound waves are
\begin{equation}
    \omega = \pm c_{\mathrm{s},g}k -i\frac{c_{\mathrm{a},g}^2}{2\frac{c^2}{4 \pi} (\eta_{\mathrm{O}}^C + \eta_{\mathrm{AD}}^C)  }.
    \label{eq:NIMHD_sound_waves_small_scales}
\end{equation}

\subsubsection{Large scales}\label{app:NIMHD_ms_large_scales}

At larger scales ($k \to 0$), \citet{1992ApJ...390..560C} recovered ideal MHD solutions with deviations due to resistivities, that are respectively two propagative modes and one purely damped mode:
\begin{equation}
    \omega = \pm \sqrt{1+ \beta} c_{\mathrm{a},g}k - i \frac{1}{2(1+\beta)} \frac{c^2}{4 \pi} (\eta_{\mathrm{O}}^C + \eta_{\mathrm{AD}}^C) k^2,
    \label{eq:NIMHD_ms_waves_large_scales}
\end{equation}
\begin{equation}
    \omega = -i \frac{\beta}{1+\beta} \frac{c^2}{4 \pi} (\eta_{\mathrm{O}}^C + \eta_{\mathrm{AD}}^C) k^2,
\end{equation}
with the plasma parameter $\beta=c_{\mathrm{s},g}^2/c_{\mathrm{a},g}^2$.

Equating the real part and the imaginary part of the propagative mode provides an estimate of the transition scale between ideal MHD and the resistive regime
\begin{equation}
    k \approx (1+ \beta)^{3/2} \frac{2c_{\mathrm{a},g}}{\frac{c^2}{4 \pi} (\eta_{\mathrm{O}}^C + \eta_{\mathrm{AD}}^C)} = (1+ \beta)^{3/2} k_{\mathrm{D}}^C,
\end{equation}
where $k_{\mathrm{D}}^C$ is a diffusion scale appearing on the solutions of Alfvén waves (Appendix \ref{app:NIMHD_Alfvén}). Equating the damping rate of the sound wave at large scales from Eq. \eqref{eq:NIMHD_ms_waves_large_scales} and at small scales from Eq. \eqref{eq:NIMHD_sound_waves_small_scales} provides the estimate $k\approx k_{\mathrm{D}}^C \sqrt{1+\beta}/2$. In practice, we found that $k_{\mathrm{D}}^C$ better fits the transition between the two regimes, including for the MHD multifluid model.  Thus, we use it as a reference.

\section{Multidust plasma modes} 

\subsection{Generalities}

The model, including the equations and the assumptions, is presented in Sect. \ref{sec:multidust_plasma_equations}. For the study of Alfvén waves and some details on magnetosonic modes that are not in Appendix \ref{app:multidust_plasma_ms}, we refer to Appendix \ref{app:multifluid_MHD_modes} on the multifluid MHD model. Indeed, the results on the multidust plasma model can be obtained by setting the effective resistivity defined in Eq. \eqref{eq:multifluid_MHD_alfven_resistivity} to $\eta_\sigma^L=-i \sigma / \hat{\sigma}_{\mathrm{H}}^L$, the effective collision rate of light species defined in Eq. \eqref{eq:multifluid_MHD_alfven_effective_collision_rate_light} to $\nu_{L, \sigma}=0$, and the contribution of the dust fluid to the generation of the electric field defined in Eq. \eqref{eq:multifluid_MHD_dust_electric_field_contribution} to $\Xi_d=\xi_d = n_dZ_d/(\sum_{i \in I} n_i Z_i)$.

\subsection{Magnetocompressive waves} \label{app:multidust_plasma_ms}

The dynamics of the $2 \mathcal{N}+4$ perturbations is
\begin{equation}
    \mathrm{d}_t \delta \rho_g = -ik \rho_g v_{g,x},
\end{equation}
\begin{equation}
  \left\{
      \begin{aligned}
         \mathrm{d}_t v_{g,x} &= - ik \frac{ c_{\mathrm{s},g}^2}{\rho_g} \delta \rho_g + \sum_{d \in I} \theta_d  \nu_d  ( v_{d,x} -v_{g,x}) , \\
             \mathrm{d}_t v_{g,y} &= \sum_{d \in I} \theta_d  \nu_d  ( v_{d,y} -v_{g,y}), \\
      \end{aligned}
    \right.
\end{equation}
\begin{equation}
  \left\{
      \begin{aligned}
         \mathrm{d}_t v_{d,x} &= - \nu_d  (v_{d,x}-v_{g,x})+ \omega_d (v_{d,y}-\sum_{i \in I} \xi_i v_{i,y}) \\ & - i k  \frac{B_0}{4 \pi \rho_d} \xi_d  \delta b  , \\
    \mathrm{d}_t v_{d,y} &= - \nu_d  (v_{d,y}-v_{g,y})  - \omega_d  (v_{d,x}-\sum_{i \in I} \xi_i v_{i,x}), \\
      \end{aligned}
    \right.
    \label{eq:multidust_plasma_vdx}
\end{equation}
\begin{equation}
    \mathrm{d}_t \delta b =  - i k B_0 \sum_{d \in I} \xi_d v_{d,x},
    \label{eq:multidust_plasma_ms_induction}
\end{equation}
with $\xi_i = n_i Z_i/\sum_{d \in I} (n_d Z_d)$ the number of charges carried by the dust fluid $i$ relative to the number of charges carried by the entire dust distribution.

At sufficiently small scales ($k \to \infty)$, if the term $- i k B_0 /(4 \pi \rho_d)  \xi_d  \delta b $ in Eq. \eqref{eq:multidust_plasma_vdx}, coming from the Hall drift and analogous to a (partial) magnetic pressure, dominates the dust dynamics, then the motion is parallel to $\mathbf{e}_x$. Equation \eqref{eq:multidust_plasma_vdx} simplifies to
\begin{equation}
    \mathrm{d}_t v_{d,x} = - i k  \frac{B_0}{4 \pi \rho_d} \xi_d  \delta b.
    \label{eq:multidust_plasma_ms_dust_pressure}
\end{equation}

Injecting this in the induction equation given by Eq. \eqref{eq:multidust_plasma_ms_induction} leads to the dispersion relation
\begin{equation}
    \omega^2 = k^2 c_{\mathrm{a},g}^2 \sum_{d \in I} \frac{\xi_d^2}{\theta_d},
\end{equation}
and thus to the value of the dust magnetosonic speed $\tilde{c}_{\mathrm{ms},d}$ in Eq. \eqref{eq:multidust_magnetosonic}, which is different from the Alfvén speed associated with the total dust mass $c_{\mathrm{a},d}= B_0/\sqrt{4 \pi \sum_{d\in I} \rho_d} = c_{\mathrm{a},g}/\sqrt{\sum_{d \in I} \theta_d}$. 
The expression of the eigenvectors describing the pairs of magnetocompressive waves is given in Eq. \eqref{eq:multidust_plasma_eigenvector}. 

For the case of a single dust fluid ($\mathcal{N}=1$), by assuming that dust velocities only couple to the magnetic field, we obtain the expression of the mode 
\begin{equation}
    \omega = \frac{1}{2} \left(\pm \sqrt{4 c_{\mathrm{a},d}^2 k^2 - \nu_d^2} - i \nu_d \right),
    \label{eq:multidust_plasma_dust_magnetosonic_1bin}
\end{equation}
 underlying the condition $k > \nu_d/(2 c_{\mathrm{a},d})$ for magnetic pressure to overcome the drag force and to enable the propagation of the mode.

We note that for the case of two dust fluids ($\mathcal{N}=2$), we obtain the inequality
\begin{equation}
\begin{aligned}
    \tilde{c}_{\mathrm{ms},d}^2 - c_{\mathrm{a},d}^2 &= c_{\mathrm{a},g}^2 \left( \frac{\xi_1^2}{\theta_1} + \frac{\xi_2^2}{\theta_2} - \frac{1}{\theta_1 + \theta_2} \right)  \\ &= c_{\mathrm{a},d}^2 \frac{(\xi_1 \theta_2 - \xi_2 \theta_1)^2}{\theta_1 \theta_2}>0.    
\end{aligned}
\end{equation}

The inequality $\tilde{c}_{\mathrm{ms},d}>c_{\mathrm{a},d}$ means that splitting a dust fluid into two fluids when sampling a distribution increases the wave speed.

\section{Multifluid MHD modes} \label{app:multifluid_MHD_modes}

The model, including the equations and the assumptions, is presented in Sect. \ref{sec:multifluid_MHD_equations}. We remind that the magnetic resistivities of the multifluid MHD model, denoted in this section as $\eta_{\mathrm{O}}^L$, $ \eta_{\mathrm{H}}^L$ and $\eta_{\mathrm{AD}}^L$, include the conductivities of the ions and the electrons that constitute $L$, but not the dust species.

\subsection{Alfvén waves} \label{app:multifluid_MHD_alfven_modes}

\begin{figure*}
    \centering
    \includegraphics[width=0.49\textwidth]{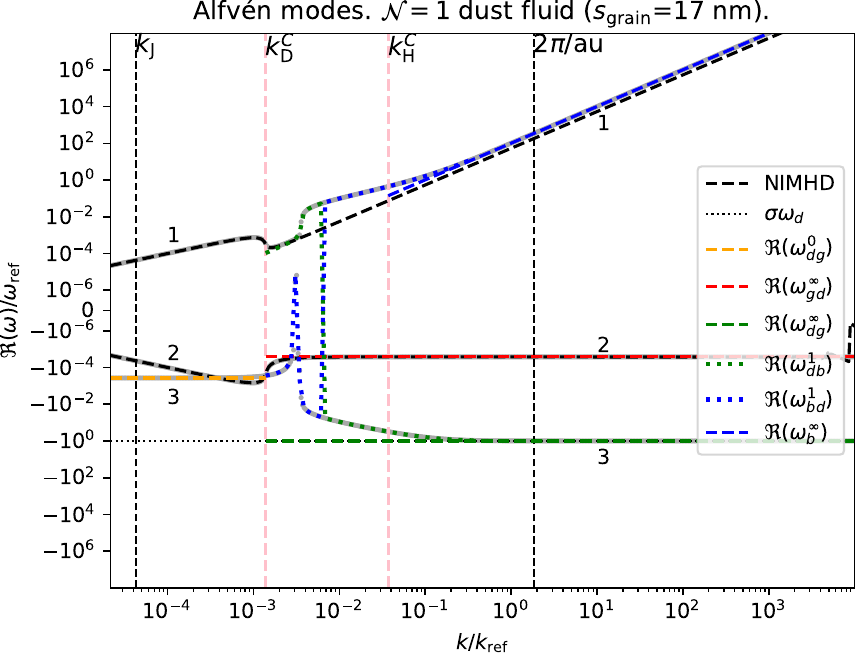}
    \includegraphics[width=0.49\textwidth]{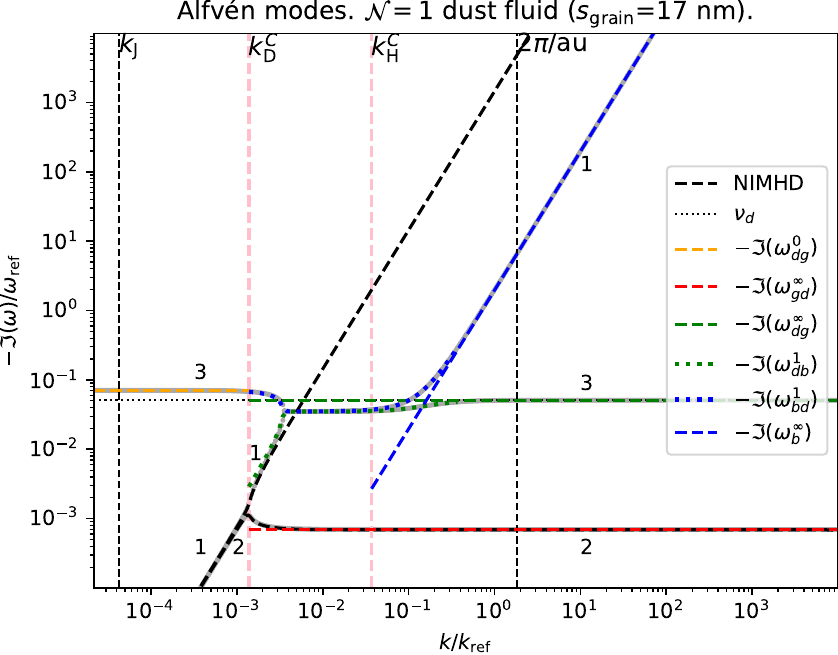}

    \includegraphics[width=0.49\textwidth]{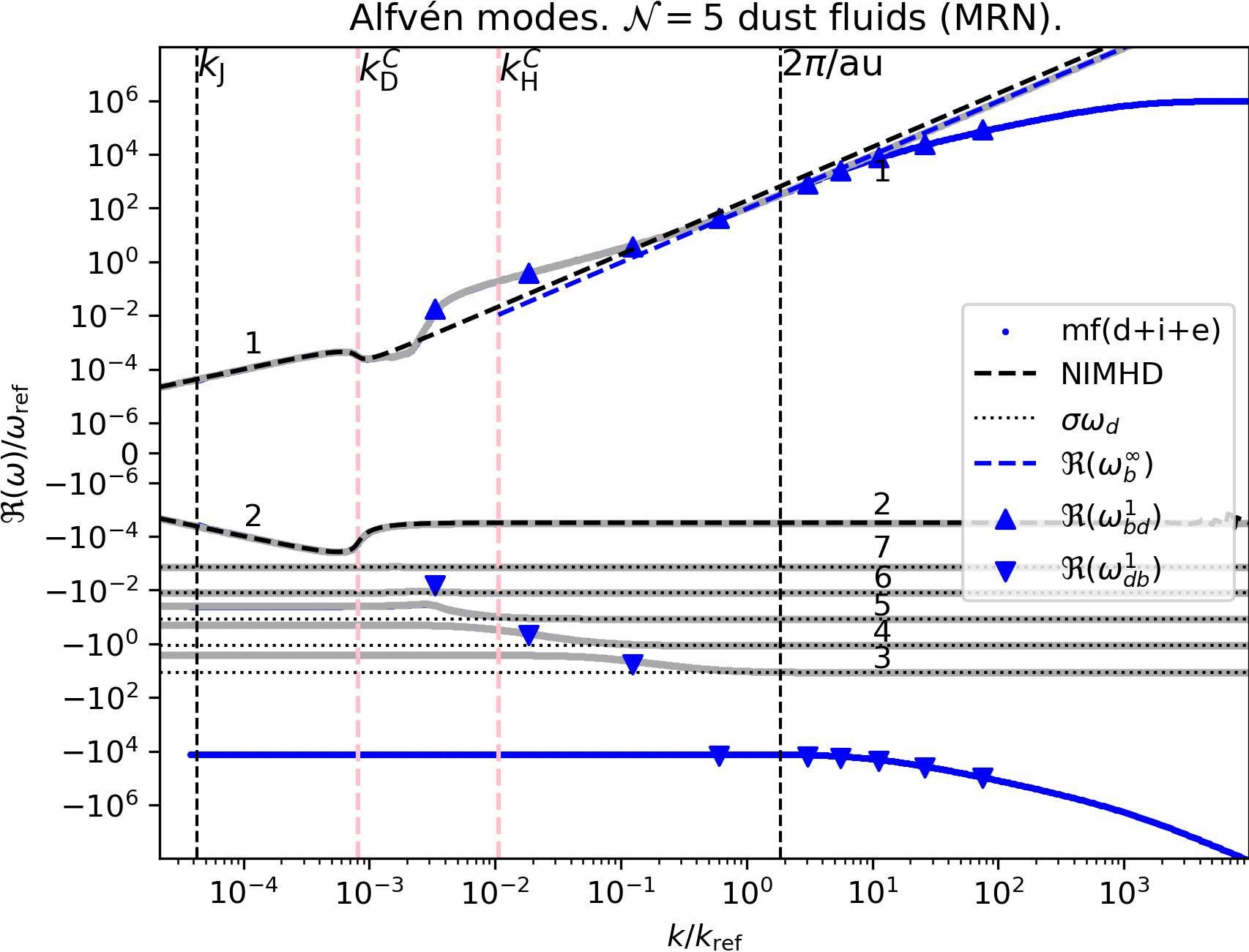}
    \includegraphics[width=0.49\textwidth]{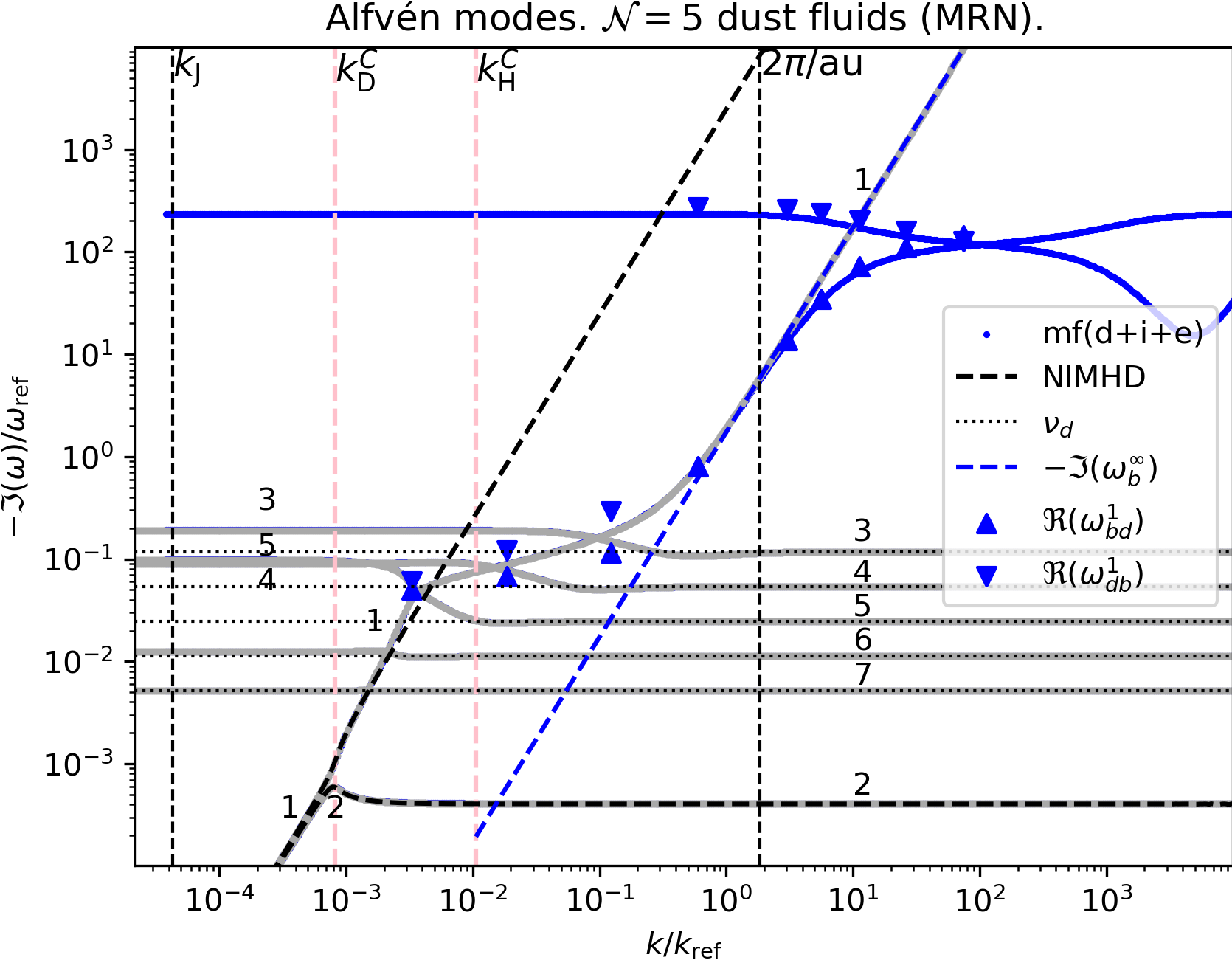}

    \caption{Dispersion relation of Alfvén modes ($\sigma=1$) for $\mathcal{N}=1$ and $5$ dust fluids (in gray). This figure complements Fig. \ref{fig:alfven_wave_dispersion_color} with analytical solutions and notations summarized in Tables \ref{table:notations} and \ref{table:notations_continuation} in Sect. \ref{app:analytical solutions and notations}. In addition, we show the modes when including the inertia of the ions and the electrons (denoted as mf(d+i+e), in blue) plotted under the solutions of the multifluid MHD (in gray). }
    \label{fig:alfven_wave_dispersion_analytical}
\end{figure*}

\subsubsection{Linearized equations}

The dynamics of the $\mathcal{N}+2$ perturbations is
\begin{equation}
    \mathrm{d}_t v_{g,\sigma} = \sum_{d \in I} \theta_d \nu_d (v_{d,\sigma} - v_{g,\sigma} ) + \frac{\sum_{l \in L} n_l Z_l e}{\rho_g} \delta e_{g,\sigma} - i\sigma \frac{B_0}{\rho_g c} j_{l,\sigma}, 
\end{equation}
\begin{equation}
    \mathrm{d}_t v_{d,\sigma} = \nu_d ( v_{g,\sigma} -v_{d,\sigma} ) + \omega_d \left( \frac{c}{B_0} \delta e_{g,\sigma} - i\sigma (v_{d,\sigma} - v_{g,\sigma}) \right),
    \label{eq:multifluid_MHD_alfven_dust_dynamics}
\end{equation}
\begin{equation}
    \mathrm{d}_t \delta b_\sigma = ikB_0 v_{g,\sigma} + ck \sigma \delta e_{g,\sigma},
\end{equation}
with the closure relations
\begin{equation}
    j_{l,\sigma}  =  - \frac{c k}{4 \pi} \sigma \delta b_\sigma - \sum_{d \in I} n_d Z_d e (v_{d,\sigma} - v_{g,\sigma} ),
\end{equation}
\begin{equation}
    \delta e_{g,\sigma} = \eta_\sigma^L  j_{l,\sigma},
\end{equation}
with 
\begin{equation}
    \eta_\sigma^L =   \eta_{\mathrm{O}}^L + \eta_{\mathrm{AD}}^L - i\sigma \eta_{\mathrm{H}}^L.
    \label{eq:multifluid_MHD_alfven_resistivity}
\end{equation}

\subsubsection{Some notations}

We found it necessary to introduce some quantities, that appear in asymptotical regimes. There is the effective collision rate of the light species (ions and electrons), which is
\begin{equation}
    \nu_{L,\sigma} = \frac{B_0^2}{\rho_g c^2 \eta_\sigma^L} +i\sigma \frac{\sum_{l \in L} n_l Z_l e B_0}{\rho_g c}.
    \label{eq:multifluid_MHD_alfven_effective_collision_rate_light}
\end{equation}

We also define for a dust fluid $d$ its contribution to the generation of the electric field
\begin{equation}
    \Xi_d = i \sigma \frac{c}{B_0} \eta_\sigma^L n_dZ_de
    \label{eq:multifluid_MHD_dust_electric_field_contribution}
\end{equation}
and the resonant frequency
\begin{equation}
    \Omega_{d, \sigma} = \sigma \omega_d - i \nu_d.
\end{equation}

\subsubsection{Larges scales: ideal MHD waves} \label{app:multifluid_MHD_alfven_modes_ideal_regime}

They are two propagative Alfvén modes at $k \to 0$ and $\omega \to 0$, driven by magnetic tension, which correspond to the ideal MHD regime for the gas and dust mixture. There is a small subtlety when comparing to the prediction of non-ideal MHD. Indeed, at large timescales, dust fluids become low-Stokes (i.e., $\omega \ll \nu_d $) and couple to the gas resulting in the mass loading of the wave, similarly to what we explain in pure hydrodynamics in Sect. \ref{sec:multidust_hydro_waves}. Thus the phase velocity of the Alfvén mode is $c_{\mathrm{a},gd}=B_0/\sqrt{4 \pi \rho}$ for the multifluid MHD instead of $c_{\mathrm{a},g}=B_0/\sqrt{4 \pi \rho_g}$.

\subsubsection{Larges scales: collision modes} \label{app:multifluid_MHD_alfven_modes_large_scales_collision}

In addition to the two propagative Alfvén modes at $k \to 0$ and $\omega \to 0$, there are collision modes. As for the hydrodynamical case, we look for modes such that $k \to 0$ and $\omega$ remaining finite. We assume that the electric current, at the origin of the magnetic perturbation, tends to zero
\begin{equation}
    j_{\sigma} = - \frac{ck}{4 \pi} \sigma \delta b_\sigma \to 0.
\end{equation}

The Ohm law becomes $\delta e_{g ,\sigma} = \eta_\sigma^L j_{l,\sigma} = - \eta_\sigma^L j_{i,\sigma}$, and the dynamics of the gas and the dust fluids become 
\begin{equation}
    \mathrm{d}_t v_{g,\sigma} = \sum_{i \in I} (\theta_i \nu_i + \Xi_i \nu_{L,\sigma} ) (v_{i,\sigma} - v_{g,\sigma}),
\end{equation}
\begin{equation}
    \mathrm{d}_t v_{i,\sigma} = (\nu_i + i \sigma \omega_i) (v_{g,\sigma} - v_{i,\sigma}) 
    -i \sigma \omega_i \sum_{d \in I} \Xi_d  (v_{g,\sigma} - v_{d,\sigma}).
\end{equation}

This system of ordinary differential equations is complex, with effective interactions between the dust fluids (via the electric field). 

For the case of a single dust fluid ($\mathcal{N}=1$) and by defining the effective collision rates $ \nu_d^\sigma = \nu_d + i\sigma \omega_d ( 1-\Xi_d)$ and $
    \nu_g^\sigma = \theta_d \nu_d + \Xi_d \nu_{L,\sigma}
$,
we can rewrite the closed system as
\begin{equation}
    \frac{\mathrm{d}}{\mathrm{d}t}
    \begin{pmatrix}
        v_{g,\sigma} \\
        v_{d,\sigma}
    \end{pmatrix}
    =
    \begin{pmatrix}
        - \nu_g^\sigma & \nu_g^\sigma \\
        \nu_d^\sigma  & -\nu_d^\sigma 
    \end{pmatrix}
    \begin{pmatrix}
        v_{g,\sigma} \\
        v_{d,\sigma}
    \end{pmatrix}
    .
\end{equation}

The eigenvalues of the matrix are $(0,-(\nu_g^\sigma +\nu_d^\sigma ))$ with the respective eigenvectors $(v_{g,\sigma},v_{d,\sigma}) = (1,1)$ and $(v_{g,\sigma},v_{d,\sigma})=(-\nu_g^\sigma,\nu_d^\sigma)$. This matrix is a drag matrix with an effective dust-to-gas ratio $\theta_d^\sigma = \nu_g^\sigma / \nu_d^\sigma$. We note that $\theta_d^\sigma \approx \theta_d$ requires $\omega_d  \ll \nu_d$ (small Hall factor). The pulsation of the collision mode is
\begin{equation}
    \omega =-i(\nu_g^\sigma +\nu_d^\sigma ),
    \label{eq:multifluid_MHD_collision_mode_large_scales}
\end{equation}
which is not purely imaginary.

\subsubsection{Small scales: collision modes} \label{app:multifluid_MHD_alfven_modes_small_scales_collision}

In non-ideal MHD, we found two modes at small scales as presented in Appendix \ref{app:NIMHD_Alfvén}. The whistler mode, with $\omega \propto k^2$, is a magnetic perturbation, whereas the other, with $\omega$ remaining finite, is a fluid velocity perturbation. In multifluid MHD, we find the same asymptotic separation, with new collision modes due to velocity drifts. 

We start by analyzing the collision modes. We assume that $\omega$ remains constant when $k \to \infty$. Then, because of the induction equation $\delta b_\sigma \propto k \delta e_\sigma$, there is no induced electric field. We note that the dust fluids can only communicate via the gas, contrary to the collision modes at large scales. For closing the equation of the gas dynamics, we use
\begin{equation}
    \delta e_{g,\sigma} = -i \sigma \frac{B_0}{c} v_{g,\sigma}.
\end{equation}

The dynamics of the gas and the dust fluids are
\begin{equation}
    \mathrm{d}_t v_{g,\sigma} = - \left(\sum_{i\in I} \theta_i \nu_i + \nu_{L,\sigma} \right)  v_{g,\sigma} +  \sum_{i\in I} \theta_i \nu_i v_{i,\sigma},
\end{equation}
\begin{equation}
   \mathrm{d}_t v_{i,\sigma} = \nu_i v_{g,\sigma} - (\nu_i + i \sigma \omega_i)  v_{i,\sigma}.
\end{equation}

The system is analogous to the one in pure hydrodynamics, but with an extra friction of the gas $\nu_{L,\sigma}$ due to ions and electrons, and more importantly, there is an extra term due to the magnetic part of the individual Lorentz forces. This is the gyration term $-i \sigma \omega_i v_{i,\sigma}$. It generates a drift relative to the gas even for low-Stokes grains, contrary to pure hydrodynamics. Indeed, the dynamical response of a dust fluid is
\begin{equation}
    v_{k,\sigma} = \frac{\nu_k}{-i \omega + \nu_k + i \sigma \omega_k} v_{g,\sigma}.
\end{equation}

The dynamical response is strong when $\omega \approx \Omega_{d, \sigma} = \sigma \omega_d - i \nu_d$. 

Even when neglecting the inertia term, the dust fluid drifts if its Hall factor $\Gamma_k=\omega_k/\nu_k $ is high compared to unity:
\begin{equation}
    v_{l,\sigma} = \frac{1}{1+i \sigma \Gamma_l} v_{g,\sigma}.
\end{equation}

For the case of a single dust fluid ($\mathcal{N}=1$), the system can be solved. The two solutions are 
\begin{equation}
    \omega= -\frac{i}{2} \left( \nu_{C,\sigma} \pm \sqrt{\nu_{C,\sigma}^2-4( \nu_{L,\sigma}(\nu_d + i \sigma \omega_d )+ i\sigma \theta_d \omega_d \nu_d)} \right),
    \label{eq:multifluid_MHD_collision_modes_small_scales_1bin}
\end{equation}
with the effective collision rate $\nu_{C,\sigma} = \nu_{L,\sigma} + \theta_d\nu_d + \nu_d + i\sigma \omega_d $. In pure hydrodynamics, $\nu_{C,\sigma} \to (1+\theta_d) \nu_d$, which is the relaxation rate of collision modes. The $-$ solution corresponds to the standard non-ideal MHD mode, and the $+$ solution to the collision mode of interest. We note that if $\theta_d \ll1$ and $\nu_{L,\sigma} \ll \nu_d , \omega_d$, then $\nu_{C,\sigma} \approx \nu_d + i\sigma \omega_d = i\Omega_{d, \sigma} $. Therefore, in practice, the solution is close to the resonant frequency $\Omega_{d, \sigma} $, implying a strong dynamical response of the dust fluid.

\subsubsection{Small scales: magnetic mode} \label{app:multifluid_MHD_alfven_whistler_small}

As mentioned in the previous section, by assuming that the gas and dust velocities vanish at $\omega \propto k^2$ when $k \to \infty$, then $j_{l,\sigma}=j_\sigma$ and 
from the induction equation,
we obtain the solution
\begin{equation}
    \omega = -i \frac{c^2}{4 \pi} \eta_\sigma^L k^2.
    \label{eq:multifluid_MHD_alfven_whistler_small}
\end{equation}

It is the same solution form as in non-ideal MHD (Appendix \ref{app:NIMHD_Alfvén} with Eq. \eqref{eq:NIMHD_Alfvén_whistler}) but here the conductivities of the dust species are not included in the resistivities. Dust couples to the magnetic field perturbation at a more intermediate scale.

\subsubsection{Intermediate scales: dust coupling to the magnetic perturbation}

The intermediate regime, at scales smaller than the diffusion scale $1/k_{\mathrm{D}}^C$, is different from non-ideal MHD. The magnetic field perturbation decouples from the gas and couples to the dust, until it is carried by the ions and the electrons in the whistler mode. We assume $v_{g,\sigma}=0$. We obtain 
\begin{equation}
    \mathrm{d}_t v_{i,\sigma} = - (  \nu_i +i\sigma \omega_i)  v_{i,\sigma} - \sigma \frac{\omega_i}{B_0} \frac{c^2}{4 \pi} \eta_\sigma^L k \delta b_\sigma +i\sigma \omega_i \sum_{d \in I} \Xi_d v_{d,\sigma},
\end{equation}
\begin{equation}
    \mathrm{d}_t \delta b_{\sigma} = -  \frac{c^2}{4 \pi} \eta_\sigma^L k^2 \delta b_\sigma + ik B_0 \sum_{i \in I} \Xi_i v_{i,\sigma}
\end{equation}

For the case of a single dust fluid ($\mathcal{N}=1$), it simplifies to
\begin{equation}
    \mathrm{d}_t 
    \begin{pmatrix}
        v_{d,\sigma} \\
        \delta b_\sigma 
    \end{pmatrix}
    =
    \begin{pmatrix}
        - \nu_d^\sigma & - \sigma \frac{\omega_d}{B_0} \frac{c^2}{4 \pi} \eta_\sigma^L k \\
        i k B_0 \Xi_d & -  \frac{c^2}{4 \pi} \eta_\sigma^L k^2
    \end{pmatrix}
    \begin{pmatrix}
        v_{d,\sigma} \\
        \delta b_\sigma 
    \end{pmatrix},
\end{equation}
where $ \nu_d^\sigma= \nu_d + i \sigma\omega_d (1-\Xi_d )$ is the collision rate we obtain when studying the collision modes at large scales (Appendix \ref{app:multifluid_MHD_alfven_modes_large_scales_collision}).

The solutions are
\begin{equation}
\begin{aligned}
    \omega =  \frac{i}{2} \left(- ( \nu_{d}^\sigma +  \frac{c^2}{4 \pi} \eta_\sigma^L k^2 ) \right. \\ \left.   \pm  \sqrt{ ( \nu_{d}^\sigma  +  \frac{c^2}{4 \pi} \eta_\sigma^L k^2 )^2 -4 (\nu_{d}+i \sigma\omega_d )   \frac{c^2}{4 \pi} \eta_\sigma^L k^2 } \right)   .
    \label{eq:multifluid_MHD_Alfven_wave_dust_intermediate_scales}
\end{aligned}
\end{equation}

We can identify the two modes by their small-scale asymptotes. The $-$ solution becomes the magnetic mode whereas the $+$ solution  becomes the collision mode on fluid velocities.

If we neglect the drag forces with the gas, the solutions of Eq. \eqref{eq:multifluid_MHD_Alfven_wave_dust_intermediate_scales} simplify to the waves propagating in a dusty plasma under Hall MHD:
\begin{equation}
    \omega= \frac{\sigma}{2} \left( -\frac{c^2}{4 \pi \hat{\sigma}_{\mathrm{H}}^L} k^2 \pm \sqrt{(\frac{c^2}{4 \pi \hat{\sigma}_{\mathrm{H}}^L}  k^2 )^2 + 4 c_{\mathrm{a},d}^2k^2} \right),
    \label{eq:dusty_Hall_MHD_modes}
\end{equation}
with $c^2/(4 \pi \hat{\sigma}_{\mathrm{H}}^L)= c_{\mathrm{a},d}^2 / \omega_d$ according to Eq. \eqref{eq:multidust_plasma_Hall_conductivity}. Equation \eqref{eq:dusty_Hall_MHD_modes}
 approximates the two large-scale asymptotes of the propagative part of Eq. \eqref{eq:multifluid_MHD_Alfven_wave_dust_intermediate_scales}:
\begin{equation}
    \omega = - \sigma \frac{c^2}{4 \pi \hat{\sigma}_{\mathrm{H}}^L} k^2=-\sigma \frac{c_{\mathrm{a},d}^2k^2}{\omega_d}, \text{ and } \omega = \sigma \frac{4 \pi}{c^2} \hat{\sigma}_{\mathrm{H}}^L c_{\mathrm{a},d}^2 = \sigma \omega_d.
\end{equation}

\subsubsection{Coupling of multiple charged fluids with the magnetic field} \label{app:multifluid_Alfven_all_inertia}

It is possible to analyze the propagation of Alfvén modes considering the inertia of all charged species (dust, ion and electrons) as shown in \citep{2023A&A...674A.149H}. Indeed, these modes are incompressible contrary to magnetosonic modes for which variations in the density of the charged species generate local deviations to electroneutrality that requires the treatment of the displacement current, leading to plasma oscillations. 

In this section, we can consider that $d \in I = C$, standing for the dust fluids, the ions and the electrons, and thus $L= \varnothing$. The linearized equations for Alfvén modes are

\begin{equation}
\mathrm{d}_t v_{g,\sigma} = \sum_{d\in I} \theta_d \nu_d (   v_{d,\sigma}-  v_{g,\sigma} ) ,
\end{equation}
\begin{equation}
\mathrm{d}_t v_{d,\sigma} = \nu_d (v_{g,\sigma} - v_{d,\sigma})-i \sigma \omega_d   v_{d,\sigma} + \omega_d \frac{c}{B_0} \delta e_{\sigma}  ,
\end{equation}
\begin{equation}
    \mathrm{d}_t \delta b_\sigma = ck \sigma \delta e_\sigma.
\end{equation}

Instead of using an Ohm law to close the system of equations, the Ampere law provides an implicit constraint by connecting the magnetic field with the electrical current, and so the velocities of the charged fluids:
\begin{equation}
    j_\sigma = - \frac{ck}{4 \pi } \sigma \delta b_\sigma = \sum_{d \in I } n_d Z_d e v_{d,\sigma}.
\end{equation}

We include this implicit closure in the linearized dynamical system as in Eq.  \eqref{eq:formal_ODE_dynamics} and we use the same numerical method. The dispersion relation is plotted in blue lines in Fig. \ref{fig:alfven_wave_dispersion_analytical} under the solution of multifluid MHD (in gray).

We recover the same modes as in multifluid MHD, but the whistler mode becomes an ion-cyclotron mode at small scales ($\omega \approx \Omega_{i,\sigma} =\sigma \omega_i - i \nu_i$) due to the ion inertia. Moreover, we obtain an additional mode that turns to an electron-cyclotron mode ($\omega \approx \Omega_{e,\sigma}  = \sigma \omega_e - i \nu_e$) at small scales (outside the figure) due to the electron inertia. These modes are analogous to the collision modes we find previously accounting for the inertia of the dust fluids.

The complete model is useful to derive the velocity of the Alfvén waves when the charged fluids couple to the magnetic field and decouple from the gas. Indeed, assuming that $v_{g,\sigma}=0$, we obtain a simple dispersion relation
\begin{equation}
    k^2 =  \omega \sum_{d \in I} \frac{4 \pi \rho_d}{B_0^2} \frac{\omega_d^2}{\Omega_{d,\sigma}-\omega},
\end{equation}
with the eigenvector
\begin{equation}
   v_{d,\sigma}= \frac{\sigma \omega_d}{\omega -\Omega_{d, \sigma }} \frac{\omega}{k} \frac{\delta b_\sigma}{B_0}.
\end{equation}

As for the sound waves, we can study this dispersion relation assuming a hierarchy of stopping times and gyration times (Appendix \ref{app:multidust_hydro_waves_collision_modes}). We split the sum with the contributions of high-Stokes fluids (more rigorously, $|\omega /\Omega_{d,\sigma} | \ll 1$) and low-Stokes fluids (i.e., $|\omega /\Omega_{d,\sigma} | \gg 1$). We can obtain a second-order polynomial in $\omega$:
\begin{equation}
\begin{aligned}
    \omega \sum_{d \in I} \frac{4 \pi \rho_d}{B_0^2} \frac{\omega_d^2}{\Omega_d-\omega} &=  -\sum_{\substack{i \in I, \\ |\omega /\Omega_{i,\sigma} | \gg 1}}  \frac{4 \pi \rho_i}{B_0^2} \omega_i^2 \\&+ \omega \sum_{ \substack{l \in I, \\ |\omega /\Omega_{l,\sigma} | \ll 1}} \frac{4 \pi \rho_l}{B_0^2} \frac{\omega_l^2}{\Omega_{l,\sigma}} \\&+ \omega^2 \sum_{ \substack{l \in I, \\ |\omega /\Omega_{l,\sigma} | \ll 1}} \frac{4 \pi \rho_l}{B_0^2} \frac{\omega_l^2}{\Omega_{l,\sigma}^2},
    \label{eq:multifluid_alfven_dispersion_relation_hierarchy}
\end{aligned}
\end{equation}    
with the zeroth-order term for the low-Stokes fluids providing the scale
\begin{equation}
    \kappa_I = \sqrt{\sum_{i \in I, |\omega /\Omega_{i,\sigma} | \gg 1}  \frac{4 \pi \rho_d}{B_0^2} \omega_d^2},
\end{equation}
the zeroth-order term for the high-Stokes fluids providing the conductivity (to compare with Eq. \eqref{eq:multidust_plasma_Hall_conductivity})
\begin{equation}
    \Sigma_L = \sum_{l \in I, |\omega /\Omega_{l,\sigma} | \ll 1} \frac{4 \pi \rho_l}{B_0^2} \frac{\omega_l^2}{\Omega_{l,\sigma}},
\end{equation}
and the first-order term for the high-Stokes fluids providing the mass-loaded Alfvén speed
\begin{equation}
    c_{\mathrm{a},L} = \frac{B_0}{\sqrt{4 \pi \sum_{l \in I, |\omega /\Omega_{l,\sigma} | \ll 1} \tilde{\rho}_l }},
\end{equation}
with the effective contribution of a fluid in the mass loading
\begin{equation}
    \tilde{\rho}_l = \rho_l \Gamma_l^2/(\Gamma_l-i \sigma )^2.
    \label{eq:mass_loading_alfven_multifluid}
\end{equation}
We note that $\tilde{\rho}_l \to 0$ if $\Gamma_l \to 0$ and $\tilde{\rho}_l \to \rho_l $ if $\Gamma_l \to \infty$. It means that only charged fluids with a significant Hall factor ($\Gamma_d \gtrsim 1$) contribute to the loading. 

Once the norm of Eq. \eqref{eq:mass_loading_alfven_multifluid} taken, it is reminiscent of the weight function chosen by \citet{2007A&A...476..263G} in their equation (12). Their estimate follows the discussion of \citet{2004ApJ...610..781C} to determine which grains should contribute to the propagation of a magnetosonic wave (based on their Hall factors) assuming a mass-loaded speed, and their estimate is probably motivated by the expression of ambipolar drift velocity (first term in Eq. \eqref{eq:light_drift}). We present a different mechanism from this intuitive mass loading for magnetosonic waves in Appendix \ref{app:multidust_plasma_ms} and Sect. \ref{sec:multifluid_MHD_ms_modes}.

The two solutions to the simplified dispersion relation (Eq. \eqref{eq:multifluid_alfven_dispersion_relation_hierarchy}) are
\begin{equation}
    \omega = - \frac{c_{\mathrm{a},L}^2 \Sigma_L}{2} \pm c_{\mathrm{a},L} \sqrt{\frac{c_{\mathrm{a},L}^2 \Sigma_L^2}{4} + \kappa_I^2 + k^2}.
    \label{eq:multifluid_Alfven_solution}
\end{equation}

We plot the two solutions in Fig. \ref{fig:alfven_wave_dispersion_analytical} with blue upward-pointing arrows for the + solution and downward-pointing arrows for the - solution. In practice, because the hierarchy assumption is not necessary satisfied for a given charge distribution, we only split the charged species based on the magnetic Stokes number $\Re(\omega)/|\omega_d|$, which is relevant for species with a high Hall factor (i.e. species that preferentially couple to the magnetic field).

At low frequencies ($\Re(\omega) < \min(\omega_i)$), once the charges carry the propagation of Alfvén waves, $\kappa_I$ tends to zero and, if high Hall-factor species mainly contribute to the charge electroneutrality that can be written as $\sum_{k \in C} \rho_k \omega_k = 0$, then $\Sigma_L$ becomes small. The main contribution in Eq. \eqref{eq:multifluid_Alfven_solution} is then $\omega \approx \pm c_{\mathrm{a},L} k$. This approximation emphasizes the symmetrization of the Alfvén waves when accounting for the grain inertia, which strengthens the preliminary observation of \citet{2023A&A...674A.149H}.

\subsection{Magnetosonic modes} \label{app:multifluid_MHD_ms_modes}

\begin{figure*}
    \centering
    \includegraphics[width=0.49\textwidth]{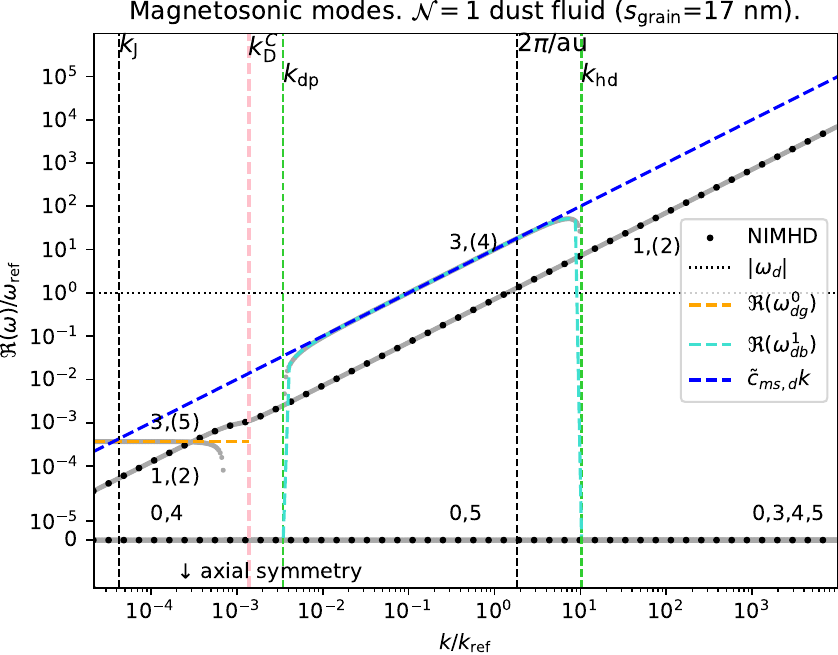}
    \includegraphics[width=0.49\textwidth]{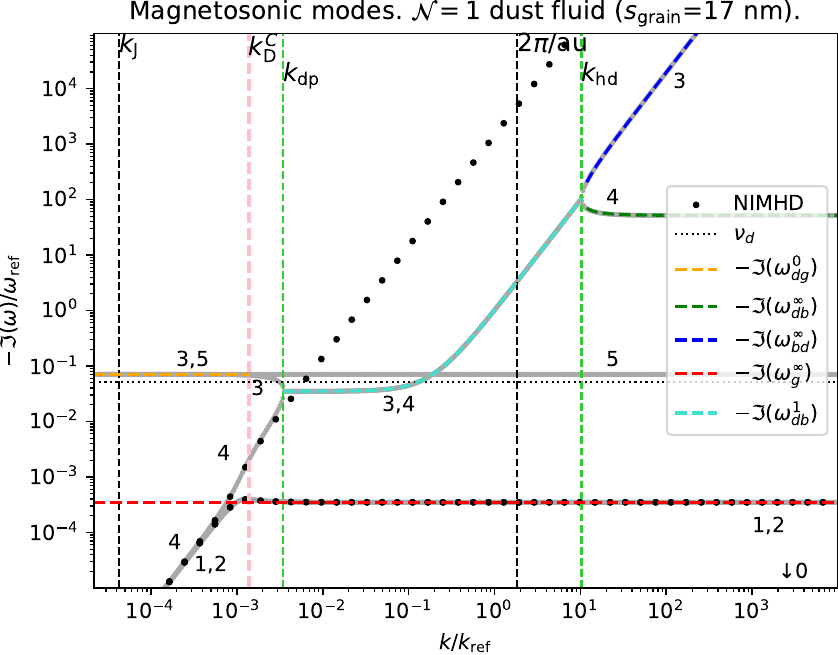}

    \includegraphics[width=0.49\textwidth]{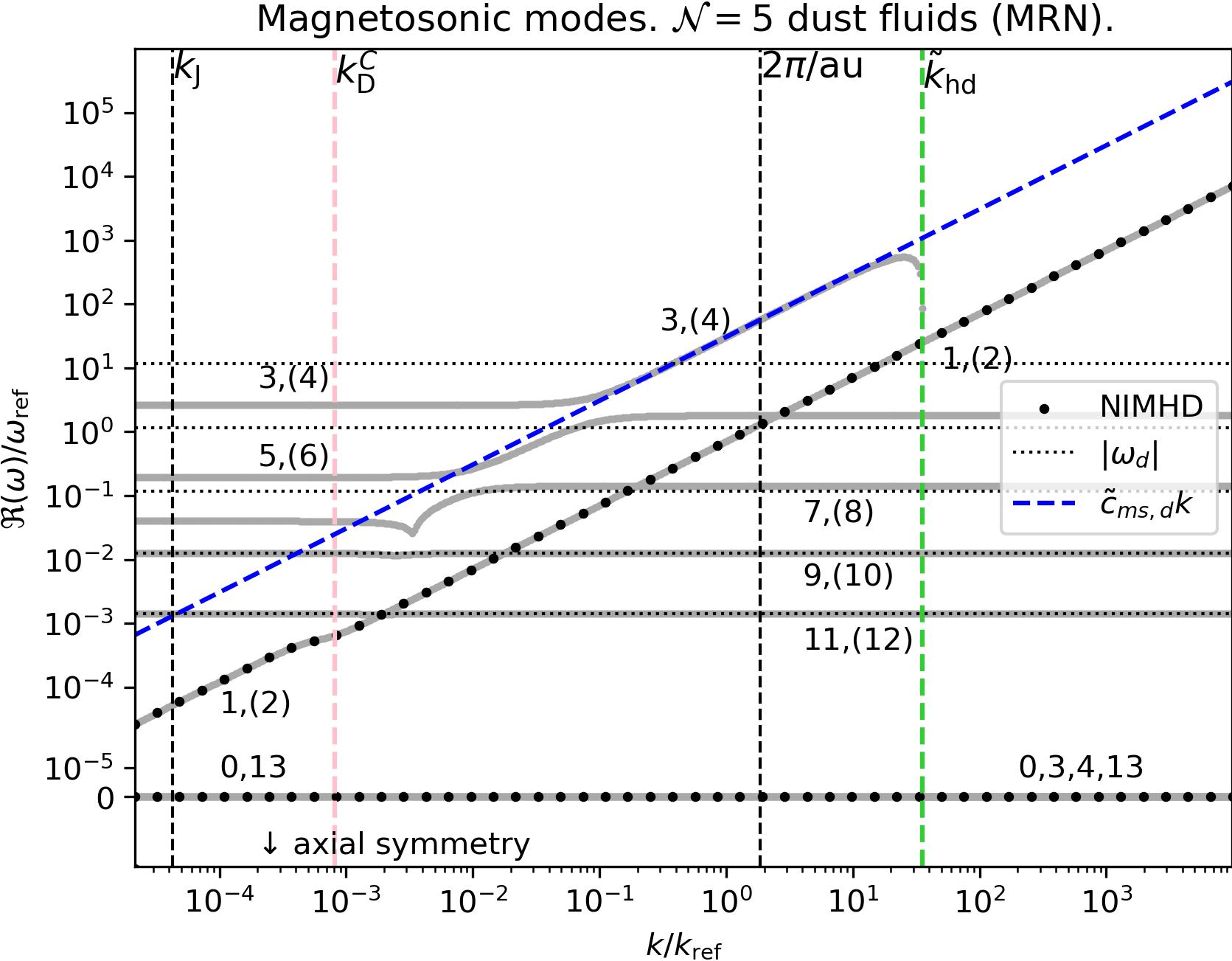}
    \includegraphics[width=0.49\textwidth]{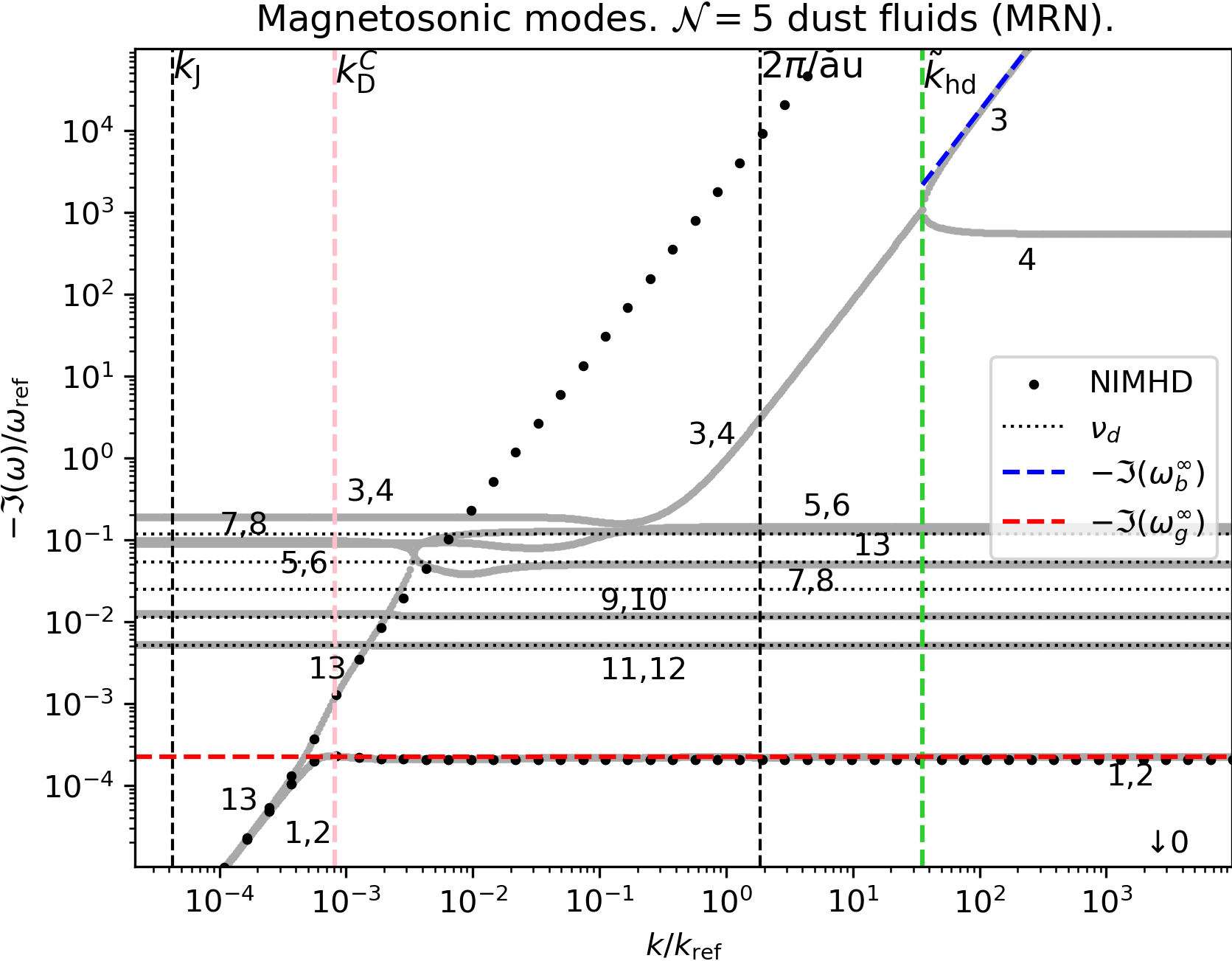}
    
    \caption{
    Dispersion relation of magnetosonic modes for $\mathcal{N}=1$ and $5$ dust fluids (in gray). This figure complements Figs. \ref{fig:ms_wave_dispersion_17nm_color} and \ref{fig:ms_wave_dispersion_5bins_100bins_color} with analytical solutions and notations summarized in Tables \ref{table:notations} and \ref{table:notations_continuation} in Sect. \ref{app:analytical solutions and notations}. }
    \label{fig:ms_wave_dispersion_analytical}
\end{figure*}

\subsubsection{Linearized equations}

The dynamics of the $2\mathcal{N}+4$ perturbations is
\begin{equation}
    \mathrm{d}_t \delta \rho_g = -ik \rho_g v_{g,x},
\end{equation}
\begin{equation}
  \left\{
      \begin{aligned}
         \mathrm{d}_t v_{g,x} =& -\frac{ik c_{\mathrm{s},g}^2}{\rho_g} \delta \rho_g + \sum_{d \in I} \theta_d \nu_d ( v_{d,x} -v_{g,x} ) \\
         &+ \frac{\sum_{l \in L} n_l Z_l e}{\rho_g} \delta e_{g,x} + \frac{B_0}{\rho_g c} j_{l,y}, \\
         \mathrm{d}_t v_{g,y} =&  \sum_{d \in I} \theta_d \nu_d ( v_{d,y} -v_{g,y} ) + \frac{\sum_{l \in L} n_l Z_l e}{\rho_g} \delta e_{g,y} -\frac{B_0}{\rho_g c} j_{l,x},  \\
      \end{aligned}
    \right.
\end{equation}
\begin{equation}
  \left\{
      \begin{aligned}
         \mathrm{d}_t v_{d,x} &=  \nu_d ( v_{g,x} -v_{d,x} ) -\omega_d (v_{g,y} - v_{d,y}   ) + \frac{\omega_d c}{B_0} \delta e_{g,x}, \\
          \mathrm{d}_t v_{d,y} &= \nu_d ( v_{g,y} -v_{d,y} ) +\omega_d (v_{g,x} - v_{d,x}   ) + \frac{\omega_d c}{B_0} \delta e_{g,y},  \\
      \end{aligned}
    \right.
    \label{eq:multifluid_MHD_ms_linearized_dust_velocities}
\end{equation}
\begin{equation}
    \mathrm{d}_t \delta b = - ik B_0 v_{g,x} - ikc \delta e_{g,y},
\end{equation}
with the closure relations
\begin{equation}
  \left\{
      \begin{aligned}
         \delta e_{g,x} &=  (\eta_{\mathrm{O}}^L + \eta_{\mathrm{AD}}^L) j_{l,x} + \eta_{\mathrm{H}} j_{l,y}, \\
          \delta e_{g,y} &=  - \eta_{\mathrm{H}}^L j_{l,x}+ (\eta_{\mathrm{O}}^L + \eta_{\mathrm{AD}}^L) j_{l,y} , \\
      \end{aligned}
    \right.
\end{equation}
\begin{equation}
  \left\{
      \begin{aligned}
         j_{l,x} &=  - \sum_{d \in I} n_d Z_d e (v_{d,x} - v_{g,x}) , \\
          j_{l,y} &=  - \sum_{d \in I} n_d Z_d e (v_{d,y} - v_{g,y}) - \frac{i c k}{4 \pi} \delta b. \\
      \end{aligned}
    \right.
\end{equation}

We note that the inversion of Eq. \eqref{eq:multifluid_MHD_ms_linearized_dust_velocities} leads to 
\begin{equation}
\begin{aligned}
        \begin{pmatrix}
        \delta v_{d,x} \\
        \delta v_{d,y} 
    \end{pmatrix}
    &= \frac{1}{(- i \omega  +\nu_d)^2 + \omega_d^2} \\
    & \times
    \begin{pmatrix}
        (-i \omega + \nu_d)\nu_d \delta v_{g,x} +\omega_d (\nu_d \delta v_{g,y} + \frac{\omega_d c}{B_0} \delta e_y ) \\
        (- i \omega  + \nu_d)(\nu_d \delta v_{g,y} + \frac{\omega_d c}{B_0} \delta e_y ) - \omega_d \nu_d \delta v_{g,x}
    \end{pmatrix},
\end{aligned}
\end{equation}
and thus we can define a typical resonant frequency
\begin{equation}
    \Omega_{d,\pm} = \pm \omega_d - i \nu_d.
\end{equation}

\subsubsection{Larges scales: ideal MHD waves} 

They are two propagative magnetosonic modes at $k \to 0$ and $\omega \to 0$, driven by magnetic pressure and thermal pressure, which correspond to the ideal MHD regime for the gas and dust mixture. As explained for Alfvén waves in Appendix \ref{app:multifluid_MHD_alfven_modes_ideal_regime}, dust fluids asymptotically couple to the gas resulting in a magnetosonic speed $c_{\mathrm{ms},gd}=(c_{\mathrm{s},gd}^2  + c_{\mathrm{a},gd}^2 )^{1/2}$, instead of the value predicted by the standard non-ideal MHD that is $c_{\mathrm{ms},g}=(c_{\mathrm{s},g}^2  + c_{\mathrm{a},g}^2 )^{1/2}$. We find that non-ideal MHD cannot capture the mass loading of the wave by multiple components and the decoupling of the dust fluids from the sonic waves at smaller scales, as predicted in Sect. \ref{sec:multidust_hydro_waves}.  

\subsubsection{Large scales: collision modes}

The situation is similar to Alfvén waves in Appendix \ref{app:multifluid_MHD_alfven_modes_large_scales_collision}. At large scales, in addition to ideal magnetosonic waves, there are collision modes, with thermal pressure gradients and magnetic pressure gradients (generated by the electric current) vanishing to zero, compared to drag forces due to velocity drifts. When assuming $\omega$ finite with $k \to 0$, we loose the problem geometry induced by the direction of $\mathbf{k}$. The solutions of the collision modes are therefore the solutions in Appendix \ref{app:multifluid_MHD_alfven_modes_large_scales_collision}, which could be recovered by setting the variables $u_\pm = u_x \pm i u_y$.

\subsubsection{Small scales: sonic modes}

We assume that the thermal pressure dominates the gas dynamics, constraining its motion along $\mathbf{e}_x$. By assuming that the gas is linearly decoupled, we obtain the dynamics
\begin{equation}
    \mathrm{d}_t
    \begin{pmatrix}
        \rho_g \\
        v_{g,x}
    \end{pmatrix}
    =
    \begin{pmatrix}
        0 & -ik \rho_g \\
        -ik c_{\mathrm{s},g}^2/\rho_g & - \nu_g
    \end{pmatrix}
        \begin{pmatrix}
        \rho_g \\
        v_{g,x}
    \end{pmatrix},
\end{equation}
with the effective collision rate
\begin{equation}
    \nu_g 
    = \sum_{d\in I} \theta_d \nu_d + \left(\sum_{l\in L} n_l Z_le \right)^2  \frac{\eta_{\mathrm{O}}^L+\eta_{\mathrm{AD}}^L}{\rho_g} 
\end{equation}

Therefore the sound wave solutions are
\begin{equation}
    \omega \approx \pm c_{\mathrm{s},g} k - i \frac{\nu_g}{2} .
    \label{eq:multifluid_MHD_sonic_wave_small_scales}
\end{equation}

This does not seem to be a generalized form of Eq. \eqref{eq:NIMHD_sound_waves_small_scales} in non-ideal MHD. This is could be related to the assumptions. In practice, because of the small mass ratio between the charged species and the gas, the damping rate is sufficiently small to neglect the effects of the inertia of charged species. It is difficult to distinguish from the prediction of non-ideal MHD and from what we obtain when including the inertia of all charged species $\omega = \pm c_{\mathrm{s},g} k - i  (\sum_C \theta_k \nu_k)/2 $.

\subsubsection{Small scales: magnetic mode} \label{app:multifluid_MHD_ms_small_scales_magnetic_mode}

At large scales, with $\omega \propto k^2$, the magnetic perturbation linearly decouples from the fluid velocities. From the induction equation we obtain
\begin{equation}
    \omega = -i \frac{c^2}{4 \pi}(\eta_{\mathrm{O}}^L+ \eta_{\mathrm{AD}}^L) k^2,
    \label{eq:multifluid_MHD_ms_small_scales_magnetic_mode}
\end{equation}
where the conductivities of the dust fluids are not included in resistivities (but dust still affects the underlying charge balance). Only ions and electrons are included.

\subsubsection{Intermediate and small scales: dust coupling to the magnetic perturbation}

We assume that the dynamics of the dust fluid is along $\mathbf{e}_x$ because of the equivalent magnetic pressure. The dynamics coupled to the magnetic perturbation is
\begin{equation}
    \mathrm{d}_t 
        \begin{pmatrix}
        v_{d,x}/c_{\mathrm{a},d} \\
        \delta b /B_0
    \end{pmatrix}
    =-i 
    \begin{pmatrix}
        - i \nu_{\mathrm{D}} & \dfrac{k}{k_{\mathrm{H}}^{L} } \omega_d \\
        \dfrac{k}{k_{\mathrm{H}}^{L} } \omega_d  & -ik^2 \dfrac{c_{\mathrm{a},d}}{k_{\mathrm{D}}^{L} }
    \end{pmatrix}
            \begin{pmatrix}
        v_{d,x}/c_{\mathrm{a},d} \\
        \delta b /B_0
    \end{pmatrix}
    \label{eq:multifluid_MHD_monograin_ODE}
\end{equation}
with the resistive scales and the effective collision rate
\begin{equation}
    k_{\mathrm{D}}^{L} = \frac{c_{\mathrm{a},d}}{ \frac{c^2}{4 \pi} (\eta_{\mathrm{O}}^L + \eta_{\mathrm{AD}}^L) } , k_{\mathrm{H}}^{L}  = \frac{c_{\mathrm{a},d}}{\frac{c^2}{4 \pi} \eta_{\mathrm{H}}^L}, \nu_{\mathrm{D}} = \nu_d + \frac{\omega_d^2}{c_{\mathrm{a},d} k_{\mathrm{D}}^{L} }.
\end{equation}

We note that $\omega_d/k_{\mathrm{H}}^L \approx c_{\mathrm{a},d}$ in the multidust plasma approximation, thanks to Eq. \eqref{eq:multidust_plasma_Hall_conductivity} (rightmost equality), underlying the ideal coupling part of the dust fluid with the magnetic field (Appendix \ref{app:multifluid_feedback_Lorentz_forces}).

The eigenvector is formally
\begin{equation}
    (\omega + i \nu_{\mathrm{D}} ) \frac{v_{d,x}}{c_{\mathrm{a},d}} = \frac{k}{k_{\mathrm{H}}^{L}} \omega_d \frac{\delta b}{B_0}.
    \label{eq:multifluid_MHD_magnetic_dust_eigenvector}
\end{equation}
where $\omega$ satisfies the dispersion relation 
\begin{equation}
    \omega^2 + i ( \nu_{\mathrm{D}} + k^2 c_{\mathrm{a},d}/k_{\mathrm{D}}^L) \omega - k^2 ( \nu_{\mathrm{D}} c_{\mathrm{a},d}/k_{\mathrm{D}}^{L} + \omega_d^2 /(k_{\mathrm{H}}^{L})^2).
\end{equation}

Its discriminant is \begin{equation}
    \Delta = -k^4 (c_{\mathrm{a},d}/k_{\mathrm{D}}^{L} )^2 + 2k^2 ( (c_{\mathrm{a},d}/k_{\mathrm{D}}^{L}) \nu_{\mathrm{D}} + 2 (\omega_1/k_{\mathrm{H}}^{L})^2 ) - \nu_{\mathrm{D}}^2,
\end{equation}
with the two annulation points $k_{\mathrm{dp}}< k_{\mathrm{hd}}$ such that
\begin{equation}
\begin{aligned}
    &k_{\mathrm{dp}}, k_{\mathrm{hd}} = \\ &\frac{k_{\mathrm{D}}^L}{c_{\mathrm{a},d}}\sqrt{ \frac{c_{\mathrm{a},d} \nu_{\mathrm{D}}}{k_{\mathrm{D}}^L} + 2 \left(\frac{\omega_d}{k_{\mathrm{H}}^L} \right)^2 \mp 2 \sqrt{\left(\frac{\omega_d}{k_{\mathrm{H}}^L}\right)^4 + \left(\frac{\omega_d}{k_{\mathrm{H}}^L}\right)^2 \frac{c_{\mathrm{a},d} \nu_{\mathrm{D}}}{k_{\mathrm{D}}^L}} } .
    \label{eq:multifluid_MHD_kdp_khd}
    \end{aligned}
\end{equation}

At intermediate scales, between $k_{\mathrm{dp}}$ et $k_{\mathrm{hd}}$, the wave propagates ($\Delta>0$):
\begin{equation}
    \omega = \pm \sqrt{\Delta} / 2 -i (\nu_{\mathrm{D}} + k^2 c_{\mathrm{a},d}/k_{\mathrm{D}}^{L})/2.
    \label{eq:multifluid_MHD_dust_ms_wave_1bin}
\end{equation}

The wave propagates more than it is damped if $|\Re(\omega)|>|\Im(\omega)|$. This is the case if $k_{\mathrm{dp}}<k_{\mathrm{dp}}^{\mathrm{p}}<k<k_{\mathrm{hd}}^{\mathrm{p}} <k_{\mathrm{hd}}$, with 
\begin{equation}
    k_{\mathrm{dp}}^{\mathrm{p}},k_{\mathrm{hd}}^{\mathrm{p}} = \frac{k_{\mathrm{D}}^{L}}{c_{\mathrm{a},d}} \sqrt{ \mp \sqrt{\left( \frac{\omega_d}{k_{\mathrm{H}}^{L}} \right)^4 - \left( \frac{c_{\mathrm{a},d} \nu_{\mathrm{D}}}{k_{\mathrm{D}}^{L}}  \right)^2} + \left( \frac{\omega_d}{k_{\mathrm{H}}^{L}} \right)^2}.
    \label{eq:multifluid_MHD_kdp_khd_propagation}
\end{equation}

At smaller scales, when $k > k_{\mathrm{hd}}$, we obtain two purely damped modes ($-\Delta>0$):
\begin{equation}
    \omega = \pm i\sqrt{-\Delta} / 2 -i (\nu_{\mathrm{D}} + k^2 c_{\mathrm{a},d}/k_{\mathrm{D}}^{L})/2.
    \label{eq:multifluid_MHD_ms_small_scales_damped_modes}
\end{equation}

From the eigenvector expression in Eq. \eqref{eq:multifluid_MHD_magnetic_dust_eigenvector} and for $k \to \infty$, we can observe the asymptotic decoupling between the dust and the magnetic perturbation. The $+$ solution implies that $\omega \propto k^0$ and so $v_{d,x} \propto k \delta b$. It is a mode on the dust velocity. The $-$ solution implies that $\omega \propto k^2$, and so $\delta b \propto k v_{d,x}$. It corresponds to the diffusion of the magnetic field (Appendix \ref{app:multifluid_MHD_ms_small_scales_magnetic_mode}).

\subsubsection{Additional geometric mode}

The dynamics of the velocity barycenter of the species (gas, dust, ions and electrons) is driven by forces along $\mathbf{e}_x$ that are induced by the gradients of the thermal pressure and the magnetic pressure. When setting these forces to zero, we find that a translation at a constant velocity along $\mathbf{e}_y$ is solution. More precisely, the eigenvector is given by $v_{g,y}=v_{1,y}=...=v_{\mathcal{N},y} \neq 0$, $v_{g,x}=v_{1,x}=...=v_{\mathcal{N},x}=0$, $\delta \rho_g=0$ and $\delta b =0$. The associated eigenvalue is $\omega =0$. This solution is referred as the mode 0 in Figs. \ref{fig:ms_wave_dispersion_17nm_color}, \ref{fig:ms_wave_dispersion_5bins_100bins_color} and \ref{fig:ms_wave_dispersion_analytical}.

\end{document}